\documentclass{article}

\usepackage{arxiv}

\usepackage[utf8]{inputenc} 
\usepackage[T1]{fontenc}    
\usepackage{hyperref}       
\usepackage{url}            
\usepackage{booktabs}       
\usepackage{amsfonts}       
\usepackage{nicefrac}       
\usepackage{microtype}      
\usepackage{lipsum}		
\RequirePackage{graphicx}
\usepackage[usenames,dvipsnames]{xcolor}
\usepackage{doi}

\RequirePackage{amsthm,amsmath,amsfonts,amssymb}
\RequirePackage[authoryear]{natbib}

\usepackage{booktabs}
\RequirePackage{multirow}   
\usepackage{comment}
\usepackage[figuresright]{rotating}
\usepackage{rotating}
\newsavebox\CBox
\def\textBF#1{\sbox\CBox{#1}\resizebox{\wd\CBox}{\ht\CBox}{\bf{#1}}}

\title{Scalable Multiple Network Inference With The Joint Graphical Horseshoe}

\date{} 					

\author{ \href{https://orcid.org/0000-0003-2701-5686}{\includegraphics[scale=0.06]{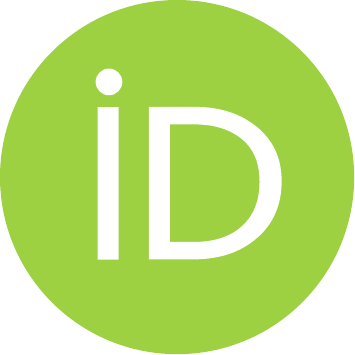}\hspace{1mm}Camilla~Lingj\ae rde}\\
	MRC Biostatistics Unit\\
	University of Cambridge\\
	\texttt{camilla.lingjaerde@mrc-bsu.cam.ac.uk} \\
	\And
	\href{https://orcid.org/0000-0001-7413-5002}{\includegraphics[scale=0.06]{orcid.pdf}\hspace{1mm}Benjamin P.~Fairfax} \\
	MRC Weatherall Institute for Molecular Medicine\\
	University of Oxford\\
	\texttt{benjamin.fairfax@oncology.ox.ac.uk} \\
	\And
	\href{https://orcid.org/0000-0003-1998-492X}{\includegraphics[scale=0.06]{orcid.pdf}\hspace{1mm}Sylvia~Richardson} \\
	MRC Biostatistics Unit\\
	University of Cambridge\\
	\texttt{sylvia.richardson@mrc-bsu.cam.ac.uk} \\
	\And
	\href{https://orcid.org/0000-0002-7113-2540}{\includegraphics[scale=0.06]{orcid.pdf}\hspace{1mm}Hélène~Ruffieux} \\
	MRC Biostatistics Unit\\
	University of Cambridge\\
	\texttt{helene.ruffieux@mrc-bsu.cam.ac.uk} \\
}

\renewcommand{\headeright}{}
\renewcommand{\undertitle}{}
\renewcommand{\shorttitle}{The Joint Graphical Horseshoe}

\hypersetup{
pdftitle={Scalable Multiple Network Inference With The Joint Graphical Horseshoe},
pdfsubject={stat.ME},
pdfauthor={Camilla~Lingj\ae rde, Benjamin P.~Fairfax,Hélène~Ruffieux,Sylvia~Richardson},
pdfkeywords={Bayesian graphical models, Cancer genomics, Expectation conditional maximisation, Gene networks, Genomics, Graphical horseshoe, Horseshoe prior, High-dimensional inference, Integrative analysis, Multi omics, Network models},
}

\begin{document}
\maketitle

\begin{abstract}
   Network models are useful tools for modelling complex associations. In statistical omics, such models are increasingly popular for identifying and assessing functional relationships and pathways. If a Gaussian graphical model is assumed, conditional independence is determined by the non-zero entries of the inverse covariance (precision) matrix of the data. The Bayesian graphical horseshoe estimator provides a robust and flexible framework for precision matrix inference, as it introduces local, edge-specific parameters which prevent over-shrinkage of non-zero off-diagonal elements. However, its applicability 
   is currently limited in statistical omics settings, which often involve high-dimensional data from multiple conditions that might share common structures. We propose (i) a scalable expectation conditional maximisation (ECM) algorithm for obtaining a posterior estimate of the precision matrix in the graphical horseshoe, and (ii) a novel joint graphical horseshoe estimator, which borrows information across multiple related networks to improve estimation. We show numerically %
   that our single-network ECM approach is more scalable than the existing graphical horseshoe Gibbs implementation, while achieving the same level of accuracy. We also show that our joint-network proposal successfully leverages shared edge-specific information between networks while still retaining differences, outperforming state-of-the-art methods at any level of network similarity. Finally, we leverage our approach to clarify gene regulation activity within and across immune stimulation conditions in monocytes, and formulate hypotheses on the pathogenesis of immune-mediated diseases.  
\end{abstract}


\section{Introduction}

In statistical omics, network models are increasingly popular for representing complex associations and assessing pathway activity. With such models, the links between genes, proteins or other types of omics data can be represented and studied, providing valuable insight into functional relationships. The progress of high-throughput genomic technologies has led to the collection of large, genome-wide data sets, and the availability of biomeasurements of different types has enabled the development of integrative modelling approaches which can increase statistical power while providing detailed insight into complex biological mechanisms (\cite{someren2002, karczewski2018integrative}).

If a Gaussian graphical model is assumed, an association (conditional independence) network can be estimated by determining the non-zero entries of the inverse covariance (precision) matrix of the data. There is significant literature on this problem, both frequentist and Bayesian. Notable frequentist methods include  the neighbourhood selection (\cite{Meinshausen06}), the graphical lasso (\cite{friedman2008}) and the graphical SCAD (\cite{fan2009}). In later years, Bayesian methods such as the Bayesian graphical lasso (\cite{wang2012bayesian}), Bayesian spike-and-slab approaches (\cite{wang2015spike}) and the graphical horseshoe (\cite{li2019graphical}) have gained popularity. The Bayesian model formulation of the graphical horseshoe leads to many desirable properties, particularly for identifying weak edges. Indeed, the global-local horseshoe prior it relies on permits the introduction of edge-specific parameters that prevent over-shrinkage of non-zero off-diagonal elements, resulting in a highly flexible framework. However, in the high-dimensional settings commonly needed for investigating biological networks, the Gibbs sampling implementation proposed by  \cite{li2019graphical} becomes computationally inefficient or even unfeasible. Moreover, the graphical horseshoe has only been formulated for a single network, whereas interest has grown in the network analysis of multiple data sets that might share common structures. In biomedical applications, such related data sets could be different tissues, conditions or patient subgroups, or different omics types, such as gene levels and the protein levels encoded by these genes. A joint approach that utilises the common information while preserving the differences will both have increased statistical power and provide insight into the different mechanisms in play. 

In the field of multiple Gaussian graphical models, notable frequentist methods include the joint graphical lasso (\cite{danaher2014}), which enforces similar graphical structures by solving a penalised likelihood problem, and a group extension of the graphical lasso to multiple networks (\cite{guo2011joint}). In a Bayesian framework, \cite{peterson2015bayesian} propose a Gibbs sampling approach which uses a Markov random field prior on multiple graphs to learn the similarity of graphical structures. 
Other notable Bayesian approaches include two expectation-maximisation (EM) approaches, namely, the Bayesian spike-and-slab joint graphical lasso (\cite{li2019bayesian}), which builds on seminal work by \cite{wang2015spike} and formulates a multiple-network model based on a Gaussian spike-and-slab prior, and  GemBag (\cite{yang2021gembag}), which relies on a spike-and-slab prior with Laplace distributions. For a more thorough review on existing Bayesian graphical methods, we refer to \cite{ni2022bayesian}.

The method presented in this paper is, to our knowledge, the first to adapt graphical modelling based on global-local priors to the multiple network setting. We propose an expectation conditional maximisation (ECM) algorithm for obtaining a posterior estimate of the precision matrix in the graphical horseshoe which allows us to tackle network inference in biological problems of realistic sizes. Building on this efficient implementation, we formulate a joint model that permits borrowing information between multiple networks using the graphical horseshoe prior. 
We provide the two R packages \texttt{fastGHS} and \texttt{jointGHS}, which implement the single and joint methods respectively. 

The paper is organised as follows. In Section \ref{sec:problem}, we recall the classical graphical horseshoe estimator of \cite{li2019graphical} and discuss the advantages of the horseshoe prior in graphical settings. %
In Section \ref{sec:motivation}, we motivate the need for further development to enable practical and meaningful inference at scale by introducing a %
multi-condition gene regulation study in monocytes. 
In Section \ref{sec:ECM}, we describe our new inference procedure for the single-network graphical horseshoe, and in Section \ref{sec:multiplenet}, we present our joint graphical horseshoe model formulation. %
In Section \ref{sec:sims}, we demonstrate the performance of our proposed methodology on simulated data, and in Section \ref{sec:applications}, we apply it to the monocyte gene regulation study. Finally, we highlight possible extensions in Section \ref{sec:discussion}. 

\section{Problem statement}
\label{sec:problem}

Consider a network model where each node is associated with some measurable attribute. Observed values of the multivariate random vector $\boldsymbol{x} = (X_1, \ldots, X_{p})^T$ of node attributes, each entry corresponding to one of $p$ variables, can then be used to infer a graph under suitable model assumptions. Given multivariate Gaussian node attributes, with $n\times p$ observation matrix $\boldsymbol{X}$ with i.i.d. rows $\boldsymbol{x_1}, \ldots, \boldsymbol{x_n} \sim \mathcal{N}(\boldsymbol{0},\boldsymbol{\Sigma})$, we can infer a \emph{partial correlation network} by estimating the inverse covariance matrix, or precision matrix, $\boldsymbol{\Theta}=\boldsymbol{\Sigma}^{-1}$. The partial correlation between nodes $i$ and $j$ conditioned upon all others is then given by 
\begin{equation}
\label{eq:conditionalcor}
\rho_{ij\mid V\backslash \{i,j\} } = - \frac{\theta_{ij}}{\sqrt{\theta_{ii}\theta_{jj}}}, \nonumber
\end{equation}
where the $\theta_{ij}$'s are the entries of $\boldsymbol{\Theta}$ and $V$ is the set of all node pairs (\cite{lauritzen1996graphical}). For Gaussian variables, correlation equal to zero is equivalent to independence, which implies that a conditional independence graph can be constructed by determining the non-zero entries of the precision matrix. The graph is assumed to be sparse, i.e., %
the number of edges in the edge set $E$ relative to the number of potential edges in the graph, $2\vert E\vert/(p^2-p)$, is small. The precision matrix must also necessarily be positive definite, $\boldsymbol{\Theta} \succ 0$. %

\cite{li2019graphical} have recently proposed the \emph{graphical horseshoe} to obtain a sparse estimate of the precision matrix $\boldsymbol{\Theta}$, 
repurposing the horseshoe prior initially introduced by \cite{carvalho2010horseshoe} in the normal means setting to a graphical setting. The graphical horseshoe model puts horseshoe priors on the off-diagonal elements of the precision matrix, encouraging sparse solutions. An uninformative prior is put on the diagonal elements, and the positive definiteness constraint is respected. Due to symmetry, it is sufficient to consider the upper off-diagonal elements of $\boldsymbol{\Theta}$. Normal scale mixtures with half-Cauchy hyperpriors are used on the off-diagonal elements. The hierarchy of the model is as follows:
\begin{align}
\label{eq:horseshoehierarchy}
    \theta_{ii} &\propto 1, \hspace{2.7cm} i = 1, \ldots, p, \nonumber\\
    \theta_{ij} \mid \lambda_{ij} &\sim \mathcal{N}(0,\lambda_{ij}^2\tau^2), \hspace{1cm} 1 \leq i < j \leq p,\\
    \lambda_{ij} &\sim \text{C}^+(0,1), \nonumber
\end{align}
with $\boldsymbol{\Theta}\succ 0$ and where $\text{C}^+(0,1)$ is the half-Cauchy distribution with density $p(x) \propto (1+x^2)^{-1}, \ x>0$. A key feature of the horseshoe prior in (\ref{eq:horseshoehierarchy}) is the presence of local shrinkage parameters, $\lambda_{ij}$, %
which are meant to flexibly capture edge-specific effects with no or very limited over-shrinkage, while the global parameter $\tau$ is set to ensure overall sparsity. In the next section, we motivate the extension of the graphical horseshoe to a multiple network setting, with a scalable implementation.

\section{Data and motivating example}
\label{sec:motivation}

A wide array of prevalent diseases, such as inflammatory bowel disease, rheumatoid arthritis and cancer, are believed to result from an inappropriate immune activity and  consequent inflammation. It is now established that exposing monocyte cell cultures to specific stimuli can create conditions that resemble certain immune-mediated disease states (\cite{biswas2010macrophage}). Indeed, different stimuli may activate different cellular or molecular pathways that contribute to disease development. Hence, systematically investigating the impact of different types of immune stimulation on gene regulatory activity – and pinpointing the mechanisms that are common to several stimuli or stimulus specific – can provide valuable insights into the pathophysiology of such diseases. 

We propose to contribute to this research by analysing graphical structures from a detailed gene expression dataset where primary CD14$^+$ monocytes in $432$ healthy European individuals were exposed to different types of immune stimulation (hereafter ``conditions'').  %
Specifically, monocyte expression was quantified using Illumina HumanHT-12 v4 BeadChip arrays before and after immune stimulation via exposition to inflammation proxies, namely, $\text{interferon-}\gamma$ ($\text{IFN-}\gamma$) or differing durations of lipopolysaccharide (LPS 2h or LPS 24h). 
The number of samples available in each condition is 
 $n_{\text{unstim}}=413$, $n_{\text{IFN-}\gamma}=366$, $n_{\text{LPS2h}}=260$, $n_{\text{LPS24h}}=321$ for unstimulated cells, and $\text{IFN-}\gamma$-, LPS 2h- and LPS 24h- stimulated cells, respectively (\cite{fairfax2014innate}).

Examining the effect of genetic variation on gene regulatory activity after stimulation can help pinpointing gene sets implicated in disease risk and development. Indeed, previous studies %
suggest that gene stimulation triggers regulatory activity that leads to a beneficial environment for \emph{hotspots} to establish  (\cite{fairfax2014innate, lee2014common, kim2014characterizing, ruffieux2020global}) – hotspots are genetic variants regulating large numbers of genes, thereby potentially representing important %
players in disease mechanisms (\cite{yao2017dynamic}). We will therefore focus our study to networks of genes under hotspot control, in the different conditions.

While \cite{fairfax2014innate} mainly report condition-specific gene regulatory activities, they also observe effects across all conditions.
The largest hotspot identified by \cite{ruffieux2020global} is persistent across all four conditions. Specifically, using their global-local hotspot modelling approach ATLASQTL, they found that the genetic variant rs6581889 was the top hotspot in the $\text{IFN-}\gamma$, unstimulated and LPS 2h studies (associated with $333$, $242$ and $96$ transcripts, respectively), and it was the second largest hotspot in the LPS 24h study (associated with $18$ transcripts). This hotspot is located on chromosome 12, only a few Kb away from two genes it controls, namely, \emph{LYZ} and \emph{YEATS4}, 
which are thought to play a central role in the pathogenesis of immune disorders (\cite{fairfax2012genetics}). %

Given the shared hotspot control in all four immune stimulation conditions, borrowing information with a joint network approach on the controlled genes across these conditions seems particularly appropriate to further investigate the gene regulation mechanisms triggered by rs6581889. %
At the same time, \cite{ruffieux2020global} found that there is only a partial overlap between the genes associated with the hotspot across the different conditions, which calls for a modelling approach that can also effectively detect stimulus-specific effects – such effects being highly relevant to understand how distinct pathways may be activated in different disease states. %
Such an analysis should help characterising the complex interplays among the genes controlled, i.e., the direct effects on the distal genes or the indirect effects, mediated via other genes controlled by the hotspot (typically via proximal genes, such as \emph{LYZ} and \emph{YEATS4}). Although graphical modelling approaches seem particularly appropriate to disentangle direct and mediated effects, they have not been employed thus far.

With its desirable theoretical properties as well as its high performance in numerical studies, 
the graphical horseshoe estimator (\cite{li2019graphical}) would be a natural choice for graph inference. %
However its applicability in the above monocyte setting is hampered by two limitations: first, %
its Gibbs sampler implementation does not scale to the problem dimensions ($p=381$ genes for $n\leq 413$ observations in each condition). Second, its model is formulated for the analysis of a single network, meaning that it can only be applied separately to each of the four conditions. %
Although relevant for identifying common structures across the conditions, the Bayesian spike-and-slab joint graphical lasso (\cite{li2019bayesian}) and the joint graphical lasso (\cite{danaher2014}) do not enjoy the flexibility granted
 by the horseshoe prior's local scales for identifying network-specific effects (whose detection is key to disentangle disease-specific mechanisms as explained above). Moreover, these methods did not reach convergence %
 within %
$48$ hours, as jointly modelling all four conditions requires inferring a total of $289\,560$ edges. %
This severe high dimensionality makes  essential to develop approaches that are specifically designed to scale to the current statistical omics problem sizes. %

Motivated by the monocyte problem, this work is concerned with proposing a new framework that addresses the two shortcomings outlined above to enable effective network inference in realistic practical settings. Namely, we aim to (i) develop an expectation conditional maximisation (ECM) algorithm for the graphical horseshoe as a %
fast yet accurate alternative to the Gibbs sampling procedure proposed by \cite{li2019graphical}, and (ii) formulate a joint model for multiple networks, leveraging the global-local horseshoe feature to borrow strength across shared patterns \emph{while} preserving differences across networks.

Equipped with this framework, we will return to the monocyte problem in Section \ref{sec:applications} to demonstrate the computational feasibility of joint network modelling on these data and exploit the advantages of the global-local formulation for inferring and interpreting condition-specific and shared gene regulation structures. %

\section{An ECM algorithm for estimating the graphical horseshoe}
\label{sec:ECM}

As a first step in detailing our proposal, we outline the expectation conditional maximisation (ECM) procedure, which adapts the spike-and-slab EM approach of \cite{rovckova2014emvs} to the graphical horseshoe setting. We first detail the updates for the single-network graphical horseshoe prior (\ref{eq:horseshoehierarchy}). 
The ECM approach, described first by \cite{meng1993maximum}, is a generalised EM algorithm (\cite{dempster1977maximum}) where a complex maximisation step (M-step) is replaced with several computationally simpler conditional maximisation steps (CM-steps). 

\subsection{Full conditional posteriors}

As in \cite{li2019graphical}, the full conditional posteriors of the local $\lambda_{ij}$'s can be derived by introducing the augmented variables $\nu_{ij}$. We next employ the following reparameterisation, introducing the latent $\nu_{ij}$ and writing
\begin{align}
    \lambda_{ij}^2  \mid \nu_{ij} & \sim \text{InvGamma}(1/2, 1/\nu_{ij}), \qquad 1 \leq i < j \leq p,\nonumber \\
    \nu_{ij} & \sim \text{InvGamma}(1/2, 1). \nonumber
\end{align}
Using a key observation from \cite{makalic2015simple}, we find the full conditional posteriors as 
\begin{align}
\label{eq:lambda_nu_fullconditionals}
    \lambda_{ij}^2 \mid \cdot & \sim \text{InvGamma}(1, 1/\nu_{ij} + \theta_{ij}^2/(2\tau^2)), \nonumber \\
    \nu_{ij} \mid \cdot & \sim \text{InvGamma}(1, 1+1/\lambda_{ij}^2),
\end{align}
where $\cdot$ denotes all other variables. The latent variables can be collected in the latent matrix $\boldsymbol{N} = (\nu_{ij})$. The global shrinkage parameter $\tau$ in (\ref{eq:horseshoehierarchy}) is for now treated as a fixed hyperparameter; its specification will be detailed in %
Section \ref{subsec:tau_select}.
To obtain conditional posteriors for the precision matrix and the local scale parameters, each column and row of the matrices $\boldsymbol{\Theta}$ and $\boldsymbol{\Lambda} = (\lambda_{ij}^2)$ are partitioned from a $p \times p$ matrix of parameters. Without loss of generality, we describe the updates for the last row and column. As in \cite{wang2012bayesian}, we write %
\begin{align}
    & \boldsymbol{\Theta} =  \begin{pmatrix} 
        &\boldsymbol{\Theta}_{(-p)(-p)} &\boldsymbol{\theta}_{(-p)p} \\
        &\boldsymbol{\theta}_{(-p)p}^T &\theta_{pp}
    \end{pmatrix}, \ 
    \boldsymbol{S} =  \begin{pmatrix} 
        &\boldsymbol{S}_{(-p)(-p)} & \boldsymbol{s}_{(-p)p} \\
        &\boldsymbol{s}_{(-p)p}^T & s_{pp}
    \end{pmatrix}, \nonumber \\
    & \boldsymbol{\Lambda} =  \begin{pmatrix} 
        &\boldsymbol{\Lambda}_{(-p)(-p)} & \boldsymbol{\lambda}_{(-p)p} \\
        &\boldsymbol{\lambda}_{(-p)p}^T & 1 \nonumber
    \end{pmatrix},
\end{align}  
where $\boldsymbol{S}=\boldsymbol{X}^T\boldsymbol{X}$ is the scatter matrix of the observed data $\boldsymbol{X}$. The diagonal elements of $\boldsymbol{\Lambda}$ are not of relevance and can be set to an arbitrary value such as $1$. The posterior distribution for the last column (and row) of $\boldsymbol{\Theta}$ can be obtained as 
\begin{align}
    p(\boldsymbol{\theta}_{(-p)p}, \theta_{pp} \mid \boldsymbol{\Theta}_{(-p)(-p)}, \boldsymbol{S}, \boldsymbol{\Lambda})  
    &\propto (\theta_{pp} - \boldsymbol{\theta}_{(-p)p}^T \boldsymbol{\Theta}_{(-p)(-p)}^{-1}\boldsymbol{\theta}_{(-p)p})^{n/2}  \nonumber \\
    & \quad \  \times \exp{\left\{ -\boldsymbol{s}_{(-p)p}^T \boldsymbol{\theta}_{(-p)p} - s_{pp} \theta_{pp}/2 - \boldsymbol{\theta}_{(-p)p}^T(\boldsymbol{\Lambda}^* \tau^2)^{-1}\boldsymbol{\theta}_{(-p)p}/2 \right\}}. \nonumber
\end{align}
With a variable change, the conditional distributions can be reformulated as
\begin{align}
    \label{eq:theta_conditionaldist}
     \boldsymbol{\theta}_{(-p)p} \mid \boldsymbol{\Theta}_{(-p)(-p)}, \boldsymbol{S}, \boldsymbol{\Lambda} & \sim \text{Normal}(-C \boldsymbol{s}_{(-p)p}, C) \nonumber, \\
     \theta_{pp} - \boldsymbol{\theta}_{(-p)p}^T \boldsymbol{\Theta}_{(-p)(-p)}^{-1}\boldsymbol{\theta}_{(-p)p} \mid \boldsymbol{\Theta}_{(-p)(-p)}, \boldsymbol{S}, \boldsymbol{\Lambda} & \sim \text{Gamma}(n/2 + 1, s_{pp}/2), 
\end{align}
where $C = \{ s_{pp}\boldsymbol{\Theta}_{(-p)(-p)}^{-1} + (\boldsymbol{\Lambda}^* \tau^2)^{-1} \}^{-1}$ and $\boldsymbol{\Lambda}^* = \text{diag}(\boldsymbol{\lambda}_{(-p)p})$. By iteratively permuting each row and column to be the last, the conditional posterior of all elements of the precision matrix can then be found row and column wise. 

\subsection{ECM algorithm}

Given the estimates from the previous iteration $l$, the objective function is obtained as
\begin{align}
\label{eq:ObjectiveFunc}
    Q(\boldsymbol{\Theta}, \boldsymbol{\Lambda} \mid \boldsymbol{\Theta}^{(l)}, \boldsymbol{\Lambda}^{(l)}) = & \text{E}_{\boldsymbol{N}\mid \boldsymbol{\Theta}^{(l)}, \boldsymbol{\Lambda}^{(l)}, \boldsymbol{S}}\left\{\log p(\boldsymbol{\Theta}, \boldsymbol{\Lambda},  \boldsymbol{N} \mid \boldsymbol{S}) \mid \boldsymbol{\Theta}^{(l)}, \boldsymbol{\Lambda}^{(l)}\right\} \nonumber \\
    = & \frac{n}{2}\log{(\det{\boldsymbol{\Theta}})} - \frac{1}{2}\text{tr}(\boldsymbol{S}\boldsymbol{\Theta}) + \sum_{i<j} \Big\{ -4 \log{(\lambda_{ij})} - \frac{\theta_{ij}^2}{2\tau^2\lambda_{ij}^2}  \\
    & - 2 \text{E}_{\cdot \mid \cdot} \left\{ \log{(\nu_{ij})}\right\} 
     - \bigg( \frac{1}{\lambda_{ij}^2} +1 \bigg)\text{E}_{\cdot \mid \cdot} \left( \frac{1}{\nu_{ij}}\right)\Big\} +\text{const.}, \nonumber
\end{align}
where $\text{E}_{\cdot \mid \cdot}(\cdot)$ denotes $\text{E}_{\boldsymbol{N} \mid \boldsymbol{\Theta}^{(l)}, \boldsymbol{\Lambda}^{(l)}, \boldsymbol{S}}(\cdot)$ and const. is a constant not depending on $\boldsymbol{\Theta}$ or $\boldsymbol{\Lambda}$. Note that the objective function accounts for the full priors of $\boldsymbol{\Theta}$ and $\boldsymbol{\Lambda}$.

In the E-step of the algorithm, the conditional expectations in (\ref{eq:ObjectiveFunc}) are computed, and the CM-step performs the maximisation with respect to $(\boldsymbol{\Theta}, \boldsymbol{\Lambda})$. 
 Similarly to the Bayesian spike-and-slab joint graphical lasso of \cite{li2019bayesian} where the objective function is maximised over both the precision matrix and sparsity parameters, this approach finds the posterior mode of $(\boldsymbol{\Theta}, \boldsymbol{\Lambda})$, accounting for prior distributions on all parameters.

From (\ref{eq:lambda_nu_fullconditionals}), the full conditional distributions of the $\nu_{ij}$'s are inverse Gamma. Therefore, the E-step updates are:
\begin{align}
\label{eq:Estep}
    &\text{E}_{\cdot \mid \cdot} \left\{ \log{(\nu_{ij})}\right\} = \log{\bigg( 1 + \frac{1}{\lambda_{ij}^{2^{(l)}}} \bigg)} - \psi(1), \nonumber \\
     &\text{E}_{\cdot \mid \cdot} \left( \frac{1}{\nu_{ij}}\right) = \frac{1}{1+1/\lambda_{ij}^{2^{(l)}}} = \frac{\lambda_{ij}^{2^{(l)}}}{\lambda_{ij}^{2^{(l)}}+1} =:\lambda_{ij}^{*^{(l)}}, 
\end{align}
where $\psi(\cdot)$ is the digamma function. 

Next, the CM-step maximises (\ref{eq:ObjectiveFunc}) with respect to $(\boldsymbol{\Theta}, \boldsymbol{\Lambda})$ in a coordinate ascent fashion, with the expectations replaced with the expressions found in (\ref{eq:Estep}). The following closed-form updates are obtained for the $\lambda_{ij}^2$'s
\begin{align}
\label{eq:lambda_update}
    \lambda_{ij}^{2^{(l+1)}} = \frac{\lambda_{ij}^{*^{(l)}} + \theta_{ij}^2/(2\tau^2)}{2}. 
\end{align}
There is no closed form for the update of the precision matrix, however, (\ref{eq:theta_conditionaldist}) gives the updates for the last row and column of $\boldsymbol{\Theta}$:
\begin{align}
\label{eq:thetaupdate}
    \theta_{pp}^{(l+1)} &= \boldsymbol{\theta}_{(-p)p}^{{(l+1)}^T}\big(\boldsymbol{\Theta}_{(-p)(-p)}^{{(l+1)}}\big)^{-1} \boldsymbol{\theta}_{(-p)p}^{{(l+1)}} + \frac{n}{s_{pp}}, \nonumber \\
    \boldsymbol{\theta}_{(-p)p}^{{(l+1)}} &= -\left\{s_{pp} \big(\boldsymbol{\Theta}_{(-p)(-p)}^{{(l+1)}}\big)^{-1} + \frac{1}{\tau^{2}} \big(\boldsymbol{\Lambda}^{*^{(l+1)}}\big)^{-1}\right\}^{-1} \boldsymbol{s}_{(-p)p},
\end{align}
setting $l+1=l$ at each iteration. By iteratively permuting each row and column to be the last, all elements of the precision matrix can be updated row and column wise. With these updates, the positive definiteness constraint for $\boldsymbol{\Theta}$ is maintained at each iteration as long as the initial value is positive definite. This can be shown with an argument equivalent to that of \cite{wang2012bayesian}: assume that the update $\boldsymbol{\Theta}^{(l)}$ is positive definite, then all its $p$ leading principal minors are positive. After updating the last row and column as in (\ref{eq:thetaupdate}), the updated precision matrix $\boldsymbol{\Theta}^{(l+1)}$ has the same leading principal minors as in $\boldsymbol{\Theta}^{(l)}$ except for the last one, which is of order $p$. The last leading principal minor is clearly equal to $\det(\boldsymbol{\Theta}^{(l+1)}) = \gamma \det(\boldsymbol{\Theta}_{(-p)(-p)}^{{(l)}})$, where $\det(\boldsymbol{\Theta}_{(-p)(-p)}^{{(l)}})$ is the $(p-1)^{\text{th}}$ leading principal minor of $\boldsymbol{\Theta}^{{(l)}}$ and thus positive, and we have, from (\ref{eq:theta_conditionaldist}), that $\gamma = \theta_{pp} - \boldsymbol{\theta}_{(-p)p}^T \boldsymbol{\Theta}_{(-p)(-p)}^{-1}\boldsymbol{\theta}_{(-p)p}>0$. Consequently, $\det(\boldsymbol{\Theta}^{(l+1)})>0$ and so the updated $\boldsymbol{\Theta}^{(l+1)}$ is positive definite.

With this CM-step update, it is ensured that $Q(\boldsymbol{\Theta}^{(l+1)}, \boldsymbol{\Lambda}^{(l+1)} \mid \boldsymbol{\Theta}^{(l)}, \boldsymbol{\Lambda}^{(l)}) \geq Q(\boldsymbol{\Theta}^{(l)}, \boldsymbol{\Lambda}^{(l)} \mid \boldsymbol{\Theta}^{(l)}, \boldsymbol{\Lambda}^{(l)})$ (\cite{meng1993maximum}). By iterating between the E-step and the CM-step until convergence, we obtain an estimator of the posterior mode of $(\boldsymbol{\Theta}, \boldsymbol{\Lambda})$. The full derivations for this section are given in Section S.1 of Supplement A. 

One of the main computational advantages of the ECM approach over stochastic search is that the posterior mode is fast to obtain. The estimates are computed directly and a full stochastic search is not necessary. Further, the entries corresponding to unidentified edges tend to converge to values close to zero and the separation with the identified edges increases as the algorithm converges – such an observation has also been reported by others in the context of EM or variational inference (\cite{kook2021bvar}).
We hereafter refer to this ECM implementation as ``fastGHS''.

\section{Multiple network inference}
\label{sec:multiplenet}

In this section, we describe the \emph{joint graphical horseshoe} for multiple network inference. By sharing information through common latent variables, the method gives more precise estimates for networks with any level of similarity. The heavy tails of the horseshoe prior permits effectively capturing network-specific edges, a property that few Bayesian methods developed for similar purposes share. The resulting joint graphical horseshoe estimator \emph{simultaneously} shares information between networks \emph{and} captures their differences. In addition, due to the scalability of the ECM implementation, our method allows for joint network modelling for a larger number of networks than existing Bayesian approaches do.

\subsection{Joint graphical  horseshoe  model  formulation}

Given $K$ networks with $p$ nodes each, and $n_k\times p$ observation matrices $\boldsymbol{X}_k$ for $k=1,\ldots, K$, we are interested in estimating the precision matrices $\{\boldsymbol{\Theta}_1, \ldots, \boldsymbol{\Theta}_K\}$. We let the $k^{\text{th}}$ precision matrix follow the hierarchical model 
\begin{align}
    \theta_{iik} &\propto 1, \hspace{2.9cm} i = 1, \ldots, p, \nonumber\\
    \theta_{ijk} \mid \lambda_{ijk} &\sim \mathcal{N}(0,\lambda_{ijk}^2\tau_k^2), \hspace{1cm} 1 \leq i < j \leq p,\nonumber\\
    \lambda_{ijk} &\sim \text{C}^+(0,1),\nonumber
\end{align}
with $\boldsymbol{\Theta_k}\succ 0$, $k = 1, \ldots, K$. 
This is the standard graphical horseshoe model for each network separately. To share information across networks, we introduce the latent variables $\nu_{ij}$ and write
\begin{align}
    \lambda_{ijk}^2  \mid \nu_{ij} & \sim \text{InvGamma}(1/2, 1/\nu_{ij}), \nonumber \\
    \nu_{ij} & \sim \text{InvGamma}(1/2, 1).\nonumber
\end{align}
We then derive the full conditional posteriors. Because the $\lambda_{ijk}$'s of the different data sets are independent given the $\nu_{ij}$'s, we have
\begin{align}
    \lambda_{ijk}^2 \mid \cdot & \sim \text{InvGamma}(1, 1/\nu_{ij} + \theta_{ijk}^2/(2\tau_k^2)), \nonumber
\end{align}
similarly to the standard graphical horseshoe. Hence, information is now shared across networks through the common latent variable $\nu_{ij}$, and the full conditional posterior of the $\nu_{ij}$'s now depends on the $\lambda_{ijk}$'s of all $K$ networks:
\begin{align}
\label{eq:nu_fullconditionalsgroups}
    p(\nu_{ij} \mid \cdot ) &\propto \text{InvGamma}\left(\frac{K+1}{2}, 1+\sum_{k=1}^K \frac{1}{\lambda_{ijk}^2}\right).
\end{align}
The derivation of (\ref{eq:nu_fullconditionalsgroups}) is given in Section S.1.4 of Supplement A. 

The global scales $\tau_k$ are network-specific to allow for different sparsity levels across networks; their specification is detailed in Section~\ref{subsec:tau_select}. Alternative approaches that directly model structured and shared sparsity, e.g., using a Markov random field prior as in the spike-and-slab graphical approach of \cite{peterson2015bayesian}, could also be considered. In practice however, the use of local scales results in considerable flexibility in adapting to the overall sparsity levels of the different networks, as our numerical experiments from Sections \ref{sec:sims} and \ref{sec:applications} below suggest. %

\subsection{ECM approach}

The E-step and CM-step of the multiple-network ECM algorithm are similar to the single network version. Since the networks are independent given the common latent variables $\nu_{ij}$, we can perform the maximisation of the $\lambda_{ijk}$'s and the $\theta_{ijk}$'s for $k=1,\ldots, K$ separately. The main difference is that the distribution of 
$\nu_{ij}$ now depends on the local shrinkage parameters of all $K$ networks. 

Using the full conditional distribution (\ref{eq:nu_fullconditionalsgroups}), the E-step updates are: %
\begin{align}
\label{eq:exp_nugroups}
    &\text{E}_{\cdot \mid \cdot} \left\{ \log{(\nu_{ij})}\right\} = \log{\bigg( 1 + \sum_{k=1}^K \frac{1}{\lambda_{ijk}^{2^{(l)}}} \bigg)} - \psi\left(\frac{K+1}{2}\right), \nonumber \\
     &\text{E}_{\cdot \mid \cdot} \left( \frac{1}{\nu_{ij}}\right) = \frac{K}{2\left(1+ \sum_{k=1}^K 1/\lambda_{ijk}^{2^{(l)}} \right)} =: \lambda_{ij\cdot}^{*^{{(l)}}}.
\end{align}

The CM-step updates for the $\lambda_{ijk}^2$'s are obtained by replacing the expectation $\text{E}_{\cdot \mid \cdot}( \nu_{ij}^{-1}) = \lambda_{ij}^{*^{{(l)}}}$ in (\ref{eq:lambda_update}) by the $\lambda_{ij\cdot}^{*^{{(l)}}}$ update of the E-step in (\ref{eq:exp_nugroups}):
\begin{align}
\label{eq:lambda_updategroups}
    \lambda_{ijk}^{2^{(l+1)}} = \frac{\lambda_{ij\cdot}^{*^{(l)}} + \theta_{ijk}^2/(2\tau_k^2)}{2}. 
\end{align}
The precision matrices $\boldsymbol{\Theta}_k$ are also updated separately for each network, as they are independent given the $\nu_{ij}$'s. Setting $l+1=l$ at each iteration, we get the ordinary graphical horseshoe block updates (\cite{li2019graphical}) given by
\begin{align}
    \theta_{ppk}^{(l+1)} &= \boldsymbol{\theta}_{(-p)pk}^{{(l+1)}^T}\big(\boldsymbol{\Theta}_{(-p)(-p)k}^{{(l+1)}}\big)^{-1} \boldsymbol{\theta}_{(-p)pk}^{{(l+1)}} + \frac{n_k}{s_{ppk}}, \nonumber \\
    \boldsymbol{\theta}_{(-p)pk}^{{(l+1)}} &= -\big(s_{ppk} \big(\boldsymbol{\Theta}_{(-p)(-p)k}^{{(l+1)}}\big)^{-1} + \frac{1}{\tau_k^{2}} \big(\boldsymbol{\Lambda}_k^{*^{(l+1)}}\big)^{-1}\big)^{-1} \boldsymbol{s}_{(-p)pk},\nonumber
\end{align}
where the matrix partitioning is analogous to (\ref{eq:thetaupdate}). We iterate between the E-step and the CM-step until convergence is achieved for all $K$ graphs, when the updates for all precision matrix elements of all $K$ graphs differ from their previous estimate in absolute value by less than some tolerance threshold. We hereafter refer to this ECM implementation of our joint graphical network model as ``jointGHS''.

\subsection{Global shrinkage parameter selection}
\label{subsec:tau_select}

The specification of the horseshoe global scale parameter $\tau$ (or $\tau_k$ in the multiple network case) has been a subject of active debate over the past years (see, e.g., \cite{carvalho2009handling, carvalho2010horseshoe, piironen2017}). The different proposals may be grouped into three strategies: (i) use a prior on $\tau$, typically a half-Cauchy prior, (ii) fix it, or (iii) use a selection criterion. Previous work has found that strategy (i) can result in degenerate solutions when using deterministic inference algorithms (such as our ECM algorithm) or other empirical Bayes procedures, in very sparse settings (\cite{scott2010bayes, polson2010shrink, bhadra2019lasso, van2019learning}). %
As \cite{li2019graphical} indicate, strategy (ii) can be employed to control the sparsity level of the graphical horseshoe estimates and avoid over-shrinkage. This is a common approach in non-graphical horseshoe settings. For instance, \cite{van2014horseshoe} fix $\tau$ based on theoretical justification, namely $\tau$ should be of the order of the proportion of non-null effects to guarantee asymptotic minimaxity, an argument that \cite{bhadra2017horseshoe+} also follow in the context of the horseshoe+ estimator. \cite{piironen2017} instead proposed to make assumptions on the ``effective model size''; their approach is however not transferable to our graphical setting due to the iterative nature of our updates (\ref{eq:thetaupdate}). We instead implement a procedure based on strategy (iii), as detailed hereafter. 

Assuming the multiple network setting, we propose to select each $\tau_k^2$ separately for each network using the AIC criterion for Gaussian graphical models (\cite{akaike1973}), before running the joint analysis. As demonstrated in Section S.5 of Supplement A, fastGHS and jointGHS typically do not over-select edges, and using more stricter criteria such as the BIC would result in severe under-selection of edges. For a given global shrinkage parameter $\tau_k^2$ and corresponding precision matrix estimate $\widehat{\boldsymbol{\Theta}}_{k,\tau_k^2}$ found with fastGHS, the AIC score is given by 

\begin{equation}
    \text{AIC}(\tau_k^2) = \frac{n_k}{n_k-1} \text{tr}(\boldsymbol{S}_k\widehat{\boldsymbol{\Theta}}_{k,\tau_k^2}) - n_k \log\left\{\text{det}(\widehat{\boldsymbol{\Theta}}_{k,\tau_k^2})\right\} +  2\vert E_{\tau_k^2} \vert, \nonumber
\end{equation}
where tr is the trace, $\boldsymbol{S}_k = \boldsymbol{X}_k^T\boldsymbol{X}_k$ is the scatter matrix and $\vert E_{\tau_k^2} \vert$ is the size of the corresponding edge set.

For small $\tau_k^2$, small increases lead to large changes in the AIC score (see Section S.5 of Supplement A). However, for sufficiently large values, the AIC score stabilises as the global shrinkage parameter increases. This can be attributed to the flexibility of the local scale parameters, which compensate for the larger global scale parameter, thus still effectively capturing the magnitude of local effects. Hence, instead of attempting to identify the globally AIC minimising value of $\tau_k^2$, which is computationally expensive, we start with a small value and increase it until the AIC has stabilised. This approach shares similarities with the ``dynamic posterior search'' of \cite{rovckova2018spike}. Formally, using a suitable grid of $M$ increasing values $\{\tau_{k,1}^2, \ldots, \tau_{k,M}^2\}$, we set $\tau_{k}^2$ to be

\begin{equation}
    \tau_{k,\text{AIC}}^2 = \min \Big \{ \tau_{k,m}^2 \colon \vert \text{AIC}(\tau_{k,m}^2)-\text{AIC}(\tau_{k,m-1}^2)  \vert  < \epsilon \Big \}, \nonumber
\end{equation}
for some convergence tolerance $\epsilon$. 

By selecting the $\tau_k$'s separately, we allow for different sparsity levels across networks. 
In practice, our jointGHS implementation runs the single-network approach on each network separately (optionally in parallel for computational efficiency), sets $\tau_k$ using the above procedure, and then uses them in the final joint network run. 

\subsection{More on the heavy horseshoe tail}
\label{subsec:heavytail}

\begin{figure}
    \centering
    \includegraphics[scale=0.35]{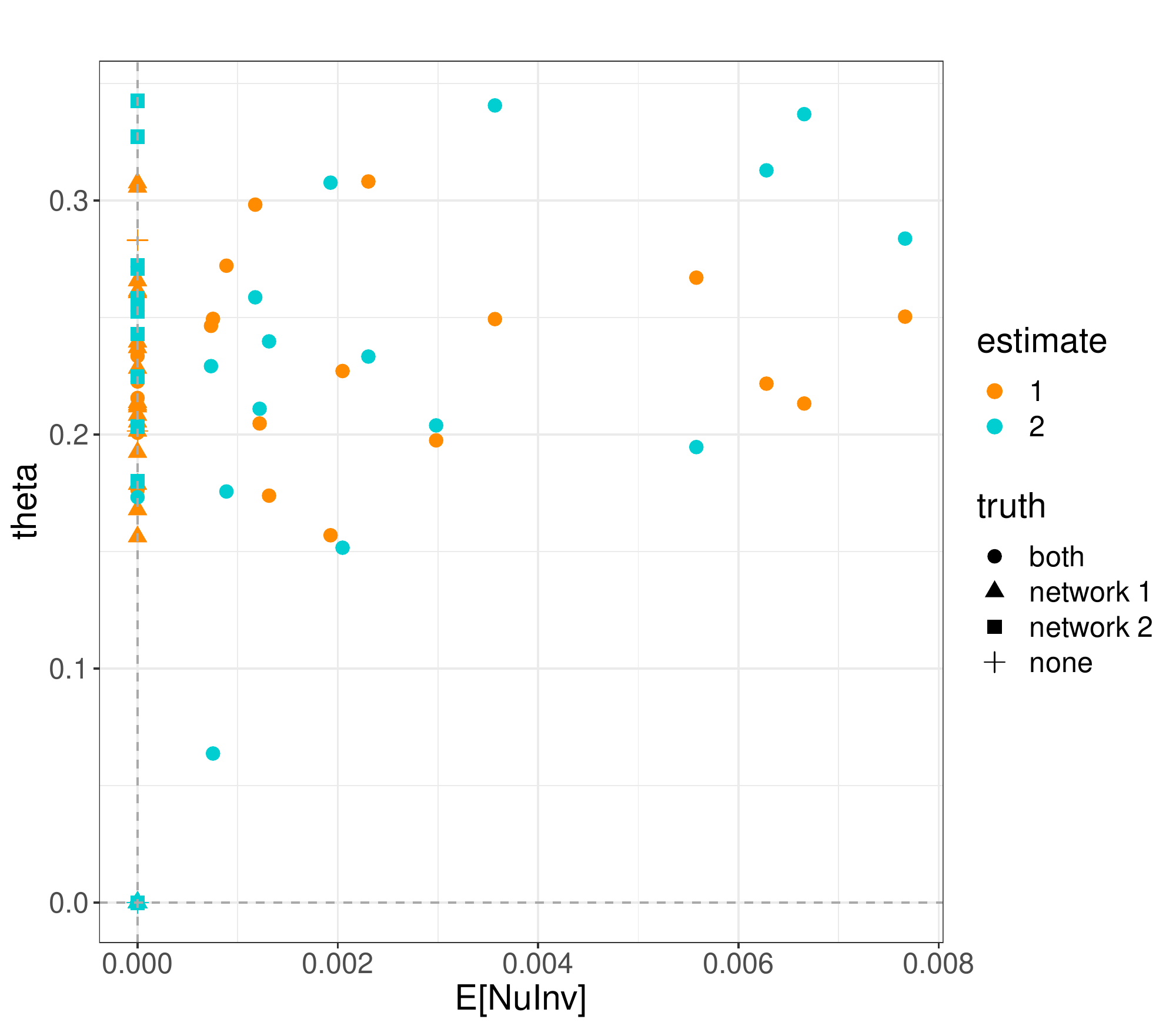}
    \caption{Off-diagonal precision matrix elements estimated by the joint graphical horseshoe, plotted against the expectation of the corresponding inverse shared latent parameters for local scales. The data sets consist of $K=2$ networks with $40\%$ edge agreement, with $p=100$ nodes, and $n_1=100$ and $n_2=150$ observations respectively. Both networks have simulated sparsity of $0.02$. The points are colored according to the network for which they are estimated, and the shape indicates whether the corresponding edge truly is present in both, one or neither networks. The details of the data generation and analysis are given in Section S.3 of Supplement A.}
    \label{fig:theta_vs_nu}
\end{figure}

In the joint graphical horseshoe, information is shared through the common latent parameter $\nu_{ij}$. In practice, only the full conditional expectation of $1/\nu_{ij}$ is used in the ECM algorithm. The larger $1/\nu_{ij}$ in (\ref{eq:exp_nugroups}), the larger the CM-updates (\ref{eq:lambda_updategroups}) for the local scales $\lambda_{ijk}$, $k=1,\ldots,K$, and hence the larger the updates for the corresponding precision matrix elements. Thus, a large posterior expected value of $1/\nu_{ij}$ signifies strong evidence for the edge $(i,j)$ being present in all networks. It is also clear from the CM-updates (\ref{eq:lambda_updategroups}) that a conditional expectation of $1/\nu_{ij}$ close to zero does not imply that the updates for all the local scales will be close to zero. That is, thanks to the heavy tail of the half-Cauchy distribution, if there is enough evidence from the data, an edge can be identified in an individual network even though the common latent parameter suggests no edge. This is illustrated by Figure \ref{fig:theta_vs_nu}, which shows the precision elements estimated by the joint graphical horseshoe for $K=2$ networks, plotted against the posterior expectations of the corresponding shared latent parameters $1/\nu_{ij}$. The true networks have $99$ edges each, of which $39$ in common.  Figure \ref{fig:theta_vs_nu} shows that the expectation of $1/\nu_{ij}$ is only far from zero when an edge is present in both networks. When this expectation is close to zero, i.e., shared information is not found, posterior output still captures edges (i.e. non-zero $\theta_{ijk}$) specific to each network. This illustrates how the joint graphical horseshoe estimator can simultaneously share information between networks and capture their differences.

\section{Simulations}
\label{sec:sims}

To evaluate the performance of our approach, we have performed comprehensive simulation studies in R (\cite{Rcite}). We have generated data as close as possible to our omics application of interest, with non-zero partial correlations between $0.1$ and $0.2$ in magnitude and with the \emph{scale-free property} (i.e. the degree distribution follows a power-law distribution), a common assumption for omics data (\cite{chen2004content}). We assess graph accuracy by the precision, i.e., the fraction of the inferred edges that are actually present in the true graph (also known as positive predictive value or complement of the false discovery rate), and by the recall, i.e., the fraction of edges in the true graph that are present in the inferred one (also known as sensitivity or true positive rate). Because the networks inferred by the different methods may result in different sparsity estimates, some consideration is needed when comparing their precision and recall. For example, the recall tends to increase as the number of edges increases, favouring methods that over-select edges. In omics applications, one rather wishes to identify few but highly reliable associations than a large number of associations, many of which will be spurious. In the discussion of the results, we therefore put more emphasis on the precision, and consider the recall to be an informative additional measure, particularly in situations when two methods have comparable precision. 

Our numerical experiments are divided into three parts. In Section \ref{subsec:singlesim}, we assess the statistical and computational performance of fastGHS in a single-network setting, comparing it to the Gibbs sampling version of \cite{li2019graphical} and to the graphical lasso (\cite{friedman2008}). In Section \ref{subsec:singleVSjointsim}, we demonstrate that, thanks to joint modelling, the accuracy of the jointGHS increases with the number of related networks. Finally, in Section \ref{subsec:jointsim}, we compare the performance of jointGHS to the Bayesian spike-and-slab joint graphical lasso (\cite{li2019bayesian}), the joint graphical lasso (\cite{danaher2014}) and GemBag (\cite{yang2021gembag}). Details on all simulation studies are given in Section S.3 of Supplement A, and the corresponding code is available on Github (\url{https://github.com/Camiling/jointGHS_simulations}).

\subsection{Comparison of fastGHS with the Gibbs sampling scheme for single networks}
\label{subsec:singlesim}

In this section we compare the performance of our ECM implementation of the graphical horseshoe to that of the Gibbs sampler by \cite{li2019graphical}. As a baseline reference, we also provide the results of the widely used graphical lasso algorithm  (\cite{friedman2008}). We consider settings with different numbers of nodes, $p \in \{50, 100\}$, and observations, $n \in \{100, 200\}$. For each setting, we construct a $p\times p$ precision matrix and sample $N=20$ data sets with $n$ observations from the corresponding multivariate Gaussian distribution. 

\subsubsection{Runtime profiling}

All above graphical settings give rise to high-dimensional problems: for instance, with $n=200$ and $p=100$, there are $(p^2-p)/2=4\,950$ potential edges. For the Gibbs sampling implementation of the graphical horseshoe, a larger $p$, such as $200$, leads to computational problems as the algorithm entails singular updates, likely a result overflow not being properly dealt with (this holds for both the original MATLAB implementation and our translation into R, where the algorithm halts as it attempts to solve a singular system); running examples for this can be found at \url{https://github.com/Camiling/jointGHS_simulations}. In this comparison we therefore only consider up to $p=100$ nodes. We emphasise that the limitation $p<200$ for the Gibbs sampler applies to our particular simulation settings. In their simulations, \cite{li2019graphical} apply the Gibbs sampler to networks with as many as $p=400$ nodes, but these networks are sparser than ours and have larger partial correlations of magnitude $0.25-0.75$ (ours are of magnitude $0.1-0.2$). This makes the networks strongly identifiable from data, and thus fewer singularity and convergence issues are encountered.

\subsubsection{Edge-selection performance}

Table \ref{table:simulationsmall} indicates the edge-selection performance of fastGHS is comparable to the Gibbs sampler. The two graphical horseshoe implementations perform better than the graphical lasso in terms of both precision and recall in all but one case. %
For this exception, the graphical lasso has the best performance in terms of precision, likely because it has the sparsest estimate and therefore its inferred edges are more accurate, yet at the expense of a lower recall. %
The horseshoe-based methods have the best overall performance for a wider range of scenarios.

\begin{table}
	\centering
	\caption{Simulation results for our ECM implementation of the graphical horseshoe (fastGHS), the Gibbs sampling implementation of the graphical horseshoe (GHS) and the graphical lasso (Glasso) applied to multivariate Gaussian data from graphs with different numbers of vertices $p$ and observations $n$. The results are averaged over $N=20$ replicates, and shows sparsity, precision and recall, as well as their standard errors in parentheses. For each case, the highest value of the precision (resp. recall) is marked in bold (resp. italic), and so is the precision (resp. recall) of any other method within one standard error of it.} 
	\renewcommand{\arraystretch}{1.4}
	\begin{tabular}{r r r r l r r r}
		\vspace{0.5mm}\\
		\toprule
		Case & True sparsity &  p & n  & Method & Estimated Sparsity& Precision & Recall \\
		\hline
        1 & 0.04 & 50 & 100& Glasso  &  0.019 (0.006)  &  0.80 (0.12)  &  \emph{0.37} (0.08) \\
        &&&& GHS  &  0.017 (0.002)  &  \textBF{0.91} (0.06)  &  \emph{0.38} (0.05) \\ 
        &&&& fastGHS &  0.017 (0.002)  &  \textBF{0.94} (0.04)  &  \emph{0.39} (0.05) \\ 
        \hline 
        2 & 0.04 & 50 & 200& Glasso  &  0.020 (0.003)  &  0.88 (0.08)  &  \emph{0.44} (0.03) \\
        &&&& GHS  &  0.017 (0.002)  &  \textBF{0.98} (0.03)  &  \emph{0.42} (0.04) \\ 
        &&&& fastGHS &  0.017 (0.002)  &  \textBF{0.99} (0.02)  & \emph{0.43} (0.04) \\ 
        \hline 
        3 & 0.02 & 100 & 100& Glasso  &  0.011 (0.002)  &  \textBF{0.61} (0.08)  &  \emph{0.33} (0.04) \\ 
        &&&& GHS  &  0.015 (0.001)  &  0.49 (0.06)  &  \emph{0.37} (0.04) \\ 
        &&&& fastGHS &  0.015 (0.001)  &  0.46 (0.05)  &  \emph{0.35} (0.03) \\ 
        \hline 
        4 & 0.02 & 100 & 200& Glasso  &  0.008 (0.001)  &  0.86 (0.06)  &  0.36 (0.02) \\
        &&&& GHS  &  0.009 (0.001)  &  \textBF{0.91} (0.05)  &  \emph{0.40} (0.03) \\ 
        &&&& fastGHS &  0.009 (0.001)  &  \textBF{0.93} (0.05)  &  \emph{0.41} (0.04) \\ 
        \hline 
	\end{tabular}
	\label{table:simulationsmall}
\end{table} 

\begin{figure}
    \centering
    \includegraphics[scale=0.3]{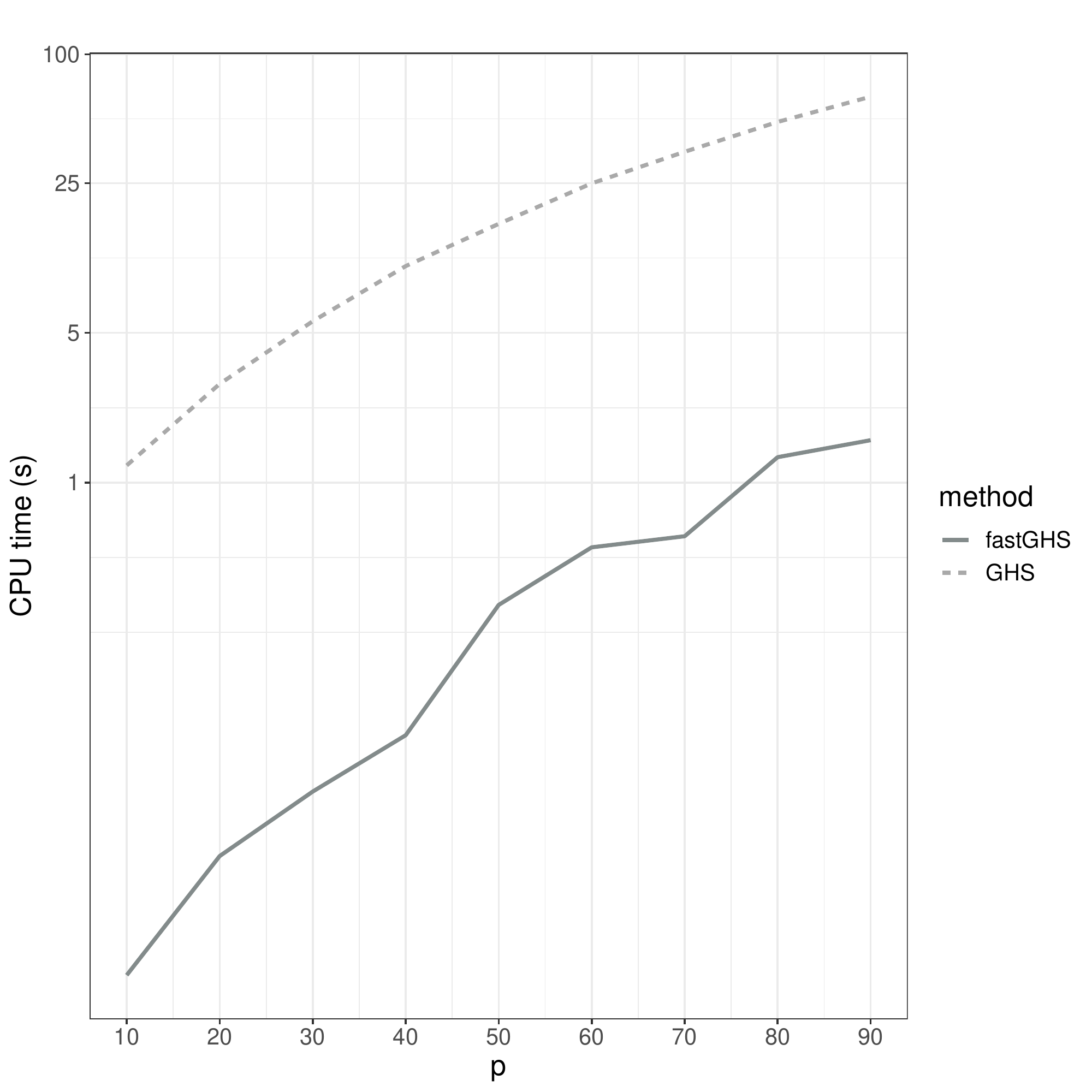}
    \caption{CPU time in seconds on a logarithmic scale to infer a network for a grid of node numbers $p$ and $n=100$ observations, for our ECM implementation (fastGHS) and the Gibbs sampling implementation (GHS) of the graphical horseshoe. Computations were performed on a 16-core Intel Xeon CPU, 2.60 GHz.}
    \label{fig:timeplot_small}
\end{figure}

As anticipated, our runtime profiling for the Gibbs sampler and our ECM implementation of the graphical horseshoe indicate striking differences. Comparing the two types of inference is not straightforward as they rely on different stopping rules and convergence diagnostics. 
To alleviate the risk of unfair comparison, we run the Gibbs sampler for a relatively small number of MCMC samples, namely $1\,000$, after  $100$ burn-in 
iterations, while for the ECM algorithm, we use a maximum of $10\,000$ iterations.
Figure \ref{fig:timeplot_small} shows on a logarithmic scale the CPU time used to infer a network for different numbers of nodes $p$, %
with $n=100$ observations.  %
For $p=90$ nodes, fastGHS is $30$ times faster than the Gibbs sampler. For larger $p$, only the ECM estimator can be used, which in practice is limited only by the available memory to store the $p\times p$ matrix updates for $\boldsymbol{\Theta}$, $\boldsymbol{\Lambda}$ and $\boldsymbol{N}$. 

\subsection{Increased accuracy with joint modelling}
\label{subsec:singleVSjointsim}

Now that we have established that the performance of our ECM implementation of the graphical horseshoe is comparable to that of the Gibbs sampler, we aim to investigate the gain in statistical power when applying our joint graphical horseshoe estimator, as a function of the number of networks modelled jointly. We use jointGHS to reconstruct $K\in\{2, 4, 10\}$ graphs with $p=50$ nodes, also applying our single-network method fastGHS on each network separately to serve as baseline.
While we simulate scenarios with different degrees of shared information across the $K$ graphs (edge disagreement), for a given scenario, the edge disagreement is the same for any pair of networks. Of note, neither the spike-and-slab joint graphical lasso  (\cite{li2019bayesian}) nor the joint graphical lasso (\cite{danaher2014}) can run within reasonable time for the setting with $K = 10$ networks ($<48$ hours, see Section S.7.3 of Supplement A). The results are averaged over $N=40$ replicates, and show the precision and recall for the first estimated graph in each setting, reconstructed from $n=80$ observations. All graphs have true sparsity~$0.04$.

\begin{figure}
    \centering
    \hspace*{-1cm}
    \includegraphics[scale=0.4]{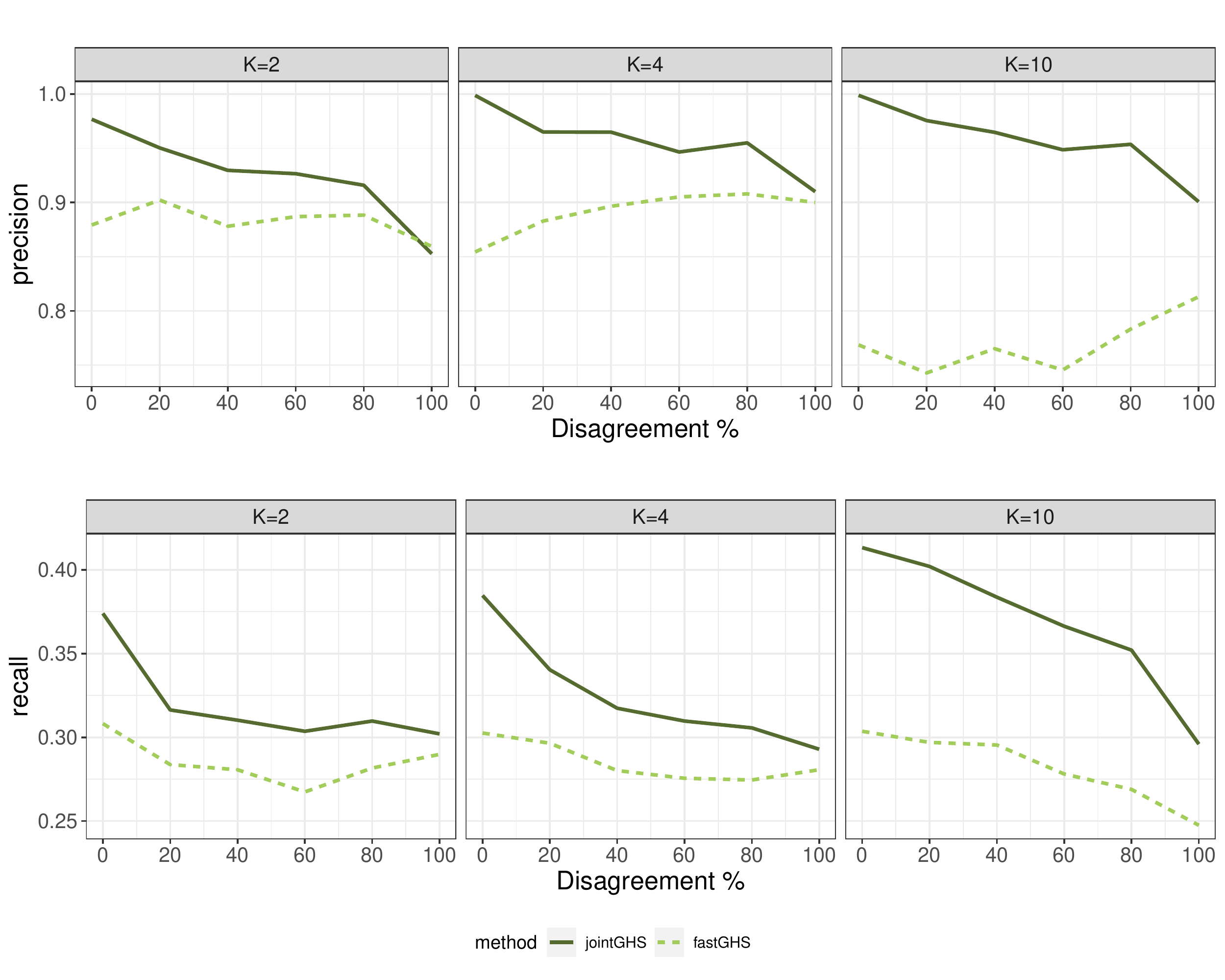}
    \caption{Performance of the joint graphical horseshoe (jointGHS) and single network graphical horseshoe (fastGHS), reconstructing $K\in\{2, 4, 10\}$ graphs with $p=50$ nodes and various similarity of the true graph structures. The edge disagreement between the two graphs is shown as the percentage of edges in one network not present in the other. The results are averaged over $N=40$ replicates, and show the precision and recall for the first estimated graph in each setting, reconstructed from $n=80$ observations. All graphs have true sparsity~$0.04$.}
    \label{fig:jointVSsingle}
\end{figure}

Figure \ref{fig:jointVSsingle} shows the precision and recall for jointGHS and fastGHS as a function of the available information (total number of graphs $K$) and level of disagreement between them. Although the simulated graph structure remains the same in all settings, the sparsity of the inferred jointGHS graphs varies with total number of graphs and their level of similarity. Hence, to ensure a fair comparison, we obtained single-network estimates with the same sparsity as the joint estimates in each setting, making the fastGHS results vary with both $K$ and the level of similarity; we refer to Section S.3 of Supplement A for details. 

As expected, %
the joint approach clearly outperforms the single network approach in terms of both precision and recall, and the improvement increases with the number of graphs $K$, since more shared information is available. This applies to all levels of edge disagreement, including when the graphs have no common edges. This may be explained by the sparsity of the graphs: while the networks do not share any edge, they do share the fact that no edge is present for a large number of pairs $(i, j)$. Thus, information is still be shared through conditional expectations of the common $\nu_{ij}^{-1}$'s being equal to zero. The larger $K$ is, the stronger this information is, leading to a large improvement compared to a single network approach.

\subsection{Comparison of jointGHS with other joint network inference methods}
\label{subsec:jointsim}

Now that we have demonstrated the benefit of joint modelling, we next assess the computational and statistical performance of our joint graphical horseshoe estimator, jointGHS, through comparisons with the Bayesian spike-and-slab joint graphical lasso (SSJGL; \cite{li2019bayesian}), the joint graphical lasso (JGL; \cite{danaher2014}) and GemBag (\cite{yang2021gembag}). %

\subsubsection{Runtime profiling}
\label{subsubsec:joint_runtime}

\begin{figure}
    \centering
    \includegraphics[width=0.75\textwidth]{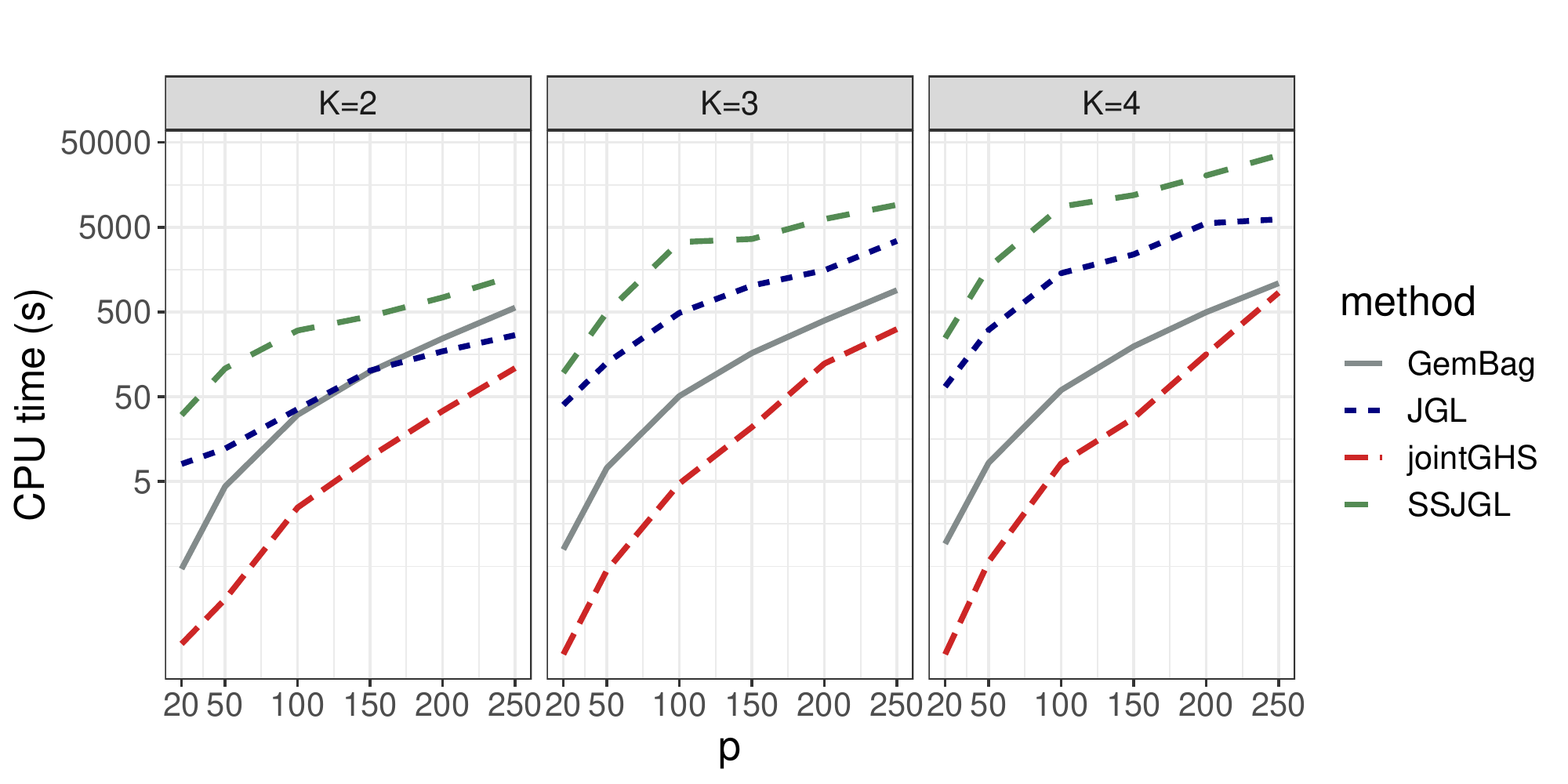}
    \caption{CPU time in seconds on a logarithmic scale to jointly infer networks for various numbers of nodes $p$ and networks $K$, using the joint graphical horseshoe (jointGHS), the spike-and-slab joint graphical lasso (SSJGL), the joint graphical lasso (JGL) and GemBag. In all settings, the networks agree on $50\%$ of their edges. For $K=2$, the networks are inferred from $n_1=100$ and $n_2=150$ observations, for $K=3$, the networks are inferred from $n_1=n_2=100$ and $n_3=150$ observations, and for $K=4$ the networks are inferred from $n_1=n_2=100$ and $n_3=n_4=150$ observations. Computations were performed on a 16-core Intel Xeon CPU, 2.60 GHz.}
    \label{fig:timeplot_joint}
\end{figure}

Figure \ref{fig:timeplot_joint} compares the runtime of all methods for a grid of node numbers $p$, $K \in \{2, 3, 4\}$ networks each with $n_k \in \{100, 150\}$: jointGHS is the fastest of the four methods, for all settings, followed by GemBag, JGL and finally SSJGL. The last two approaches become computationally prohibitive as the number of networks $K$ increases. Indeed, \cite{danaher2014} highlight that the JGL algorithm scales well in problems with only two classes ($K = 2$), as a closed-form solution to the generalised fused lasso problem can be obtained in that case (this also holds for SSJGL). Both SSJGL and JGL use the same alternating direction method of multipliers (ADMM) algorithm to update the precision matrix estimates. 

Notably, in their respective simulation studies, \cite{li2019bayesian} and \cite{danaher2014} apply their methods to as many as $p=400$ and $p=500$ nodes. However, in the SSJGL numerical experiments, precision matrix elements are sampled from the G-Wishart distribution with $3$ degrees of freedom, giving very strong partial correlations ($0.5-0.9$ in absolute value). Similarly, following the precision matrix construction described in the JGL experiments gives partial correlations of $\approx 0.6$ in absolute value. This renders the networks strongly identifiable from data, leading to faster convergence than in our simulations where partial correlations and between $0.1$ and $0.2$. Thus, motivated by realistic biological network strengths, we are considering a more challenging inference problem.

The lower runtime of jointGHS and GemBag can be explained by their EM/ECM implementations, although their higher scalability compared to SSJGL – which also implements an EM algorithm – could be due to their computationally efficient C++ subroutines. Because of these computational limitations, we use relatively small numbers of nodes $p$ to make so all methods can run within reasonable time ($<48$ hours). Note however that jointGHS successfully completes within this timeframe on examples with $p > 1\,000$ nodes (see Section S.7.3 of Supplement~A). %

\subsubsection{Edge-selection performance}\label{subsubsec:joint_res}

We next compare the edge-selection performance of all four methods on problems with $K=2$ graphs of $p=50$ nodes each, and $n_1 = 50$ and $n_2 = 80$ observations, respectively. We consider six settings, with different levels of graph similarity, i.e., proportion of edges present in both graphs. Namely, we simulate data with similarity varying between $0\%$ edge disagreement (i.e., the same edge set) to $100\%$ edge disagreement (i.e., no common edges). For each setting, we construct two $p\times p$ precision matrices with the desired level of similarity and we sample $N=100$ data sets from each of the two corresponding multivariate Gaussian distributions. In all settings, both graphs have true sparsity $0.04$, corresponding to $49$ edges. We report the precision and recall of the final estimate of each method. A threshold-free comparison based on precision-recall curves and corresponding areas under the curves is given in Section S.7.2 of Supplement~A.

Table \ref{table:simulationjoint} shows the performance of the joint network approaches. The joint graphical lasso JGL, applied with its default AIC-based selection criteria for sparsity- and similarity-selection, has low precision in all settings. The method tends to severely over-select edges, reporting nearly ten times more edges as the number simulated edges: this leads to high recall values but very low precision. 

For GemBag, over-selection of edges is not as severe as for (JGL), but the method still reports more edges than the joint graphical horseshoe (jointGHS) and the spike-and-slab joint graphical lasso (SSJGL) in all settings, which results in a lower precision but a higher recall than the two methods. Interestingly, while the first network with $n_1=50$ seems to benefit from GemBag's joint modelling, it is not the case of the second network with the largest sample size $n_2=80$, as its estimation does not improve as the level of similarity between the two networks increases. Further, due to the larger sample size, GemBag selects more edges for the second network in all settings, which yields a higher recall yet a lower precision than for the first network. Remarkably, the disagreement level (percentage of edges present in only one of the two networks) of the estimated GemBag networks remains almost the same in all settings, and hence does not reflect the different simulated similarity between the two graphs: the method does not appear to adapt to varying network similarity levels, with possibly too much information shared between unrelated networks, yet too little information shared between highly related ones.

The precision of jointGHS is either higher or comparable to that of SSJGL in all settings. In general, jointGHS is more conservative than the other methods, and hence is most suitable for detecting edges with high confidence. This is further exemplified in our extended simulations, particularly in our threshold-free comparisons for sparse edge selection where jointGHS in all settings considered had the highest precision for a given recall level (Section S.7.1 of Supplement~A). %
The graphs estimated by SSJGL are denser, which tends to result in a large recall values: this holds for very similar networks, where SSJGL has the highest recall, yet as network dissimilarity increases, the recall of SSJGL decreases and becomes similar to that of jointGHS. This happens because SSJGL shrinks all precision matrices towards a common structure, thereby over-selecting edges that are absent in some networks while being present in others, as exemplified further in Section \ref{subsubsec:heavytail}. Table \ref{table:simulationjoint} indicates that the two networks are estimated by jointGHS as being increasingly different from each other as the simulated level of disagreement increases, while they are invariably estimated as almost identical by SSJGL, even when the true networks are completely unrelated.
Of note, the precision of SSJGL deteriorates in settings where the simulated networks share little information, while jointGHS effectively adapts to this setting, maintaining relatively high values of both the precision and recall for completely unrelated networks. As further demonstrated in Section \ref{subsubsec:heavytail} hereafter, this clear advantage can be attributed to the local scales $\lambda_{ijk}$ of the graphical horseshoe, which flexibly capture isolated effects thanks to their heavy Cauchy tails. 

\begin{table}
	\centering
	\renewcommand{\arraystretch}{1.4}
	\caption{Performance of the joint graphical horseshoe (jointGHS), the spike-and-slab joint graphical lasso (SSJGL), the joint graphical lasso (JGL) and GemBag for reconstructing $K=2$ graphs with various similarity of the true graph structures. Results are averaged over $N=100$ replicates. The edge disagreement between the two simulated graphs is shown as the percentage of edges in one network not present in the other, along with the average edge disagreement of the graphs estimated by each method.
	The graphs are simulated from a multivariate Gaussian distribution with $p=50$ variables and with $n_1=50$ and $n_2=80$ observations; both graphs have simulated sparsity of $0.04$. The estimated sparsity, precision and recall for both graphs is reported, with standard errors in parentheses. For each case the highest value of the precision is marked in bold, and so is the precision of any other method within one standard error of it.  }
    \hspace*{-2cm}
	\begin{tabular}{r l r @{\hskip -0.1cm} r @{\hskip 0.8cm}l l l r @{\hskip 0.8cm}l l l}
        \multicolumn{9}{c}{ }\\
	    \toprule
	    &&& &\multicolumn{3}{c}{$n_1=50$}&&\multicolumn{3}{c}{$n_2=80$} \\
		\cmidrule(lr){5-7} \cmidrule(lr){9-11}
		Disagr. $\%$ & Method & $\widehat{\text{Disagr.}}$ $\%$ && Sparsity& Precision & Recall&&Sparsity& Precision & Recall \\
		\hline
        0  & JGL  & 63 &&  0.312 (0.011)  &  0.11 (0.01)  &  0.83 (0.05) &&  0.269 (0.011)  &  0.13 (0.01)  &  0.91 (0.04) \\ 
        & GemBag  & 41 &&  0.039 (0.006)  &  0.65 (0.07)  &  0.62 (0.06) 
        &&  0.064 (0.008)  &  0.49 (0.05)  &  0.77 (0.06) \\ 
        & SSJGL  & 0 &&  0.025 (0.003)  &  \textBF{0.85} (0.09)  &  0.53 (0.06) &&  0.025 (0.003)  &  0.85 (0.09)  &  0.53 (0.06) \\ 
        & jointGHS & 17 &&  0.017 (0.002)  &  \textBF{0.82} (0.08)  &  0.34 (0.04) &&  0.016 (0.001)  &  \textBF{0.93} (0.06)  &  0.38 (0.03) \\ 
        \hline 
        20  & JGL  & 65 &&  0.310 (0.02)  &  0.11 (0.01)  &  0.83 (0.05) &&  0.266 (0.017)  &  0.14 (0.01)  &  0.90 (0.04) \\ 
        & GemBag  & 42 &&  0.041 (0.007)  &  0.61 (0.07)  &  0.61 (0.06) &&  0.064 (0.01)  &  0.47 (0.06)  &  0.73 (0.06) \\ 
        & SSJGL  & 1 &&  0.025 (0.003)  &  0.77 (0.10)  &  0.48 (0.05) &&  0.025 (0.003)  &  0.79 (0.10)  &  0.49 (0.05) \\ 
        & jointGHS & 25 &&  0.016 (0.002)  &  \textBF{0.79} (0.10)  &  0.32 (0.04) &&  0.016 (0.001)  &  \textBF{0.93} (0.06)  &  0.38 (0.03) \\ 
        \hline 
        40  & JGL  & 67 &&  0.291 (0.048)  &  0.12 (0.02)  &  0.82 (0.06) &&  0.252 (0.059)  &  0.14 (0.05)  &  0.83 (0.08) \\ 
        & GemBag & 49 &&  0.038 (0.005)  &  0.57 (0.08)  &  0.54 (0.05) &&  0.066 (0.007)  &  0.41 (0.04)  &  0.66 (0.05) \\
        & SSJGL  & 1 &&  0.020 (0.002)  &  \textBF{0.76} (0.09)  &   0.38 (0.04) &&  0.020 (0.002)  &  0.77 (0.09)  &  0.39 (0.05) \\ 
         & jointGHS & 34 &&  0.016 (0.002)  & \textBF{0.76} (0.08)  &  0.30 (0.04) &&  0.010 (0.002)  &  \textBF{0.98} (0.04)  &  0.25 (0.05) \\ 
         \hline 
        60  & JGL &69 &&  0.291 (0.049)  &  0.12 (0.02)  &  0.82 (0.07) &&  0.247 (0.058)  &  0.15 (0.05)  &  0.84 (0.07) \\ 
        & GemBag  & 48 &&  0.039 (0.004)  &  0.52 (0.07)  &  0.50 (0.05) &&  0.064 (0.009)  &  0.43 (0.04)  &  0.67 (0.06) \\ 
         & SSJGL & 2 &&  0.021 (0.003)  &  0.63 (0.09)  & 0.33 (0.05) &&  0.021 (0.003)  &  0.70 (0.08)  &  0.37 (0.05) \\ 
        & jointGHS & 47 &&  0.016 (0.002)  &  \textBF{0.74} (0.08)  &  0.29 (0.04) &&  0.010 (0.002)  &  \textBF{0.98} (0.04)  &  0.25 (0.05) \\ 
        \hline 
        80  & JGL  & 71 &&  0.299 (0.040)  &  0.11 (0.02)  &  0.81 (0.06) &&  0.259 (0.046)  &  0.14 (0.04)  &  0.89 (0.06) \\ 
        & GemBag  & 47 &&  0.040 (0.005)  &  0.46 (0.07)  &  0.46 (0.06) &&  0.068 (0.008)  &  0.42 (0.05)  &  0.70 (0.06) \\
         & SSJGL  & 2  &&  0.022 (0.004)  &  0.52 (0.08)  &  0.27 (0.04) &&  0.021 (0.003)  &  0.60 (0.09)  &  0.32 (0.05) \\ 
        & jointGHS & 59 &&  0.015 (0.002)  &  \textBF{0.70} (0.11)  &  0.26 (0.04) &&  0.010 (0.002)  &  \textBF{0.98} (0.04)  &  0.25 (0.05) \\  
        \hline 
        100  & JGL & 72 &&  0.313 (0.012)  &  0.11 (0.01)  &  0.83 (0.05) &&  0.251 (0.009)  &  0.16 (0.01)  &  0.98 (0.02) \\ 
        & GemBag  & 43&&  0.050 (0.004)  &  0.34 (0.05)  &  0.43 (0.06) &&  0.074 (0.006)  &  0.48 (0.04)  &  0.89 (0.04) \\ 
        & SSJGL  & 3 &&  0.029 (0.003)  &  0.29 (0.06)  &  0.21 (0.03) &&  0.029 (0.003)  &  0.55 (0.06)  &  0.40 (0.05) \\ 
        & jointGHS & 78 &&  0.015 (0.002)  &  \textBF{0.65} (0.12)  &  0.24 (0.04) &&  0.013 (0.002)  &  \textBF{0.95} (0.06)  &  0.31 (0.05) \\
		\toprule
	\end{tabular}
	\label{table:simulationjoint}
\end{table} 

\subsubsection{Ability to capture edges on the individual-graph level}
\label{subsubsec:heavytail}

We next provide a more detailed illustration of the benefits of the horseshoe heavy-tailed local scales for capturing graph-specific edges by comparing precision matrix estimates obtained with our joint graphical horseshoe, jointGHS, and the spike-and-slab joint graphical lasso, SSJGL. We use the same data generation procedure as in Section \ref{subsec:heavytail}, with networks reconstructed from two data sets corresponding to $K=2$ graphs with $40\%$ edge agreement, $p=100$ nodes, and $n_1=100$ and $n_2=150$ observations, respectively. Both graphs have a true sparsity of $0.02$. 

\begin{figure}
    \centering
    \includegraphics[scale=0.3]{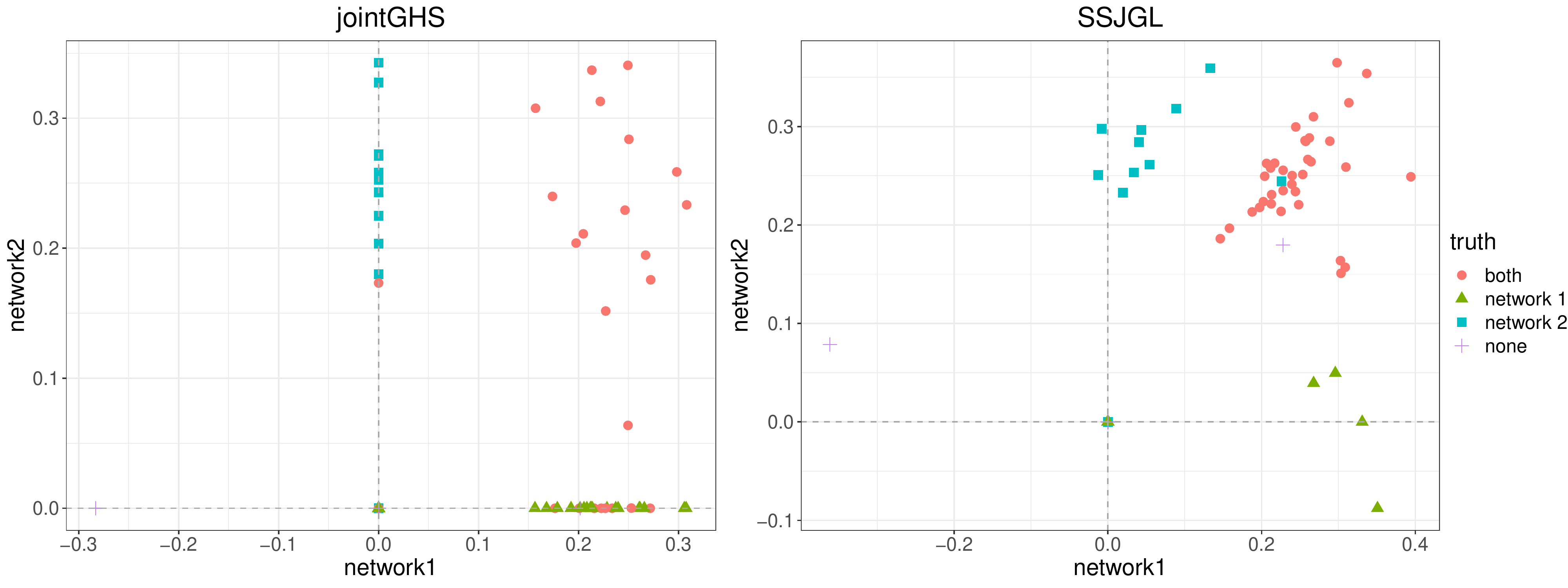}
    \caption{Comparison of estimated precision matrix elements (scaled by the diagonal as when finding partial correlations) of two networks with $40\%$ edge agreement and $p=100$ nodes, for the joint graphical horseshoe and for the spike-and-slab joint graphical lasso separately. The points' shape and color indicates whether the corresponding edge is simulated as present is both, one or neither network.}
    \label{fig:SSvsjoint}
\end{figure}

Figure \ref{fig:SSvsjoint} indicates that jointGHS effectively identifies both common and graph-specific edges. Thanks to its network-specific local scales, when an edge is  present in only one of the graphs, the corresponding precision matrix element in the other graph is correctly estimated as null, hence no false positive is reported due to excess shrinkage towards a common graph. 
As a result, jointGHS is more inclined to false negatives than false positives; in a few instance, it %
reports edges in only one of the two networks, while they actually present in both.
SSJGL displays the opposite behaviour: it shrinks excessively towards a common graph and therefore largely fails to identify the network-specific edges. While the SSJGL does well in capturing edges common to both graphs, an edge in one graph tends to be reported as present in both, resulting in a large number of false positives. This explains its excellent performance for very similar networks, but poorer performance as the similarity between two networks decreases, as discussed in Table \ref{table:simulationjoint}. Regardless of how similar the networks may be, jointGHS effectively borrows shared information across them, while successfully avoiding over-shrinkage towards a common structure to preserve graph-specific information. %

While our simulations have illustrated the flexibility of the joint graphical horseshoe, performing joint modelling of $K$ data sets consisting of many highly similar networks and a few unrelated or a priori more loosely similar networks would make little sense. While, thanks to the horseshoe local scales, the risk of the many similar networks dominating the analysis is lower with jointGHS compared to other joint graphical methods – including the spike-and-slab joint graphical lasso (\cite{li2019bayesian}) and the joint graphical lasso (\cite{danaher2014}) – it may be helpful to investigators to rule out this scenario. To this end, we outline a Bayesian bootstrapping procedure (\cite{rubin1981bayesian}) in Section S.6 of Supplement~A to check whether the joint network estimates are in strong contradiction with each of the single network estimates; this optional routine is implemented in our R package \texttt{jointGHS}.

\section{Application to a study of hotspot activity with stimulated monocyte expression}
\label{sec:applications}

Returning to the monocyte data set from Section \ref{sec:motivation}, we now apply our proposed methodology to estimate conditional independence among the gene levels under genetic control. Specifically, the finding of \cite{ruffieux2020global} about the top hotspot genetic variant (rs6581889, on chromosome $12$) being persistent across all four monocytic conditions (unstimulated cells, $\text{IFN-}\gamma$-, LPS 2h- \& LPS 24h-stimulated cells) makes a joint graphical approach particularly relevant to study the interplay within and across the different gene networks. The number of genes associated with the top hotspot \emph{in each condition} was $294$, $88$, $16$ and $215$ respectively (permutation-based FDR $<0.05$); hereafter we focus on the $p = 381$ genes associated with the hotspot in at least one condition. Further information on the data and preprocessing steps is available in \cite{ruffieux2020global}, and details on the analysis presented below can be found in Section S.4 of Supplement~A.

We first present and interpret the results obtained by applying jointGHS to jointly estimate the precision matrices, and hence network structures, of the genes in the four conditions.  We then describe a comparative study with (i) the classical graphical horseshoe applied separately to each network (using our fastGHS ECM implementation, which scales to this problem), and (ii) a competing joint modelling approach. For the latter comparison, we use GemBag (\cite{yang2021gembag}), as neither the spike-and-slab joint graphical lasso (\cite{li2019bayesian}) nor the joint graphical lasso (\cite{danaher2014}) runs within $48$ hours on the data; as previously discussed in Section \ref{subsubsec:joint_runtime}, both algorithms become substantially slower in problems with $K>2$ classes, due to the absence of closed-form solution for $K \neq 2$.

\begin{figure}
    \centering
    \includegraphics[scale=0.3]{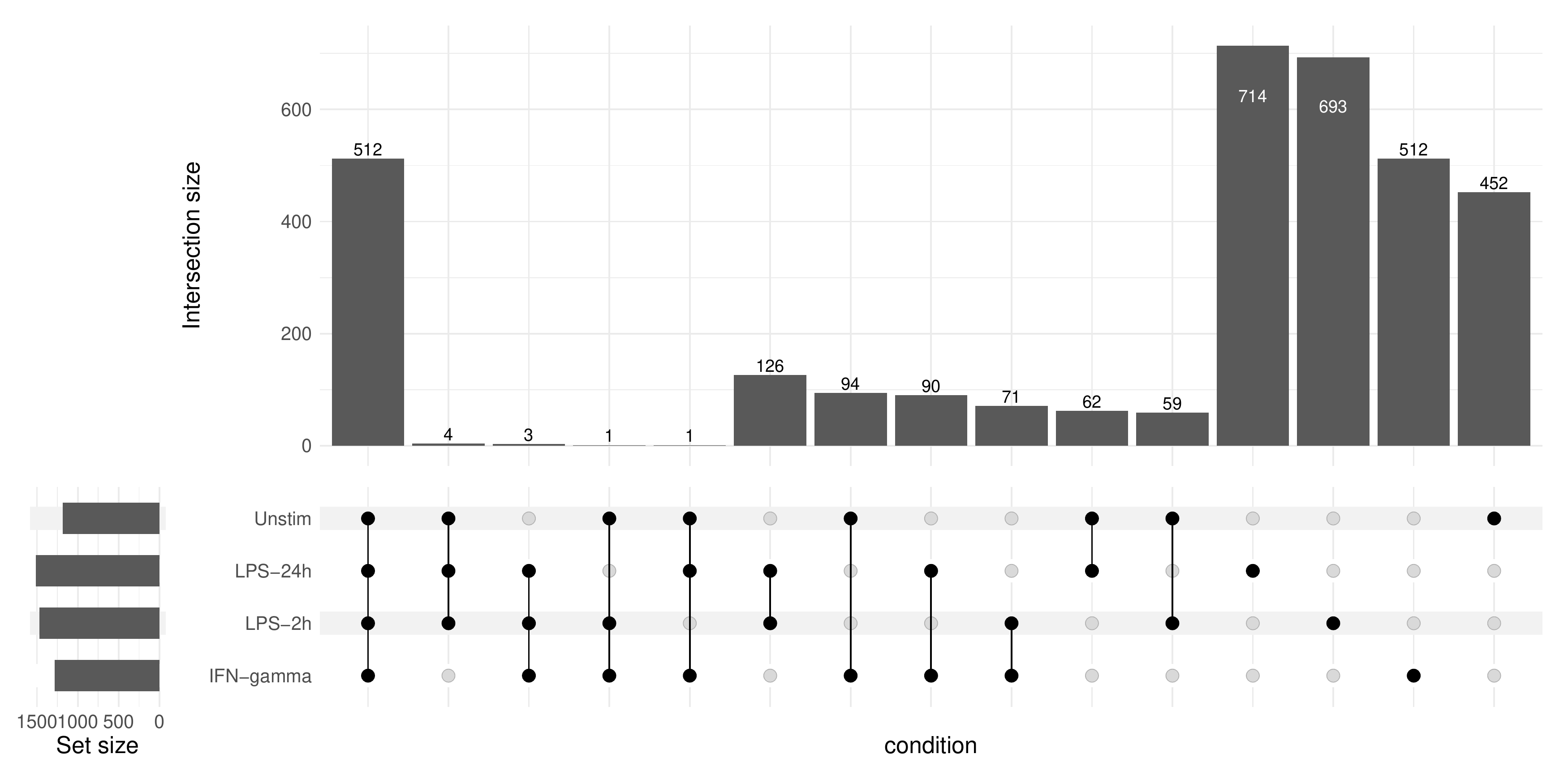}
    \caption{Upset plot of the joint graphical horseshoe graphs of the monocyte data, an alternative to a Venn diagram showing the number of edges shared strictly between given conditions (\cite{conway2017upsetr}). For each intersection, the number of edges shared only by the corresponding conditions is shown. The total number of edges for each condition is represented on the left barplot. Every possible intersection is represented by the bottom plot, and their occurrence is shown in the top barplot.}
    \label{fig:Monocyte_intersection}
\end{figure}

Figure \ref{fig:Monocyte_intersection} shows the number of edges shared between the different conditions, as inferred by jointGHS. Many edges are common to all four networks, suggesting a high degree of similarity across all monocytic conditions, likely reflecting the effect of the shared hotspot control. Very few edges are shared across three conditions only, but many pairs of conditions have edges that are shared strictly between them. In particular, LPS 2h and LPS 24h have the largest number of shared edges that are not present in the other two conditions, which is expected as they correspond to an exposition to differing durations of a same lipopolysaccharide activation. While LPS is a component of gram-negative bacterial cell walls, $\text{IFN-}\gamma$ is a cytokine important in myobacterial and viral infections (\cite{fairfax2014innate}).
In addition, and in line with the results of our simulation studies, jointGHS is able to capture many condition-specific edges, with LPS 24h having the highest number of unique edges, possibly because it corresponds to the densest graph across all conditions. %
These observations call for further biological investigations, which may motivate new mechanistic studies, such as: whether groups of edges shared by two or more conditions pertain to known pathway activation, or whether pathways of genes involved in edges unique to one stimulated condition are indicative of some condition-specific functional mechanisms. We explore such questions in the next sections. %

\subsubsection*{Network-specific activity}

A number of the network-specific structures identified by jointGHS warrant close inspection. For example, %
the %
Cytochrome C Oxidase Subunit 6A1 (\emph{COX6A1}) gene has large degree in both LPS 2h and LPS 24h, moderately large degree in $\text{IFN-}\gamma$, but low degree in the unstimulated condition. The oxidative phosphorylation pathway and immune system processes both include \emph{COX6A1} (\cite{wang2019label}), and the gene has been shown to have key functions in the replication of influenza A viruses (\cite{hao2008drosophila}), making it noteworthy that this gene's activity is found elevated only in the stimulated conditions. Another example pertains to the PHD-finger 1 protein encoding gene (\emph{PHF1}), which has high degree only in the $\text{IFN-}\gamma$ condition, where it is also found to be controlled by the top hotspot rs6581889. %
The PHD-finger 1 protein is an essential factor for epigenetic regulation and genome maintenance, and contains two kinds of histone reader modules, a Tudor domain and two PHD fingers %
(\cite{baker2008phd, liu2018phd}). The centrality of \emph{PHF1} in the network of the $\text{IFN-}\gamma$ stimulated monocytes suggests a potential role in the immune reaction, and provides a relevant alley for further studies. %

\subsubsection*{Hub genes}

Investigating hub genes in the jointGHS networks for the different conditions can help gain better understanding of the immune response driving disease mechanisms. %
Remarkably, the Autoimmune Regulator (\emph{AIRE}) gene, which is highly expressed in monocytes, has by far the most links to other genes in all conditions but $\text{IFN-}\gamma$, where it has the second most (Section S.4 of Supplement~A). This gene is known to play an important role in immunity through gene and autoantigen activation and regulation, and negative selection of autoreactive T-cells in the thymus (\cite{liston2003aire, kyewski2006central, peterson2008transcriptional}). Mutations in this gene have been associated with autoimmune polyendocrinopathy-candidiasis-ectodermal dystrophy (\emph{APECED}), distinguished by multi-organ autoimmunity (\cite{mathis2007decade, akirav2011role}). Similarly, the arylformamidase (\emph{AFMID}) gene, also known as kynurenine formamidase, is found to have amongst the higher degrees in all four conditions in addition to being associated with the top hotspot in all; it also has a link to \emph{LYZ} in all four conditions. Arylformamidase is a rate-limiting enzyme in tryptophan conversion, and deficiency is associated with immune system abnormalities (\cite{hugill2015loss, dobrovolsky2005effect}). Additional biological results and discussion can be found in Section S.4 of Supplement~A.

\subsubsection*{Hotspot control}

We next explore the extent to which the top genetic hotspot rs6581889 influences the conditional independence structure of the genes it controls. %
Using permutation testing to derive empirical p-values, we find that the subnetwork of genes associated with rs6581889 in each condition has significantly more links than the overall network ($p < 0.01$), except in the IFN-$ \gamma$ network, suggesting a hotspot-induced increase in activity. We similarly find that, in all conditions, there is a significant enrichment of genes associated with the top hotspot among the neighbours of the \emph{LYZ} gene. As  \emph{LYZ} is located a few Kb away from rs6581889, this may suggest a mediation of the hotspot effect on other genes via \emph{LYZ} – a hypothesis already examined in different studies (\cite{fairfax2012genetics, ruffieux2021epispot}) but which would require experimental validation or dedicated inspection, e.g., with Mendelian randomisation analysis. The findings are summarised in Table \ref{table:monocyte_stats}. 

\begin{table}
\caption{Sparsity and hotspot control with jointGHS. Sparsity of the jointGHS subnetwork of top-hotspot-controlled genes, as well as the overall sparsity of the full network in each condition. Proportion of neighbour genes of the two genes proximal to the hotspot (LYZ and YEATS4), as well as the overall proportion, in the different conditions. For each condition, a proportion higher than the overall proportion is marked in bold, with statistical significance (compared to randomly sampled subsets of genes of the same size) marked by $*$ (empirical p-value $<0.05$) or $**$ (empirical p-value $<0.01$) .}
\centering
\renewcommand{\arraystretch}{1.4}
\begin{tabular}[t]{l l r l l l l}
\hline
 &&& IFN-gamma & LPS 2h & LPS 24h & Unstim\\
 \hline
Sparsity & Overall && 0.018 & 0.020   & 0.021   & 0.016   \\
& Controlled by top hotpot&& \bf{0.019}   & \bf{0.042}** & \bf{0.183}** & \bf{0.022}*\\
\hline
Controlled by top hotspot & Overall && 0.77 & 0.23   & 0.04   & 0.56   \\
& \emph{LYZ} neighbourhood && \bf{0.78}   & \bf{0.42}* & \bf{0.33}** & \bf{0.77}*\\
& \emph{YEATS4} neighbourhood && 0.62   & 0.17   & \bf{0.22}** & 0.43 \\
\hline
\end{tabular}
\label{table:monocyte_stats}
\end{table}

\subsubsection*{Comparison to the single-network analysis}

We next aim to assess the possible added value of joint modelling for increasing biological insight %
by comparing the jointGHS results to those of our single-network ECM implementation of the graphical horseshoe, fastGHS.  %
To obtain comparable networks, we use for fastGHS the same sparsity levels as in the jointGHS estimates, for each condition separately (see Section S.4 of Supplement~A for details). 
The subgraph of hotspot-controlled genes inferred by jointGHS is denser as that inferred by fastGHS in all conditions; a dense graph agrees with the expectation that the hotspot triggers substantial activity among the controlled genes (\cite{ruffieux2020global}). Moreover, both \emph{LYZ} and \emph{YEATS4} have a more central role in the joint network, with more edges to other genes as well as more edges among their neighbours. This lends further support to the mediation hypothesis formulated above. Finally, many genes directly associated with \emph{LYZ} and \emph{YEATS4} have very high degree, suggesting that their interplay with other genes could be relevant for disease-driving mechanisms. 
All these observations highlight the biological insight gained by sharing information across networks with jointGHS. 

\subsubsection*{Comparison to the GemBag analysis}

The comparison of jointGHS with the other joint modelling approach GemBag (\cite{yang2021gembag}) is also informative. Strikingly, GemBag identifies a strict subset of the edges identified by jointGHS. Moreover, almost all edges are estimated as shared by all four conditions and very few are condition-specific edges; this stands in strong contrast to jointGHS which finds many condition-specific edges. Similarly, the four conditions have an almost complete overlap of top hubs (genes with node degree larger than the $90^{\text{th}}$ percentile) in the GemBag networks, again contrasting the jointGHS networks where the top hubs are mainly network-specific. The biological plausibility of the network specificities discussed above lends support to the argument that jointGHS succeeds at capturing network-specific effects thanks to its horseshoe local scales. 
Further, \emph{LYZ} and \emph{YEATS4} have very low node degrees in all conditions in the GemBag networks, whereas the two genes are central in the jointGHS networks, with a large number of neighbours, many central neighbours (i.e., hubs) and many top hotspot controlled neighbours, which aligns with evidence from previous studies (\cite{fairfax2012genetics, ruffieux2021epispot}). Further results and details on comparison of GemBag and jointGHS are given in Section S.4 of Supplement~A.

While it is reassuring that our method identifies genes known from literature to be relevant, this type of validation is biased towards gene and protein functions that have already been explored. We believe though that jointGHS could serve to generate further unexplored hypotheses about genetic co-regulation and co-expression across the stimulated monocyte networks – this would deserve further follow-up research.  More generally, our findings illustrate the potential of the joint graphical horseshoe for gaining deeper insight into the mechanisms at play among large networks of cellular and/or molecular variables for multiple conditions or tissues. 

\section{Conclusion}
\label{sec:discussion}

We have introduced an efficient ECM algorithm for estimating the precision matrix in the graphical horseshoe, fastGHS, and a novel joint graphical horseshoe estimator for multiple-network inference, jointGHS. Through simulations, we have shown that fastGHS achieves equivalent performance to the fully Bayes graphical horseshoe Gibbs sampler while being substantially more scalable. In the multiple-network setting, we have also shown that jointGHS successfully shares information between networks while capturing their differences, outperforming competing methods such as the joint graphical lasso, GemBag and the spike-and-slab joint graphical lasso, which can be very anti-conservative. This holds for any level of network similarity, even when there is little or no information to share between networks. This clear advantage of the jointGHS can be attributed to the horseshoe heavy-tailed local scales, which are able to adapt even in the absence of shared information, favouring the detection of isolated network-specific edges %
– to date, no existing joint graphical modelling approach enjoys this property. Hence, jointGHS stands out as a joint approach capable to also pinpoint differences across networks, which, in practice, is often of great interest, sometimes even more than the identification of shared structures. Finally, while our ECM implementation does not provide a fully Bayesian solution, we have demonstrated that this does not affect performance for our primary inference goal, namely edge selection, but now enables estimation for problems of realistic dimensions, which was a key ambition of our work. If desired, parameter uncertainty could still be quantified using an additional bootstrapping procedure.

We have taken advantage of jointGHS to study the gene regulation mechanisms underpinning immune-mediated diseases, using monocyte expression from four immune stimulation conditions. Joint inference on these data identified biologically-supported links, and allowed us to formulate sound mechanistic hypotheses. %

There are many possible extensions. For example, given the increasing prevalence of longitudinal studies, a natural continuation would be to propose a time-variant version of the model. Such an extension could be particularly profitable for studies aimed at understanding disease progression, and may also be relevant for the omics application of this paper, where two of the monocyte conditions involved exposure to differing durations of lipopolysaccharide (LPS 2h and LPS 24h). In settings with large numbers of timepoints, autoregression-like approach could be developed, where information would be shared between successive time points. 

To conclude, our approach is, to our knowledge, the first to extend graphical models based on global-local priors to the multiple network setting. 
 Additionally, thanks to its remarkable scalability, jointGHS effectively bridges the gap between Bayesian joint network modelling and large-scale inference for real-world studies such as encountered in statistical omics.

\section*{Software}
 The ECM graphical horseshoe approach for single or multiple networks has been implemented in the R packages fastGHS (\url{https://github.com/Camiling/fastGHS}) and jointGHS (\url{https://github.com/Camiling/jointGHS}), and has all subroutines implemented in C++ for computational efficiency. R code for the simulations and data analyses in this paper is available at \url{https://github.com/Camiling/jointGHS_simulations} and \url{https://github.com/Camiling/jointGHS_analysis}.

\section*{Data}
The CD14$^+$ monocyte gene expression data used in this study has been generated with HumanHT-12 v4 arrays and freely available for downloading from ArrayExpress45 \cite[accession E-MTAB-2232,][]{fairfax2012genetics, fairfax2014innate}.


\section*{Funding}
This research is funded by the UK Medical Research
Council programme MRC MC UU 00002/10 (C.L., H.R. and S.R.), Aker Scholarship (C.L.), Lopez--Loreta Foundation (H.R.) and Wellcome Intermediate Clinical Fellowship 01488/Z/16/Z (B.P.F.).



\bibliographystyle{unsrtnat} 
\bibliography{bibliography}       

\appendix
\renewcommand\thefigure{\thesection.\arabic{figure}}    
\renewcommand\thetable{\thesection.\arabic{table} }
\section{Appendix}
\setcounter{figure}{0}
\setcounter{table}{0}

\section{Derivation of the ECM algorithm with $\tau$ fixed}
\label{supp:sec:derive_ECM_taufixed}

\subsection{Deriving the objective function}
Assuming $\tau$ is fixed, the objective function for the graphical horseshoe ECM algorithm is as follows: 

\begin{align}
\label{Supp:eq:ObjectiveFuncFull_taufixed}
    Q(\boldsymbol{\Theta}, \boldsymbol{\Lambda} \vert \boldsymbol{\Theta}^{(l)}, \boldsymbol{\Lambda}^{(l)}) = & \text{E}_{\boldsymbol{N}\vert \boldsymbol{\Theta}^{(l)}, \boldsymbol{\Lambda}^{(l)}, \boldsymbol{S}}[\log p(\boldsymbol{\Theta}, \boldsymbol{\Lambda}, \boldsymbol{N}\vert \boldsymbol{S}) \vert \boldsymbol{\Theta}^{(l)}, \boldsymbol{\Lambda}^{(l)}] \nonumber \\
    =& \text{E}_{\cdot \vert \cdot}\big[\log \big \{ p ( \boldsymbol{S}\vert \boldsymbol{\Theta}) p(\boldsymbol{\Theta} \vert \boldsymbol{\Lambda}) p(\boldsymbol{\Lambda}\vert \boldsymbol{N}) p(\boldsymbol{N}) \big \} \vert \boldsymbol{\Theta}^{(l)}, \boldsymbol{\Lambda}^{(l)}\big]\nonumber \\ & +\text{const.} \nonumber \\
    =& \text{E}_{\cdot \vert \cdot} \bigg[ \frac{n}{2}\log{(\det{\boldsymbol{\Theta}})} - \frac{1}{2}\text{tr}(\boldsymbol{S}\boldsymbol{\Theta}) + \sum_{i<j} \Big\{ -\frac{1}{2} \log(\lambda_{ij}^2 \tau^2) - \frac{\theta_{ij}^2}{2\lambda_{ij}^2\tau^2}
    \nonumber \\
    & +\frac{1}{2}  \log{\bigg(\frac{1}{\nu_{ij}}\bigg)} + \frac{3}{2} \log{\bigg(\frac{1}{\lambda_{ij}^2}\bigg)} - \frac{1}{\nu_{ij}\lambda_{ij}^2} + \frac{3}{2} \log{\bigg(\frac{1}{\nu_{ij}}\bigg)} - \frac{1}{\nu_{ij}} \Big\}\bigg] \nonumber \\ 
    &    +\text{const.} \nonumber \\
    =& \frac{n}{2}\log{(\det{\boldsymbol{\Theta}})} - \frac{1}{2}\text{tr}(\boldsymbol{S}\boldsymbol{\Theta}) + \sum_{i<j} \Big\{ -\log{(\lambda_{ij})} - \frac{\theta_{ij}^2}{2\tau^2\lambda_{ij}^2}
     \nonumber \\
    & - \frac{1}{2} \text{E}_{\cdot \vert \cdot} \big[\log{(\nu_{ij})}\big] - 3\log{(\lambda_{ij})} - \frac{1}{\lambda_{ij}^2} \text{E}_{\cdot \vert \cdot} \bigg[\frac{1}{\nu_{ij}}\bigg] - \frac{3}{2} \text{E}_{\cdot \vert \cdot} \big[\log{(\nu_{ij})}\big] \nonumber \\
    & - \text{E}_{\cdot \vert \cdot} \bigg[ \frac{1}{\nu_{ij}}\bigg] \Big\} +\text{const.} \nonumber \\
    = & \frac{n}{2}\log{(\det{\boldsymbol{\Theta}})} - \frac{1}{2}\text{tr}(\boldsymbol{S}\boldsymbol{\Theta}) + \sum_{i<j} \Big\{ -4 \log{(\lambda_{ij})} - \frac{\theta_{ij}^2}{2\tau^2\lambda_{ij}^2}  \\
    & - 2 \text{E}_{\cdot \vert \cdot} \big[ \log{(\nu_{ij})} \big] - \text{E}_{\cdot \vert \cdot} \bigg[ \frac{1}{\nu_{ij}} \bigg] - \frac{1}{\lambda_{ij}^2}\text{E}_{\cdot \vert \cdot} \bigg[\frac{1}{\nu_{ij}} \bigg] \Big\}  +\text{const.}  \nonumber \\
    = & \frac{n}{2}\log{(\det{\boldsymbol{\Theta}})} - \frac{1}{2}\text{tr}(\boldsymbol{S}\boldsymbol{\Theta}) + \sum_{i<j} \Big\{ -4 \log{(\lambda_{ij})} - \frac{\theta_{ij}^2}{2\tau^2\lambda_{ij}^2} - 2 \text{E}_{\cdot \vert \cdot} \big[ \log{(\nu_{ij})}\big] \nonumber \\
    & - \bigg( \frac{1}{\lambda_{ij}^2} +1 \bigg)\text{E}_{\cdot \vert \cdot} \bigg[ \frac{1}{\nu_{ij}}\bigg]\Big\}  +\text{const.} \nonumber
\end{align}

\subsection{Deriving the E-step}

Given the updates for the $\nu_{ij}$'s 

\begin{align}
\label{Supp:eq:Estep_taufixed}
    &\text{E}_{\cdot \vert \cdot} \left\{ \log{(\nu_{ij})}\right\} = \log{\bigg( 1 + \frac{1}{\lambda_{ij}^{2^{(l)}}} \bigg)} - \psi(1), \nonumber \\
     &\text{E}_{\cdot \vert \cdot} \left( \frac{1}{\nu_{ij}}\right) = \frac{1}{1+1/\lambda_{ij}^{2^{(l)}}} = \frac{\lambda_{ij}^{2^{(l)}}}{\lambda_{ij}^{2^{(l)}}+1} =:\lambda_{ij}^{*^{(l)}}, 
\end{align}
the objective function becomes

\begin{align}
    \label{Supp:eq:ObjectiveFuncAfterE_taufixed}
    Q(\boldsymbol{\Theta}, \boldsymbol{\Lambda} \vert \boldsymbol{\Theta}^{(l)}, \boldsymbol{\Lambda}^{(l)}) =& \frac{n}{2}\log{(\det{\boldsymbol{\Theta}})} - \frac{1}{2}\text{tr}(\boldsymbol{S}\boldsymbol{\Theta}) + \sum_{i<j} \Big\{ -4 \log{(\lambda_{ij})} - \frac{\theta_{ij}^2}{2\tau^2\lambda_{ij}^2} - 2\log{\bigg( 1 + \frac{1}{\lambda_{ij}^{2^{(l)}}} \bigg)} \nonumber \\
    & - \bigg( \frac{1}{\lambda_{ij}^2} +1 \bigg) \bigg(\frac{\lambda_{ij}^{2^{(l)}}}{\lambda_{ij}^{2^{(l)}}+1} \bigg) \Big\} +\text{const.} \nonumber \\
    = & \frac{n}{2}\log{(\det{\boldsymbol{\Theta}})} - \frac{1}{2}\text{tr}(\boldsymbol{S}\boldsymbol{\Theta}) + \sum_{i<j} \Big\{ -4 \log{(\lambda_{ij})} - \frac{\theta_{ij}^2}{2\tau^2\lambda_{ij}^2} - 2\log{\bigg( \frac{\lambda_{ij}^{2^{(l)}}+1}{\lambda_{ij}^{2^{(l)}}} \bigg)} \nonumber \\
    & - \bigg( \frac{1}{\lambda_{ij}^2} +1 \bigg) \bigg(\frac{\lambda_{ij}^{2^{(l)}}}{\lambda_{ij}^{2^{(l)}}+1} \bigg) \Big\}  +\text{const.} \nonumber \\
    = & \frac{n}{2}\log{(\det{\boldsymbol{\Theta}})} - \frac{1}{2}\text{tr}(\boldsymbol{S}\boldsymbol{\Theta}) + \sum_{i<j} \Big\{ -4 \log{(\lambda_{ij})} - \frac{\theta_{ij}^2}{2\tau^2\lambda_{ij}^2} + 2\log{\Big( \lambda_{ij}^{*^{(l)}}\Big)}  \\
    & - \bigg( \frac{1}{\lambda_{ij}^2} +1 \bigg) \lambda_{ij}^{*^{(l)}}  \Big\} +\text{const.} \nonumber \\
    = & \frac{n}{2}\log{(\det{\boldsymbol{\Theta}})} - \frac{1}{2}\text{tr}(\boldsymbol{S}\boldsymbol{\Theta}) + \sum_{i<j} \Big\{ -4 \log{(\lambda_{ij})} - \frac{\theta_{ij}^2}{2\tau^2\lambda_{ij}^2} 
     -  \frac{\lambda_{ij}^{*^{(l)}}}{\lambda_{ij}^2} \Big\}+\text{const.} \nonumber
\end{align}

\subsection{Deriving the CM-step}

The maximisation of the objective function (\ref{Supp:eq:ObjectiveFuncAfterE_taufixed}) with respect to the $\lambda_{ij}^2$'s is easily found by solving

\begin{align}
    \frac{\partial Q}{\partial \lambda_{ij}^2} &= 0 \nonumber \\
    -\frac{2}{\lambda_{ij}^2} + \frac{\theta_{ij}}{2\tau^2 (\lambda_{ij}^2)^2} + \frac{\lambda_{ij}^{*^{(l)}}}{(\lambda_{ij}^2)^2} &= 0 \nonumber \\
    2\lambda_{ij}^2 - \frac{\theta_{ij}^2}{2\tau^2} - \lambda_{ij}^{*^{(l)}} &= 0 \nonumber
\end{align}
which has a closed-form solution that gives the updates for $\boldsymbol{\Lambda}$:

\begin{align}
    \lambda_{ij}^{2^{(l+1)}} = \frac{\lambda_{ij}^{*^{(l)}} + \theta_{ij}^2/(2\tau^2)}{2}. \nonumber
\end{align}

\subsection{Deriving full conditional posteriors for multiple networks}

The full conditional posteriors of the $\nu_{ij}$'s depends on the $\lambda_{ijk}$'s of all $K$ networks. They are given by 

\begin{align}
    p(\nu_{ij} \vert \cdot ) &= p(\nu_{ij} \vert  \{ \lambda_{ijk}^2 \}_{k=1}^K ) \nonumber \\ 
    &\propto p(\{\lambda_{ijk}^2 \}_{k=1}^K \vert \nu_{ij}) p(\nu_{ij}) \nonumber \\ 
    &= \left(\prod_{k=1}^K p(\lambda_{ijk}^2 \vert \nu_{ij})\right) p(\nu_{ij}) \nonumber \\ 
    &= \left(\prod_{k=1}^K \frac{1}{\nu_{ij}^{1/2}} \exp{\left( -\frac{1}{\nu_{ij}\lambda_{ijk}^2}\right)} \right)\frac{1}{\nu_{ij}^{3/2}}\exp{\left( -\frac{1}{\nu_{ij}}\right)} \nonumber\\
    &= \frac{1}{\nu_{ij}^{(K+3)/2}} \exp{\left( -\frac{1}{\nu_{ij}} \left( 1+\sum_{k=1}^K \frac{1}{\lambda_{ijk}^2} \right) \right)} \nonumber \\
    &\propto \text{InvGamma}\left(\frac{K+1}{2}, 1+\sum_{k=1}^K \frac{1}{\lambda_{ijk}^2}\right). \nonumber
\end{align}

\section{Derivation of the ECM algorithm without $\tau$ fixed}

\subsection{Full model}
When including $\tau$ in the updating scheme, we let it follow a half Cauchy distribution $\tau \sim \text{C}^+(0,1)$, like the local scales.

\subsection{Full conditional posterior}
When including $\tau$ in the updating scheme, we use the same reparameterisation trick as for the $\lambda_{ij}$'s, i.e. we let $\tau^2 \vert \xi \sim \text{InvGamma}(1/2, 1/\xi)$ and $\xi \sim \text{InvGamma}(1/2, 1)$, and get inverse Gamma conditional posteriors for them as well. All other conditional posteriors are as in the setting when $\tau$ is fixed. 

\subsection{Deriving the objective function}

The objective function for the graphical horseshoe ECM algorithm is then as follows: 

\begin{align}
\label{Supp:eq:ObjectiveFuncFull}
    Q(\boldsymbol{\Theta}, \boldsymbol{\Lambda}, \tau  \vert \boldsymbol{\Theta}^{(l)}, \boldsymbol{\Lambda}^{(l)}, \tau^{(l)} ) = & \text{E}_{\boldsymbol{N}, \xi \vert \boldsymbol{\Theta}^{(l)}, \boldsymbol{\Lambda}^{(l)}, \tau^{(l)}, \boldsymbol{S}}[\log p(\boldsymbol{\Theta}, \boldsymbol{\Lambda}, \tau, \boldsymbol{N}, \xi \vert \boldsymbol{S}) \vert \boldsymbol{\Theta}^{(l)}, \boldsymbol{\Lambda}^{(l)}, \tau^{(l)}] \nonumber \\
    =& \text{E}_{\cdot \vert \cdot}\big[\log \big \{ p ( \boldsymbol{S}\vert \boldsymbol{\Theta}) p(\boldsymbol{\Theta} \vert \boldsymbol{\Lambda}, \tau) p(\boldsymbol{\Lambda}\vert \boldsymbol{N}) p(\boldsymbol{N}) p(\tau \vert \xi) p(\xi) \big \} \vert \boldsymbol{\Theta}^{(l)}, \boldsymbol{\Lambda}^{(l)}, \tau^{(l)}\big]\nonumber \\ & +\text{const.} \nonumber \\
    =& \text{E}_{\cdot \vert \cdot} \bigg[ \frac{n}{2}\log{(\det{\boldsymbol{\Theta}})} - \frac{1}{2}\text{tr}(\boldsymbol{S}\boldsymbol{\Theta}) + \sum_{i<j} \Big\{ -\frac{1}{2} \log(\lambda_{ij}^2 \tau^2) - \frac{\theta_{ij}^2}{2\lambda_{ij}^2\tau^2}
    \nonumber \\
    & +\frac{1}{2}  \log{\bigg(\frac{1}{\nu_{ij}}\bigg)} + \frac{3}{2} \log{\bigg(\frac{1}{\lambda_{ij}^2}\bigg)} - \frac{1}{\nu_{ij}\lambda_{ij}^2} + \frac{3}{2} \log{\bigg(\frac{1}{\nu_{ij}}\bigg)} - \frac{1}{\nu_{ij}} \Big\} \nonumber \\ 
    & + \frac{1}{2}\log{\bigg(\frac{1}{\xi}\bigg)} + \frac{3}{2} \log{\bigg(\frac{1}{\tau^2}\bigg)} -\frac{1}{\tau^2\xi} + \frac{3}{2} \log{\bigg(\frac{1}{\xi}\bigg)} - \frac{1}{\xi}  \bigg] +\text{const.} \nonumber \\
    =& \frac{n}{2}\log{(\det{\boldsymbol{\Theta}})} - \frac{1}{2}\text{tr}(\boldsymbol{S}\boldsymbol{\Theta}) + \sum_{i<j} \Big\{ -\log{(\lambda_{ij})} - \log{(\tau)} - \frac{\theta_{ij}^2}{2\tau^2\lambda_{ij}^2}
     \nonumber \\
    & - \frac{1}{2} \text{E}_{\cdot \vert \cdot} \big[\log{(\nu_{ij})}\big] - 3\log{(\lambda_{ij})} - \frac{1}{\lambda_{ij}^2} \text{E}_{\cdot \vert \cdot} \bigg[\frac{1}{\nu_{ij}}\bigg] - \frac{3}{2} \text{E}_{\cdot \vert \cdot} \big[\log{(\nu_{ij})}\big] \nonumber \\
    & - \text{E}_{\cdot \vert \cdot} \bigg[ \frac{1}{\nu_{ij}}\bigg] \Big\} - 2 \text{E}_{\cdot \vert \cdot} \big[ \log{(\xi)} \big] - 3 \log{(\tau)} - \bigg(1+\frac{1}{\tau^2}\bigg)\text{E}_{\cdot \vert \cdot} \bigg[ \frac{1}{\xi} \bigg]+\text{const.} \nonumber \\
    = & \frac{n}{2}\log{(\det{\boldsymbol{\Theta}})} - \frac{1}{2}\text{tr}(\boldsymbol{S}\boldsymbol{\Theta}) + \sum_{i<j} \Big\{ -4 \log{(\lambda_{ij})} - \frac{\theta_{ij}^2}{2\tau^2\lambda_{ij}^2}  \\
    & - 2 \text{E}_{\cdot \vert \cdot} \big[ \log{(\nu_{ij})} \big] - \text{E}_{\cdot \vert \cdot} \bigg[ \frac{1}{\nu_{ij}} \bigg] - \frac{1}{\lambda_{ij}^2}\text{E}_{\cdot \vert \cdot} \bigg[\frac{1}{\nu_{ij}} \bigg] \Big\} - \bigg( \frac{p(p-1)}{2} + 3 \bigg) \log{(\tau)} \nonumber \\
    & - 2  \text{E}_{\cdot \vert \cdot} \big[ \log{(\xi)} \big] - \bigg( 1 + \frac{1}{\tau^2}\bigg) \text{E}_{\cdot \vert \cdot} \bigg[\frac{1}{\xi} \bigg] +\text{const.}  \nonumber \\
    = & \frac{n}{2}\log{(\det{\boldsymbol{\Theta}})} - \frac{1}{2}\text{tr}(\boldsymbol{S}\boldsymbol{\Theta}) + \sum_{i<j} \Big\{ -4 \log{(\lambda_{ij})} - \frac{\theta_{ij}^2}{2\tau^2\lambda_{ij}^2} - 2 \text{E}_{\cdot \vert \cdot} \big[ \log{(\nu_{ij})}\big] \nonumber \\
    & - \bigg( \frac{1}{\lambda_{ij}^2} +1 \bigg)\text{E}_{\cdot \vert \cdot} \bigg[ \frac{1}{\nu_{ij}}\bigg]\Big\} - \bigg(\frac{p(p-1)}{2}+3 \bigg)\log{(\tau)} - 2 \text{E}_{\cdot \vert \cdot} \big[ \log{(\xi)} \big] \nonumber \\
    & - \bigg( 1 + \frac{1}{\tau^2}\bigg)\text{E}_{\cdot \vert \cdot} \bigg[ \frac{1}{\xi}\bigg] +\text{const.} \nonumber
\end{align}
where $\text{E}_{\cdot \vert \cdot}(\cdot)$ denotes $\text{E}_{\boldsymbol{N}, \xi \vert \boldsymbol{\Theta}^{(l)}, \boldsymbol{\Lambda}^{(l)}, \tau^{(l)}, \boldsymbol{S}}(\cdot)$ and const. is a constant not depending on $\boldsymbol{\Theta}, \boldsymbol{\Lambda}$ or $\tau$. 

In the E-step of the algorithm, the conditional expectations in (\ref{Supp:eq:ObjectiveFuncFull}) are computed. After the E-step is computed, the CM-step performs the maximisation with respect to $(\boldsymbol{\Theta}, \boldsymbol{\Lambda}, \tau)$. 

\subsection{Deriving the E-step}

Given the updates for the $\nu_{ij}$'s and $\tau$

\begin{align}
\label{Supp:eq:Estep}
    &\text{E}_{\cdot \vert \cdot} \left\{ \log{(\nu_{ij})}\right\} = \log{\bigg( 1 + \frac{1}{\lambda_{ij}^{2^{(l)}}} \bigg)} - \psi(1), \nonumber \\
     &\text{E}_{\cdot \vert \cdot} \left( \frac{1}{\nu_{ij}}\right) = \frac{1}{1+1/\lambda_{ij}^{2^{(l)}}} = \frac{\lambda_{ij}^{2^{(l)}}}{\lambda_{ij}^{2^{(l)}}+1} =:\lambda_{ij}^{*^{(l)}}, \nonumber \\
    &\text{E}_{\cdot \vert \cdot} \left\{ \log{(\xi)}\right\} = \log{\bigg( 1 + \frac{1}{\tau^{{(l)}^2}} \bigg)} - \psi(1), \\
    &\text{E}_{\cdot \vert \cdot} \left( \frac{1}{\xi}\right) = \frac{1}{1+1/\tau^{{(l)}^2}} = \frac{\tau^{{(l)}^2}}{\tau^{{(l)}^2}+1} =: \tau^{*^{(l)}},\nonumber 
\end{align}
the objective function becomes

\begin{align}
    \label{Supp:eq:ObjectiveFuncAfterE}
    Q(\boldsymbol{\Theta}, \boldsymbol{\Lambda}, \tau  \vert \boldsymbol{\Theta}^{(l)}, \boldsymbol{\Lambda}^{(l)}, \tau^{(l)} ) =& \frac{n}{2}\log{(\det{\boldsymbol{\Theta}})} - \frac{1}{2}\text{tr}(\boldsymbol{S}\boldsymbol{\Theta}) + \sum_{i<j} \Big\{ -4 \log{(\lambda_{ij})} - \frac{\theta_{ij}^2}{2\tau^2\lambda_{ij}^2} - 2\log{\bigg( 1 + \frac{1}{\lambda_{ij}^{2^{(l)}}} \bigg)} \nonumber \\
    & - \bigg( \frac{1}{\lambda_{ij}^2} +1 \bigg) \bigg(\frac{\lambda_{ij}^{2^{(l)}}}{\lambda_{ij}^{2^{(l)}}+1} \bigg) \Big\} - \bigg(\frac{p(p-1)}{2}+3 \bigg)\log{(\tau)} - 2 \log{\bigg( 1 + \frac{1}{\tau^{{(l)}^2}} \bigg)} \nonumber \\
    & - \bigg( 1 + \frac{1}{\tau^2}\bigg) \bigg(\frac{\tau^{{(l)}^2}}{\tau^{{(l)}^2}+1}\bigg) +\text{const.} \nonumber \\
    = & \frac{n}{2}\log{(\det{\boldsymbol{\Theta}})} - \frac{1}{2}\text{tr}(\boldsymbol{S}\boldsymbol{\Theta}) + \sum_{i<j} \Big\{ -4 \log{(\lambda_{ij})} - \frac{\theta_{ij}^2}{2\tau^2\lambda_{ij}^2} - 2\log{\bigg( \frac{\lambda_{ij}^{2^{(l)}}+1}{\lambda_{ij}^{2^{(l)}}} \bigg)} \nonumber \\
    & - \bigg( \frac{1}{\lambda_{ij}^2} +1 \bigg) \bigg(\frac{\lambda_{ij}^{2^{(l)}}}{\lambda_{ij}^{2^{(l)}}+1} \bigg) \Big\} - \bigg(\frac{p(p-1)}{2}+3 \bigg)\log{(\tau)} - 2 \log{\bigg( \frac{\tau^{{(l)}^2}+1}{\tau^{{(l)}^2}} \bigg)} \nonumber \\
    & - \bigg( 1 + \frac{1}{\tau^2}\bigg) \bigg(\frac{\tau^{{(l)}^2}}{\tau^{{(l)}^2}+1}\bigg) +\text{const.} \nonumber \\
    = & \frac{n}{2}\log{(\det{\boldsymbol{\Theta}})} - \frac{1}{2}\text{tr}(\boldsymbol{S}\boldsymbol{\Theta}) + \sum_{i<j} \Big\{ -4 \log{(\lambda_{ij})} - \frac{\theta_{ij}^2}{2\tau^2\lambda_{ij}^2} + 2\log{\Big( \lambda_{ij}^{*^{(l)}}\Big)}  \\
    & - \bigg( \frac{1}{\lambda_{ij}^2} +1 \bigg) \lambda_{ij}^{*^{(l)}}  \Big\} - \bigg(\frac{p(p-1)}{2}+3 \bigg)\log{(\tau)} + 2 \log{\Big( \tau^{*^{(l)}} \Big)} \nonumber \\
    & - \bigg( 1 + \frac{1}{\tau^2}\bigg)\tau^{*^{(l)}}  +\text{const.} \nonumber \\
    = & \frac{n}{2}\log{(\det{\boldsymbol{\Theta}})} - \frac{1}{2}\text{tr}(\boldsymbol{S}\boldsymbol{\Theta}) + \sum_{i<j} \Big\{ -4 \log{(\lambda_{ij})} - \frac{\theta_{ij}^2}{2\tau^2\lambda_{ij}^2} 
     -  \frac{\lambda_{ij}^{*^{(l)}}}{\lambda_{ij}^2} \Big\}\nonumber \\
     & - \bigg(\frac{p(p-1)}{2}+3 \bigg)\log{(\tau)} - \frac{\tau^{*^{(l)}}}{\tau^2} +\text{const.} \nonumber
\end{align}

\subsection{Deriving the CM-step}

The maximisation of the objective function (\ref{Supp:eq:ObjectiveFuncAfterE}) with respect to $\tau$ is easily found by solving

\begin{align}
    \frac{\partial Q}{\partial \tau} &= 0 \nonumber \\
    \sum_{i<j}\bigg\{ \frac{\theta_{ij}^2}{\lambda_{ij}^2\tau^3} \bigg\} - \bigg( \frac{p(p-1)}{2}+3 \bigg) \frac{1}{\tau} + \frac{2}{\tau^3}\tau^{*^{(l)}} &= 0 \nonumber \\
    2\sum_{i<j}\bigg\{\frac{\theta_{ij}^2}{\lambda_{ij}^2} \bigg\} - (p(p-1)+6)\tau^2+4\tau^{*^{(l)}} &= 0  \nonumber
\end{align}
which has a closed-form solution that gives the update

\begin{align}
    \tau^{{(l+1)}^2} &= \frac{2\sum_{i<j}\bigg\{\frac{\theta_{ij}^2}{\lambda_{ij}^2}\bigg\} + 4\tau^{*^{(l)}}}{p(p-1)+6}. 
    \label{supp:tau_update}
\end{align}
Similarly, we can maximise (\ref{Supp:eq:ObjectiveFuncFull}) with respect to the $\lambda_{ij}^2$'s to find the updates for $\boldsymbol{\Lambda}$: 

\begin{align}
    \frac{\partial Q}{\partial \lambda_{ij}^2} &= 0 \nonumber \\
    -\frac{2}{\lambda_{ij}^2} + \frac{\theta_{ij}}{2\tau^2 (\lambda_{ij}^2)^2} + \frac{\lambda_{ij}^{*^{(l)}}}{(\lambda_{ij}^2)^2} &= 0 \nonumber \\
    2\lambda_{ij}^2 - \frac{\theta_{ij}^2}{2\tau^2} - \lambda_{ij}^{*^{(l)}} &= 0 \nonumber
\end{align}
which is solved by
\begin{align}
    \lambda_{ij}^{2^{(l+1)}} = \frac{\lambda_{ij}^{*^{(l)}} + \theta_{ij}^2/(2\tau^2)}{2}. \nonumber
\end{align}

The updates for $\boldsymbol{\Theta}$ are as in \ref{supp:sec:derive_ECM_taufixed}, but with the fixed $\tau$ replaced by the latest update $\tau^{(l)}$.

\subsection{Deriving full conditional posteriors for multiple networks, with $\tau_k$'s fixed}

The full conditional posteriors of the $\nu_{ij}$'s depends on the $\lambda_{ijk}$'s of all $K$ networks. They are given by 

\begin{align}
    p(\nu_{ij} \vert \cdot ) &= p(\nu_{ij} \vert  \{ \lambda_{ijk}^2 \}_{k=1}^K ) \nonumber \\ 
    &\propto p(\{\lambda_{ijk}^2 \}_{k=1}^K \vert \nu_{ij}) p(\nu_{ij}) \nonumber \\ 
    &= \left(\prod_{k=1}^K p(\lambda_{ijk}^2 \vert \nu_{ij})\right) p(\nu_{ij}) \nonumber \\ 
    &= \left(\prod_{k=1}^K \frac{1}{\nu_{ij}^{1/2}} \exp{\left( -\frac{1}{\nu_{ij}\lambda_{ijk}^2}\right)} \right)\frac{1}{\nu_{ij}^{3/2}}\exp{\left( -\frac{1}{\nu_{ij}}\right)} \nonumber\\
    &= \frac{1}{\nu_{ij}^{(K+3)/2}} \exp{\left( -\frac{1}{\nu_{ij}} \left( 1+\sum_{k=1}^K \frac{1}{\lambda_{ijk}^2} \right) \right)} \nonumber \\
    &\propto \text{InvGamma}\left(\frac{K+1}{2}, 1+\sum_{k=1}^K \frac{1}{\lambda_{ijk}^2}\right). \nonumber
\end{align}

\subsection{Deriving full conditional posteriors for multiple networks, without $\tau_k$'s fixed}

When including the $\tau_k$'s in the updating scheme, we let them follow half Cauchy distributions $\tau_k \sim \text{C}^+(0,1)$, like the local scales. Similarly, like for the local scales, we introduce the latent variables $\xi_k$ for the $\tau_k^2$'s with a similar parameterisation $\tau_k^2 \vert \xi_k \sim \text{InvGamma}(1/2, 1/\xi_k)$ and $\xi_k \sim \text{InvGamma}(1/2, 1)$, and get inverse Gamma full conditional posteriors for them as well. The global scales are kept network-specific to allow for different sparsity levels across networks.

The full conditional posteriors of the $\nu_{ij}$'s depends on the $\lambda_{ijk}$'s of all $K$ networks. They are given by 

\begin{align}
    p(\nu_{ij} \vert \cdot ) &= p(\nu_{ij} \vert  \{ \lambda_{ijk}^2 \}_{k=1}^K ) \nonumber \\ 
    &\propto p(\{\lambda_{ijk}^2 \}_{k=1}^K \vert \nu_{ij}) p(\nu_{ij}) \nonumber \\ 
    &= \left(\prod_{k=1}^K p(\lambda_{ijk}^2 \vert \nu_{ij})\right) p(\nu_{ij}) \nonumber \\ 
    &= \left(\prod_{k=1}^K \frac{1}{\nu_{ij}^{1/2}} \exp{\left( -\frac{1}{\nu_{ij}\lambda_{ijk}^2}\right)} \right)\frac{1}{\nu_{ij}^{3/2}}\exp{\left( -\frac{1}{\nu_{ij}}\right)} \nonumber\\
    &= \frac{1}{\nu_{ij}^{(K+3)/2}} \exp{\left( -\frac{1}{\nu_{ij}} \left( 1+\sum_{k=1}^K \frac{1}{\lambda_{ijk}^2} \right) \right)} \nonumber \\
    &\propto \text{InvGamma}\left(\frac{K+1}{2}, 1+\sum_{k=1}^K \frac{1}{\lambda_{ijk}^2}\right). \nonumber
\end{align}

In the E-step, the $\xi_k$'s follow the same distribution as in the standard graphical horseshoe within each network. This means that

\begin{align}
    &\text{E}_{\cdot \vert \cdot} \left\{ \log{(\xi_k)}\right\} = \log{\bigg( 1 + \frac{1}{\tau_k^{{(l)}^2}} \bigg)} - \psi(1), \nonumber \\
    &\text{E}_{\cdot \vert \cdot} \left( \frac{1}{\xi_k}\right) = \frac{1}{1+1/\tau_k^{{(l)}^2}} = \frac{\tau_k^{{(l)}^2}}{\tau_k^{{(l)}^2}+1} =: \tau_k^{*^{(l)}}. \nonumber
\end{align}

In the CM-step for multiple networks, the global shrinkage parameter for each graph $k$ is updated as the maximising value (\ref{supp:tau_update}) within the network:

\begin{align}
    \tau_k^{{(l+1)}^2} &= \frac{2\sum_{i<j}\bigg\{\frac{\theta_{ijk}^2}{\lambda_{ijk}^2}\bigg\} + 4\tau_k^{*^{(l)}}}{p(p-1)+6}.\nonumber
\end{align}

\section{Simulation study details}
\label{sec:simstudy}

To evaluate the performance our proposed methodology, we have done comprehensive simulation studies in R (\cite{Rcite}). We have used our R packages \texttt{fastGHS} and \texttt{jointGHS} to perform the single-network graphical horseshoe ECM routine and the joint graphical horseshoe ECM routine respectively, and the code for the simulations is available on Github (\url{https://github.com/Camiling/jointGHS_simulations}). 

\subsection{Single network inference}
\label{Supp:subsec:singlesim}
The first part of the simulation studies compares our graphical horseshoe ECM implementation (fastGHS) to the graphical horseshoe Gibbs sampler and the graphical lasso. The performance of the methods is assessed on multivariate Gaussian simulated data sets. 

\subsubsection*{Generating networks and data}

We aim to generate data similar to our omic application of interest. We have done this using the R package \texttt{huge}. The package provides the function \texttt{huge.generator()}, which generates multivariate Gaussian data with the \emph{scale-free property} as it is a known characteristic of multiomic data (\cite{kolaczyk09}). Given a number $p$ of vertices, \texttt{huge} constructs a precision matrix $\boldsymbol{\Theta}$ with $p$ edges. For a desired number of observations $n$, a data set can then be generated from the resulting multivariate Gaussian distribution with covariance matrix $\boldsymbol{\Sigma} = \boldsymbol{\Theta}^{-1}$ and expectation vector $\mathbf{0}$. This results in a $n$ by $p$ data matrix $\boldsymbol{X}$, where each column corresponds to a node. As for the values of the non-zero partial correlations, we let them be in $[0.1, 0.2]$, a range we found to reflect the values observed in our omic application. Such data is generated $N=20$ times. In our simulations, we consider different settings, i.e., different combinations of $p$ and $n$. When $p=50$, the final graph has $p$ edges and thus a \emph{sparsity} of $0.04$. When $p=100$, the sparsity of the graph is $0.02$.

\subsubsection*{Data analysis details}

In our simulation study, we wish to compare the graph reconstruction accuracy of the Gibbs sampling implementation of \cite{li2019bayesian} to our ECM implementation of the graphical horseshoe (fastGHS). We also include the graphical lasso of \cite{friedman2008}, a state-of-the-art frequentist method for Gaussian graph reconstruction. For each setting we consider, and each generated data set, we obtain a precision matrix estimate $\widehat{\boldsymbol{\Theta}}$ from the data $\boldsymbol{X}$ of interest for each of the methods. 

We perform the ordinary graphical lasso using the R function \texttt{huge}, while we perform the Gibbs sampling procedure using the code of \cite{li2019bayesian} translated by us from MATLAB to R. A translation is necessary for us to perform the simulation study in R, but we have ensured that it gives identical results to those of the original MATLAB implementation. 

The sparsity level of the graphical lasso is selected using the Stability Approach to Regularisation Selection (StARS), a selection method based on model stability (\cite{liu2010stability}), with variability threshold $\beta=0.03$. For the Gibbs sampling procedure of \cite{li2019bayesian}, we draw $nmc=1000$ MCMC samples after $burnin=100$ burn-in iterations. In fastGHS, we use a convergence tolerance threshold of $0.001$ for the precision matrix estimate and an AIC convergence threshold of $0.1$ for the global scale parameter selection. The results are averaged over the $N=20$ simulations for all settings, and we report the precision and recall, as well as the sparsity of the estimates. 

\subsubsection*{Computational time simulation details}

To illustrate the scalability of our ECM implementation for the graphical horseshoe, we save the CPU time used to infer a network for different numbers of nodes $p$, using Gaussian graphical data sets with $n=100$ observations. We limit ourselves to $p\leq90$, to permit comparison to the less scalable Gibbs sampler of \cite{li2019bayesian}. We run all methods on a 16-core Intel Xeon CPU, 2.60 GHz. We use the same parameters as for the single network simulation study, and consider the time used for one run, i.e. for a given value of $\tau^2$ in the joint graphical horseshoe or one initial value in the Gibbs sampler.

\subsection{Multiple network inference}

\subsubsection*{Generating Gaussian graphs and data of specific similarity}

We generate data for the joint graphical horseshoe simulation study using the same method as in 6.1 in the main manuscript, with the \texttt{huge} package in R to generate graphs and drawing data from the corresponding multivariate Gaussian distributions. The main difference is that we now have $K=2$ graphs instead of one. Because we wish to investigate different settings with various similarity of the $K=2$ true graph structures, we can not generate the graph structures independently. Starting by generating the first graph using \texttt{huge}, we instead modify its precision matrix in order to obtain a second graph with the desired level of similarity. Specifically, for the initial precision matrix we create, we permute a certain percentage of the edges by randomly reallocating them. This allows us to generate graphs of various similarity, ranging from $0\%$ edge disagreement (i.e. the same edge set) to $100\%$ edge disagreement (i.e. no common edges). In all settings, both graphs have true sparsity $0.04$. For each pair of graphs, we sample $N=100$ data sets from each of the two corresponding multivariate Gaussian distributions, with $n_1=50$ observations for the first graph and $n_2=80$ observations for the second. 

\subsubsection*{Data analysis details}

For each pair of generated data sets, we use the different joint graph reconstruction methods to obtain precision matrix estimates and assess their performance. In addition to our joint graphical horseshoe approach, we consider the Bayesian spike-and-slab joint graphical lasso of \cite{li2019bayesian} and the joint graphical lasso of \cite{danaher2014}. \cite{li2019bayesian} provide R code for the Bayesian spike-and-slab at \url{https://github.com/richardli/SSJGL}, and the joint graphical lasso is implemented by \cite{danaher2014} in the R package \texttt{JGL}. \cite{yang2021gembag} provide R code for GemBag at \url{https://github.com/xinming104/GemBag}. 

The spike-and-slab joint graphical lasso (SSJGL) has a wide range of parameters that must be selected, and we use the values \cite{li2019bayesian} use for their own simulations, including an EM convergence tolerance of $0.0001$. In the joint graphical lasso, the sparsity- and similarity controlling parameters must be selected. We do this using the approach suggested by \cite{danaher2014}, considering a grid of two parameters to find the values minimising their proposed adapted AIC score. For GemBag, we use the BIC to select the hyperparameters $v_0$ and $v_1$, as suggested by \cite{yang2021gembag}. For the other hyperparameters, we use the values suggested and used by \cite{yang2021gembag} in their simulations. 

In our joint graphical horseshoe approach, we use a convergence tolerance threshold of $0.001$ for the precision matrix estimate and an AIC convergence threshold of $0.1$ for the global scale parameter selection. For both graphs, the results are averaged over the $N=100$ simulations for all settings and methods, and we report the precision and recall as well as the sparsity of the estimates. When comparing jointGHS and fastGHS, we get the fastGHS estimate of each network to the same sparsity as the corresponding jointGHS estimate by setting the global scale parameter to a small value, and slowly increasing it until the same sparsity is achieved.

\section{Monocyte data}

\subsection{Data analysis details}

For the jointGHS application, we use a convergence tolerance of $0.001$ for the precision matrices. We use the same tolerance when applying fastGHS to each condition separately. Due to the large size of the problem, with $289\ 560$ potential links to be inferred, we use a looser AIC convergence threshold of $5$ for the global scale parameter selection to speed up computations. As previously discussed, inference is not very sensitive to this choice: a wide range of global scale parameter values tend to give similar results. When comparing the results from jointGHS and fastGHS, we get the fastGHS estimate of each condition to the same sparsity as the corresponding jointGHS estimate by setting the global scale parameter to a small value, and slowly increasing it until the same sparsity is achieved. 

\subsection{Additional results}

\subsubsection*{Model parameters}

\begin{figure}
    \centering
    \includegraphics[scale=0.6]{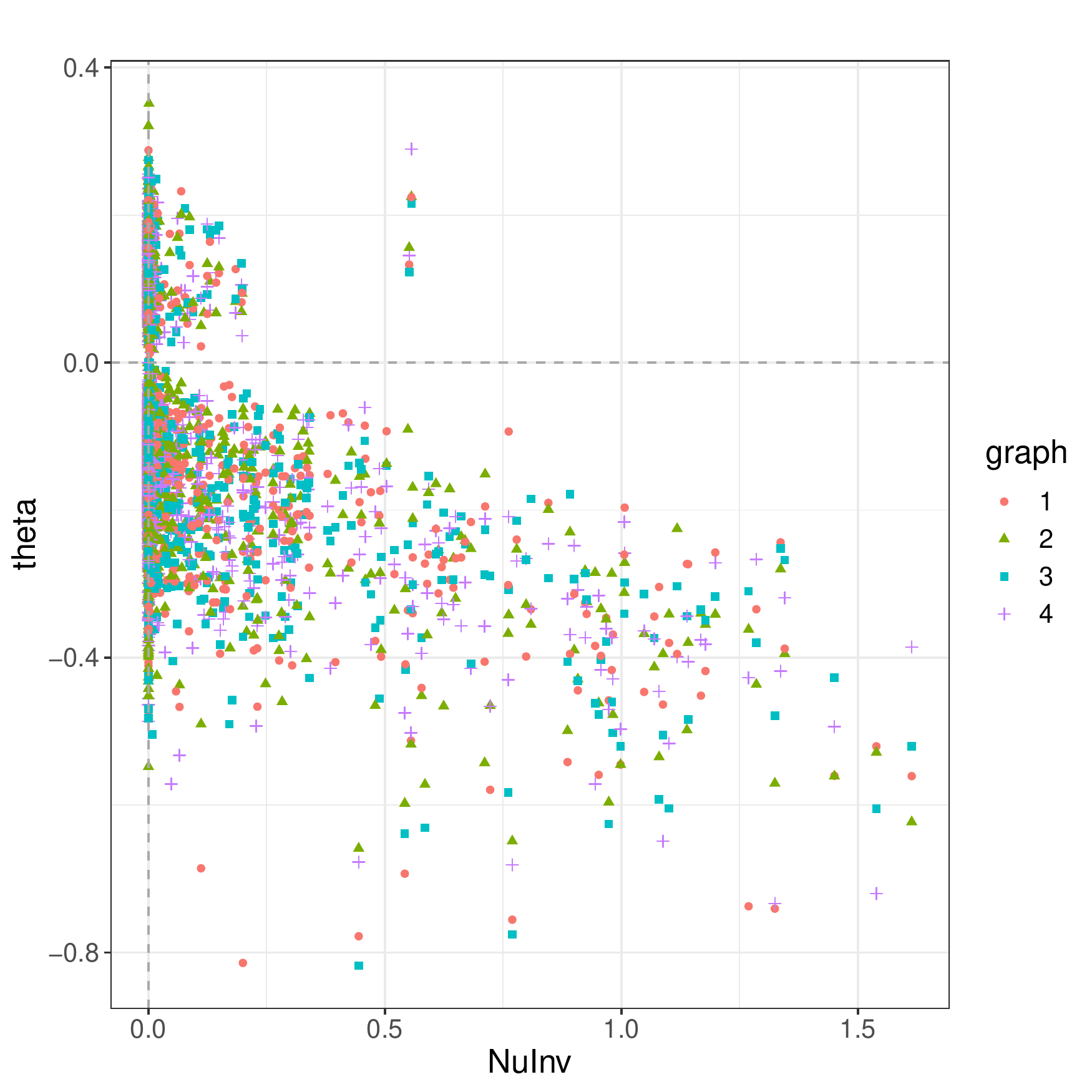}
    \caption{Scaled precision matrix elements of the jointGHS graph plotted against the common latent parameters of all conditions.}
    \label{fig:Monocyte_thetaVSNu}
\end{figure}

Figure \ref{fig:Monocyte_thetaVSNu} shows the scaled precision matrix elements of the jointGHS estimates plotted against the common latent parameters in all four conditions. We see that while many edges are found to be in common meaning information has been shared, there are still many network-specific edges that have been inferred. Thanks to the heavy horseshoe tail, we are able to capture these edges even though no common information about them is found between the conditions.

\subsubsection*{Degree distribution}

Table \ref{table:monocyte_topgenes} shows the node degree of the genes with degree larger than the $90^{\text{th}}$ percentile in the respective jointGHS networks of the different conditions. A complete list of all node degrees is given in Supplement B, and can be used for further investigation.

\begin{table}
	\centering
	\renewcommand{\arraystretch}{1.4}
	\caption{The genes with node degree larger than the $90^{\text{th}}$ percentile in the respective jointGHS networks of the different conditions. The genes that have node degree in the upper $10\%$ in all four conditions are marked in bold. The genes that only have node degree in the upper $10\%$ in one condition are marked in red. The genes that are controlled by the top hotspot in the condition in question are marked with $*$}
        \vspace{-0.5cm}
	\hspace*{-1cm}
	\resizebox{1.1\textwidth}{!}{
	\begin{tabular}{l r  r @{\hskip 0.8cm}l r r @{\hskip 0.8cm}l r r @{\hskip 0.8cm}l r}
        \multicolumn{11}{c}{}\\
        \toprule
	    \multicolumn{2}{c}{IFN-gamma}&&\multicolumn{2}{c}{LPS-2h}&&\multicolumn{2}{c}{LPS-24h} &&\multicolumn{2}{c}{Unstimulated}  \\
		\cmidrule(lr){1-2} \cmidrule(lr){4-5} \cmidrule(lr){7-8} \cmidrule(lr){10-11}
		Gene & Degree && Gene & Degree && Gene & Degree && Gene & Degree \\
		\hline
		ACBD5*  &  33 && \bf{AIRE}  &  37 &&\bf{AIRE}  &  38 &&\bf{AIRE}*  &  40  \\ 
        \bf{AIRE}*  &  25 &&B3GNT5  &  28 &&TNK2  &  25 &&\bf{AFMID}*  &  33  \\ 
        \bf{AFMID}*  &  24 &&\bf{AFMID}*  &  26 &&\bf{AFMID}*  &  25 &&PHRF1  &  24  \\ 
        \bf{LGALS3}  &  23 &&\bf{LGALS3}  &  24 &&COX6A1  &  24 &&\textbf{LGALS3}  &  23  \\ 
        \bf{SLC3A2}  &  20 &&COX6A1  &  24 &&\bf{LGALS3}  &  22 &&NBPF8  &  20  \\ 
        {\color{red}PHF1*}  &  20 &&ACBD5  &  24 &&TBC1D9  &  21 &&{\color{red}AKR1D1*}  &  20  \\ 
        \emph{YEATS4}*  &  18 &&\bf{AFTPH}  &  22 &&EPSTI1  &  21 &&\bf{AFTPH}  &  20  \\ 
        PTPLAD2*  &  18 &&PRPF8  &  21 &&CYP27A1  &  21 &&GIMAP1  &  19  \\ 
        GIMAP1  &  18 &&EPSTI1  &  21 &&YEATS4*  &  20 &&STAG3L3  &  18  \\ 
        COPZ1*  &  18 &&\bf{IMPDH1}  &  19 &&STK24  &  20 &&\bf{SLC3A2}  &  18  \\ 
        TNK2  &  17 &&GIMAP1  &  19 &&STAG3L3  &  20 &&ADAMTS4*  &  18  \\ 
        STK24  &  17 &&BIRC3  &  19 &&\bf{AFTPH}  &  20 &&{\color{red}VHL*}  &  17  \\ 
        \bf{SORL1}  &  17 &&{\color{red}TBC1D15}  &  18 &&{\color{red}CCL20}  &  19 &&SGK3*  &  17  \\ 
        SNX17  &  17 &&\bf{SORL1}  &  18 &&{\color{red}TNFSF14}  &  18 &&GSDM1*  &  17  \\ 
        \bf{RELB}  &  17 &&\bf{SLC3A2}  &  18 &&\bf{SORL1}  &  18 &&{\color{red}ZNF845}  &  16  \\ 
        CYP27A1  &  17 &&CYP27A1  &  18 &&\bf{SLC3A2}  &  18 &&YEATS4*  &  16  \\ 
        NLRP3  &  16 &&{\color{red}BFAR}  &  18 &&\bf{RELB}  &  18 &&TBC1D9  &  16  \\ 
        NBPF8  &  16 &&SNX17  &  17 &&PTPLAD2  &  18 &&KIAA0101*  &  16  \\ 
        KIAA0101*  &  16 &&{\color{red}PION}  &  17 &&{\color{red}GNB4}  &  18 &&{\color{red}JARID2}  &  16  \\ 
        EPSTI1  &  16 &&{\color{red}NOD1}  &  17 &&{\color{red}CRTC3}  &  18 &&{\color{red}GLTSCR1}  &  16  \\ 
        COX6A1  &  16 &&NLRP3  &  17 &&{\color{red}CD72}  &  18 &&PTPLAD2  &  15  \\ 
        B3GNT5  &  16 &&ALPP  &  17 &&BATF3  &  18 &&{\color{red}LYZ*}  &  15  \\ 
        SMC4  &  15 &&{\color{red}ZYX}  &  16 &&TLK1  &  17 &&FLJ38717  &  15  \\ 
        PRPF8*  &  15 &&SGK3  &  16 &&SMC4  &  17 &&{\color{red}COX19}  &  15  \\ 
        {\color{red}MAFF*}  &  15 &&\bf{RELB}  &  16 &&\bf{IMPDH1}  &  17 &&ALPP*  &  15  \\ 
        {\color{red}LOC728457}  &  15 &&NBPF8  &  16 &&HNRPR  &  17 &&AGTRAP*  &  15  \\ 
        \bf{IMPDH1}*  &  15 &&{\color{red}LOC641522}  &  16 &&{\color{red}DENND4C}  &  17 &&ACBD5*  &  15  \\ 
        GSDM1*  &  15 &&KIAA0101*  &  16 &&COPZ1  &  17 &&TLK1  &  14  \\ 
        BATF3  &  15 &&HNRPR  &  16 &&{\color{red}CHRNA5*}  &  17 &&\bf{SORL1}  &  14  \\ 
        {\color{red}TNFSF15*}  &  14 &&{\color{red}GTF2IRD2B}  &  16 &&C4orf34  &  17 &&\bf{RELB}  &  14  \\ 
        STAG3L3*  &  14 &&AGTRAP  &  16 &&{\color{red}TPM4}  &  16 &&{\color{red}PCDHB9}  &  14  \\ 
        SGK3*  &  14 && & &&SNX17  &  16 &&NLRP3  &  14  \\ 
        NFIL3  &  14 && & &&{\color{red}PLEKHB2}  &  16 &&{\color{red}KLHL8*}  &  14  \\ 
        HNRPR  &  14 && & &&PHRF1  &  16 &&\bf{IMPDH1}  &  14  \\ 
        FLJ38717*  &  14 && & &&NFIL3  &  16 &&{\color{red}IL12RB1}  &  14  \\ 
        BIRC3*  &  14 && & && & &&C4orf34*  &  14  \\ 
        \bf{AFTPH}*  &  14 && & && & &&{\color{red}C12orf43}  &  14  \\ 
        ADAMTS4*  &  14 && & && & &&BIRC3*  &  14  \\ 
        \toprule
	\end{tabular}
	}
	\label{table:monocyte_topgenes}
\end{table} 

\subsubsection*{Hub genes}

This discussion complements the hub gene subsection found in the main manuscript. A gene with many links consistent across all conditions is Galectin-3 (\emph{LGALS3}), whose expression is found to modulate T-cell growth and apoptosis (\cite{yang1996expression}). Studies indicate that it affects numerous biological processes through specific interactions with a variety of intra- and extracellular proteins, and has a regulatory role in both innate and adaptive immunity (\cite{dumic2006galectin, bernardes2006toxoplasma}). It has been found to be a negative regulator of lipopolysaccharide (LPS) mediated inflammation (\cite{li2008galectin, fermino2011lps}), and lack of Galectin-3 has been directly linked to higher $\text{IFN-}\gamma$ levels (\cite{nishi2007role, radosavljevic2012roles}). Finally, the Aftiphilin (\emph{AFTPH}) gene has many links in all conditions, including to \emph{YEATS4}. Aftiphilin has been found to be differentially expressed in the immune cells of tumor patients, suggesting it might affect tumor development through immune cell regulation (\cite{huang2021single}).

\subsubsection*{LYZ and YEATS4 and their neighbours}

\begin{figure}
    \centering
    \includegraphics[scale=0.4]{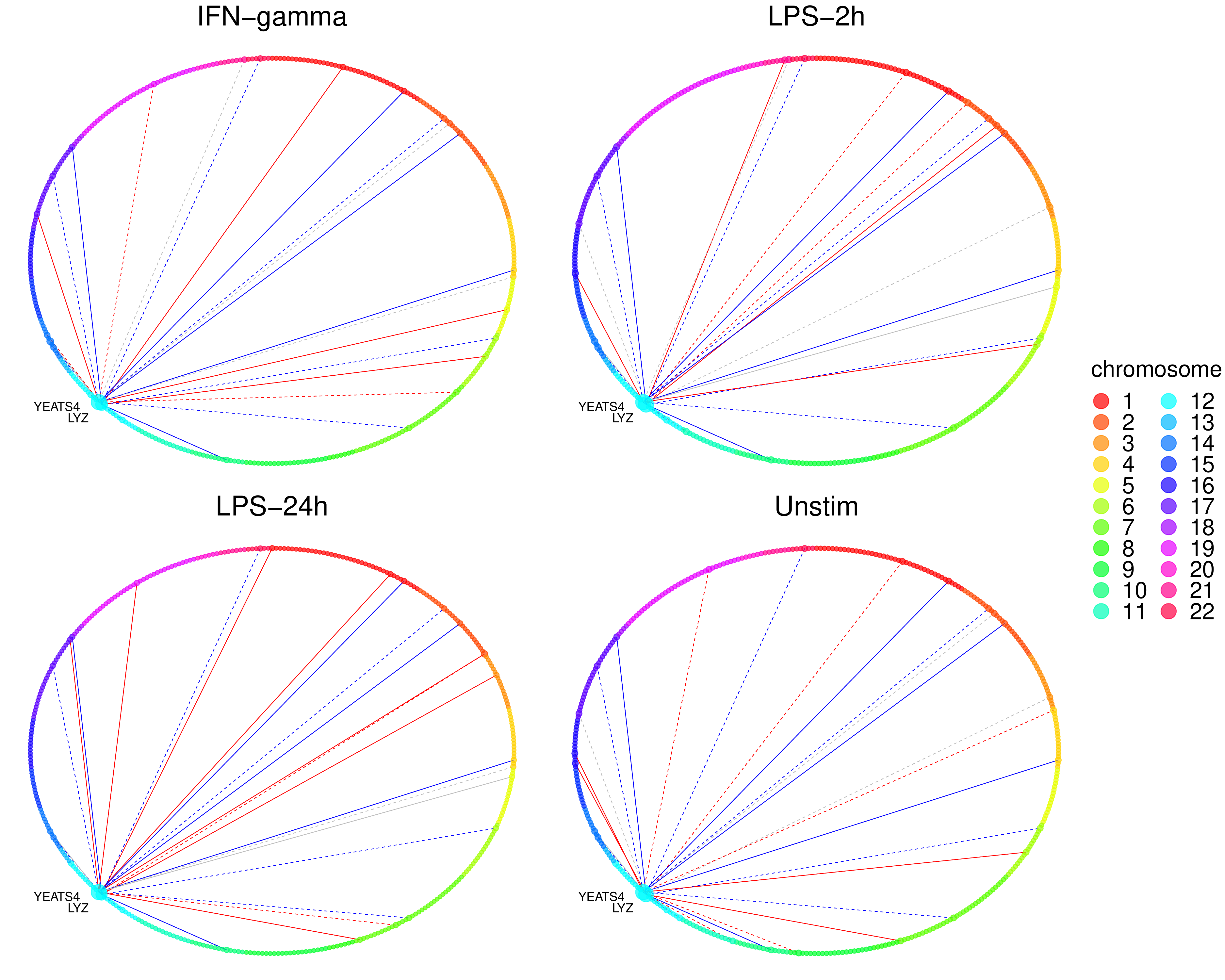}
    \caption{Subnetworks consisting of the edges of YEATS4 and LYZ in each condition of the monocyte data. A red edge indicates that the edge is unique to that condition, while a blue edge is common across all networks and a grey edge is present in two or three networks. Solid lines represent positive partial correlation while a dashed line represents negative partial correlation, and nodes are sized according to their edge degree in the subnetwork, with larger size representing a higher degree.}
    \label{fig:Monocyte_cisgraph}
\end{figure}

To investigate the role of \emph{LYZ} and \emph{YEATS4} in the different conditions, Figure \ref{fig:Monocyte_cisgraph} shows their edges in the four conditions. It is clear that the two genes have a number of neighbours, and thus are relatively influential in the networks. These neighbours are fairly spread out on the chromosomes, in accordance with our expectations about the hotspot-induced effect of \emph{LYZ} and \emph{YEATS4} on distal genes (\cite{ruffieux2020global}). While several edges are common to all conditions, we see that jointGHS identifies many edges on the specific condition level. Further, we see that a lot of the edges represent negative partial correlations, implying that the two genes could have a role in the down-regulation as well as the up-regulation of the other genes. 

Investigating the neighbourhood of \emph{LYZ} and \emph{YEATS4} in each condition can also be useful for understanding the possible mediation effect of \emph{LYZ} and \emph{YEATS4}. Table \ref{table:monocyte_cisneighbours} shows the node degree of all neighbours in the jointGHS networks of the different conditions. As we see, \emph{AFMID} is a neighbour of \emph{LYZ} in all four conditions. Given its central role in each network, this potential interplay between \emph{LYZ} and \emph{AFMID} is noteworthy. Another notable observation is that \emph{PHF1} is a neighbour of \emph{LYZ} or \emph{YEATS4} only in $\text{IFN-}\gamma$. Given its importance only in this network, the possible association between \emph{LYZ} and \emph{PHF1} may provide further insight into the role of this gene. Another neighbour of \emph{LYZ} in all conditions is MAF BZIP Transcription Factor F (\emph{MAFF}), which is a regulator for growth factor signaling (\cite{amit2007module}). We also observe that \emph{AFTPH}, which we found to have a high degree in all conditions, is a neighbour of \emph{YEATS4} in all four conditions.

\begin{table}
	\centering
	\renewcommand{\arraystretch}{1.4}
	\caption{The node degree of the neighbours of LYZ and YEATS4 in the jointGHS networks of the different conditions. For each of the two genes, the genes that are a neighbour in all four conditions are marked in bold. The genes that are only a neighbour in one of the conditions are marked in red. The genes that are controlled by the top hotspot rs6581889 in the condition in question are marked with $*$.}
	\hspace*{-2cm}
	\resizebox{1.2\textwidth}{!}{
	\begin{tabular}{l r  r @{\hskip 0.8cm}l r r @{\hskip 0.8cm}l r r @{\hskip 0.8cm}l r}
        \multicolumn{11}{c}{\emph{LYZ}}\\
        \toprule
	    \multicolumn{2}{c}{IFN-gamma}&&\multicolumn{2}{c}{LPS-2h}&&\multicolumn{2}{c}{LPS-24h} &&\multicolumn{2}{c}{Unstimulated}  \\
		\cmidrule(lr){1-2} \cmidrule(lr){4-5} \cmidrule(lr){7-8} \cmidrule(lr){10-11}
		Gene & Degree && Gene & Degree && Gene & Degree && Gene & Degree \\
		\hline
	    \bf{AFMID}*  &  24 &&\bf{AFMID}*  &  26 &&\bf{AFMID}*  &  25 &&\bf{AFMID}*  &  33  \\ 
        LGALS3  &  23 &&LGALS3  &  24 &&{\color{red}YEATS4*}  &  20 &&{\color{red}PHRF1}  &  24  \\ 
        {\color{red}PHF1*}  &  20 &&{\color{red}NDUFV3*}  &  13 &&\bf{MAFF}  &  14 &&LGALS3  &  23  \\ 
        {\color{red}NBPF8*}  &  16 &&\bf{MAFF}*  &  12 &&\bf{SNRNP48}  &  12 &&{\color{red}LOC202781}  &  13  \\ 
        \bf{MAFF}*  &  15 &&{\color{red}TPM3}  &  11 &&{\color{red}MBD4}  &  12 &&\bf{KLHL28}*  &  11  \\ 
        \bf{KLHL28}*  &  11 &&\bf{SNRNP48}  &  11 &&{\color{red}LRRFIP1}  &  10 &&\bf{SNRNP48}*  &  10  \\ 
        \bf{SNRNP48}*  &  9 &&{\color{red}LOC653086}  &  11 &&\bf{KLHL28}*  &  10 &&{\color{red}NCAPD2*}  &  10  \\ 
        {\color{red}LIN52*}  &  9 &&{\color{red}MBOAT2}  &  10 &&\bf{LOC100128098}  &  8 &&\bf{MAFF}*  &  8  \\ 
        \bf{LOC100128098}*  &  7 &&\bf{KLHL28}*  &  10 &&{\color{red}KIAA1751}  &  8 &&\bf{LOC100128098}* & 7\\
        & &&TBCCD1  &  7 && & &&TBCCD1*  &  6  \\ 
        & &&MINK1  &  7 && & &&{\color{red}RAG1AP1*}  &  6  \\ 
        & &&\bf{LOC100128098}*  &  6 && & &&{\color{red}MEFV*}  &  5  \\ 
        & && & && & &&MINK1*  &  4  \\ 
		\toprule
		\multicolumn{11}{c}{\emph{YEATS4}}\\
        \toprule
	    \multicolumn{2}{c}{IFN-gamma}&&\multicolumn{2}{c}{LPS-2h}&&\multicolumn{2}{c}{LPS-24h} &&\multicolumn{2}{c}{Unstimulated}  \\
		\cmidrule(lr){1-2} \cmidrule(lr){4-5} \cmidrule(lr){7-8} \cmidrule(lr){10-11}
		Gene & Degree && Gene & Degree && Gene & Degree && Gene & Degree \\
		\hline
        AIRE*  &  25 &&AIRE  &  37 &&LGALS3  &  22 &&\bf{AFTPH}  &  20 \\
        LGALS3  &  23 &&\bf{AFTPH}  &  22 &&\bf{TBC1D9}  &  21 &&\bf{TBC1D9}  &  16 \\
        \bf{COPZ1}*  &  18 &&\bf{IMPDH1}  &  19 &&\bf{AFTPH}  &  20 &&{\color{red}GLTSCR1}  &  16 \\
        {\color{red}RELB}  &  17 &&{\color{red}SORL1}  &  18 &&\bf{TLK1}  &  17 &&\bf{TLK1}  &  14 \\
        \bf{IMPDH1}*  &  15 &&\bf{TBC1D9}  &  15 &&\bf{IMPDH1}  &  17 &&\bf{IMPDH1}  &  14 \\
        \bf{AFTPH}*  &  14 &&{\color{red}HIST1H2BD}  &  15 &&\bf{COPZ1}  &  17 &&{\color{red}USP49*}  &  13 \\
        \bf{TLK1}  &  13 &&\bf{TLK1}  &  14 &&{\color{red}AKR1D1}  &  15 &&\bf{COPZ1}  &  13 \\
        SLC4A5*  &  13 &&\bf{COPZ1}  &  13 &&SC4MOL  &  14 &&{\color{red}BFAR}  &  13 \\
        {\color{red}TBC1D15}  &  12 &&ZNF131*  &  11 &&ZNF131*  &  11 &&SLC4A5*  &  11 \\
        {\color{red}CHST12*}  &  12 &&{\color{red}MEFV}  &  10 &&{\color{red}LYZ*}  &  11 &&\bf{TP53BP2}*  &  8 \\
        {\color{red}SMCR5*}  &  11 &&\bf{TP53BP2}*  &  9 &&{\color{red}ZNF738}  &  10 &&DDX51*  &  7 \\
        {\color{red}LYRM7}  &  11 &&\bf{TMEM106A}  &  9 &&{\color{red}LRRFIP1}  &  10 &&\bf{TMEM106A}*  &  6 \\
        \bf{TP53BP2}*  &  10 && & &&{\color{red}XRCC2}  &  9 &&{\color{red}TMEM128*}  &  5 \\
        SC4MOL*  &  10 && & &&\bf{TP53BP2}*  &  9 &&{\color{red}CDK5RAP2}  &  5 \\
        \bf{TMEM106A}*  &  9 && & &&\bf{TMEM106A}  &  9 && & \\ 
        \bf{TBC1D9}  &  9 && & &&{\color{red}MFSD11}  &  8 && & \\ 
        & && & &&DDX51*  &  7 && & \\ 
        & && & &&{\color{red}TIPRL}  &  6 && & \\ 
        \toprule
	\end{tabular}
	}
	\label{table:monocyte_cisneighbours}
\end{table} 

It can be relevant to investigate whether the neighbours of \emph{LYZ} and \emph{YEATS4} tend to be have more associations with the top hotspot. Table 3 in the main manuscript shows the proportion of the neighbours of \emph{LYZ} and \emph{YEATS4} respectively that are mediated by the top hotspot in each condition, and compares it to the total proportion of top hotspot controlled genes. We use permutation testing to assess whether more of the neighbours of each gene are controlled by the top hotspot than we would expect from a gene set of the same size; by randomly sampling $10 000$ gene sets with the same size as the neighbourhood, we calculate empirical \emph{p}-values for the observed proportion of controlled neighbours. We see that \emph{YEATS4} does not have significantly more mediated neighbours than the overall fraction, except in LPS 24h. \emph{LYZ}, on the other hand, has, at a significance level $0.05$, more mediated neighbours in all conditions.

\begin{figure}
    \centering
    \includegraphics[scale=0.5]{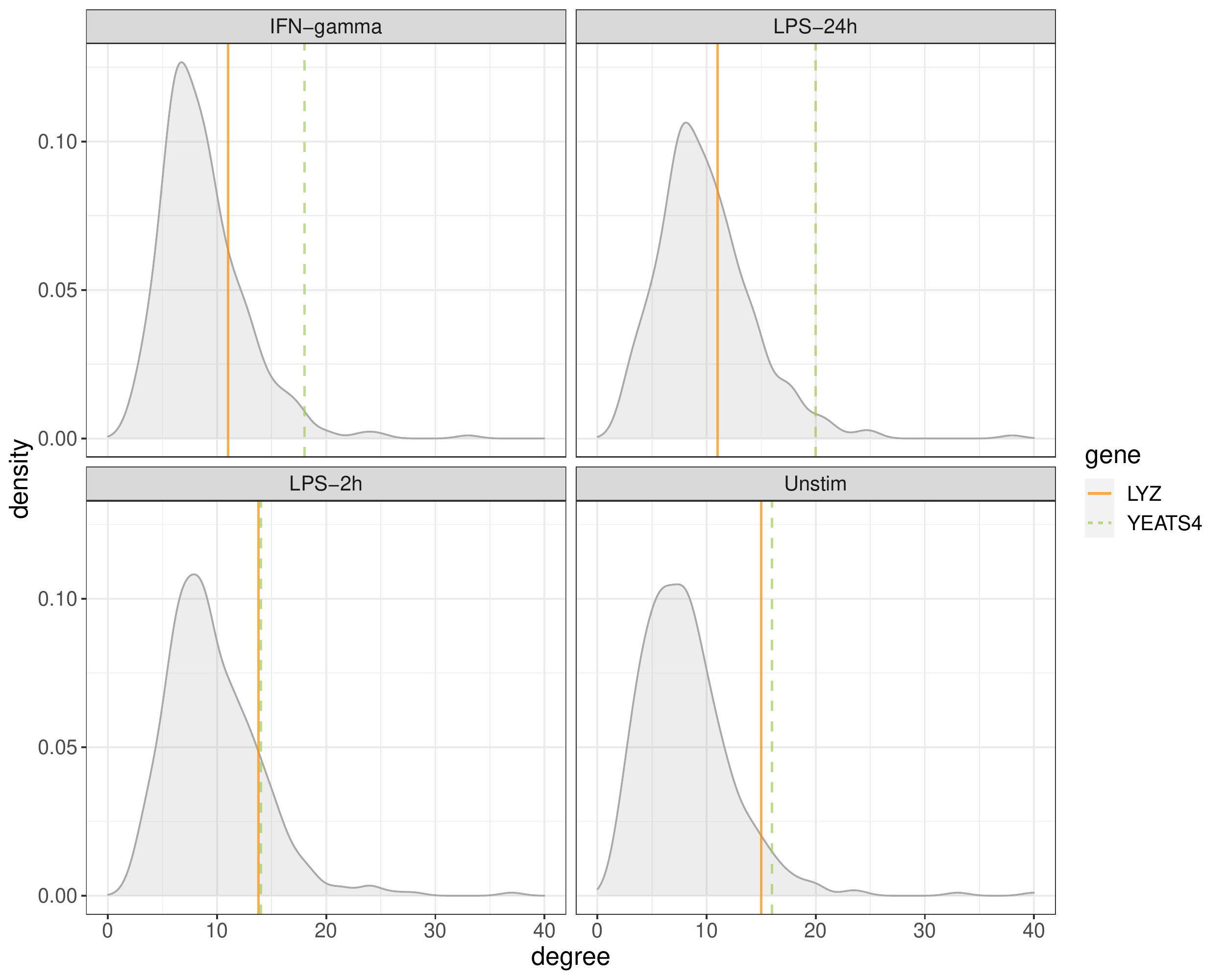}
    \caption{Density plot of the degree distribution of the jointGHS graph of each condition, with the edge degree of LYZ and YEATS4 shown as vertical lines.}
    \label{fig:Monocyte_density_cis}
\end{figure}

Figure \ref{fig:Monocyte_density_cis} shows the density of the degree distribution of the jointGHS graph of each condition, as well as the degree of \emph{LYZ} and \emph{YEATS4}. It appears that the degree distribution of the unstimulated network has more density on lower degrees, which is reasonable considering the unstimulated network is the sparsest among the four. As we see, the two genes have a relatively high degree in all conditions.

\subsubsection*{Top-hotspot mediated genes}

Investigating whether the genes controlled by the top hotspot have more associations with each other can be insightful. In Table 3 in the main manuscript, we find the sparsity of the subnetwork of genes that are mediated by the top hotspot in each condition, and compares it to the sparsity of the whole network. We use permutation testing to assess whether the subnetwork of top hotspot controlled genes is denser than we would expect from a subnetwork of the same size; by randomly sampling $10 000$ subnetworks with the same size, we calculate empirical \emph{p}-values for the observed sparsity. We see that the empirical \emph{p}-value is $0.001$ for all conditions except $\text{IFN-}\gamma$, implying the subnetworks are indeed denser in these conditions. While the difference is relatively small in the unstimulated network, it is larger in the LPS ones and in particular in LPS 24h. 


\subsubsection*{Comparison to results from single network analysis}

\begin{figure}
    \centering
    \includegraphics[width=\textwidth]{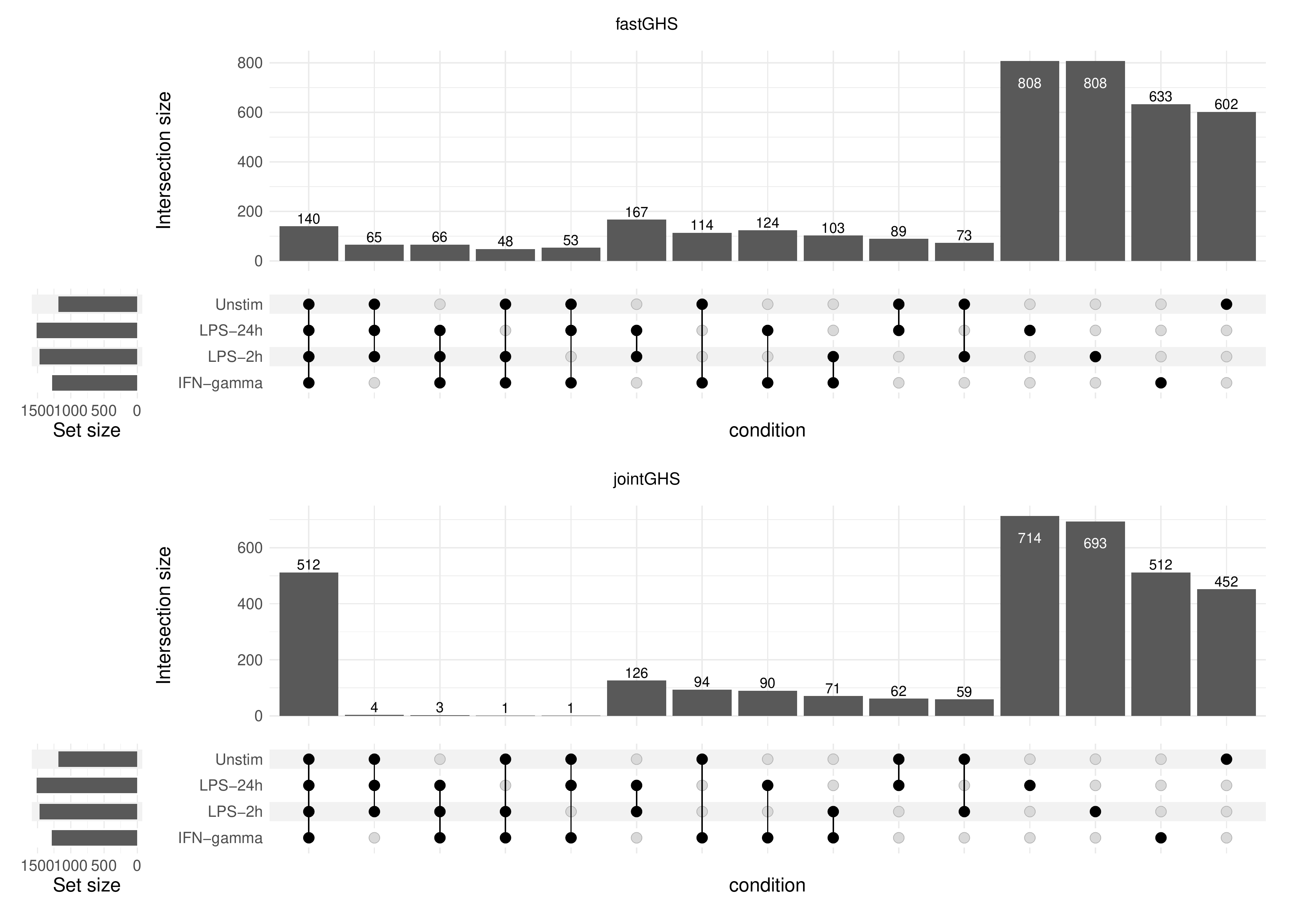}
    \caption{Upset plot for the inferred edges of the fastGHS and jointGHS graphs of the monocyte data, an alternative to a venn diagram showing the number of shared edges between conditions. For each method, the total number of edges for each condition is represented on the left barplot. Every possible intersection is represented by the bottom plot, and their occurence is shown in the top barplot.}
    \label{fig:Monocyte_intersection_singleVSjoint}
\end{figure}

Table \ref{table:sparsity_comparetosingle_monocyte} shows the sparsity of the estimated jointGHS graphs of the different conditions, as well as the percentage of the inferred edges that the jointGHS and fastGHS estimates at the same sparsity agree on. Figure \ref{fig:Monocyte_intersection_singleVSjoint} shows an upset plot (\cite{conway2017upsetr}), an alternative to a venn diagram that shows the number of edges shared between the inferred networks of each condition, for fastGHS and jointGHS separately. For each intersection, the number of edges shared only by the corresponding conditions is shown. It indicates that many edges are common to all four networks, meaning that the joint method has identified a fair amount of shared information. Compared to the fastGHS networks, we see that the intersection of edges present in all four conditions is larger in the jointGHS networks. This is as expected, as we, with the joint method, are better equipped to identify what is common and borrow information.

\begin{table}
	\centering
	\caption{The sparsity of the joint graphical horseshoe (jointGHS) estimated graphs of the different conditions in the monocyte data, as well as the percentage of the inferred edges that the jointGHS and the fastGHS estimate at the same sparsity agree on.} 
	\renewcommand{\arraystretch}{1.4}
	\hspace*{0cm}
	\resizebox{0.8\textwidth}{!}{%
	\begin{tabular}{l r r r r}
		\multicolumn{5}{c}{ }\\
	    \toprule
		 & IFN-gamma & LPS-2h & LPS-24h & Unstimulated \\
		\hline
		Sparsity & 0.018 & 0.020 & 0.021 & 0.016 \\
		$\%$ Edge agreement with fastGHS & 66.8 & 64.6 & 65.9 & 62.3 \\
		[0.1cm] 
		\hline 
	\end{tabular}}
	\label{table:sparsity_comparetosingle_monocyte}
\end{table} 


\begin{figure}
    \centering
    \includegraphics[scale=0.4]{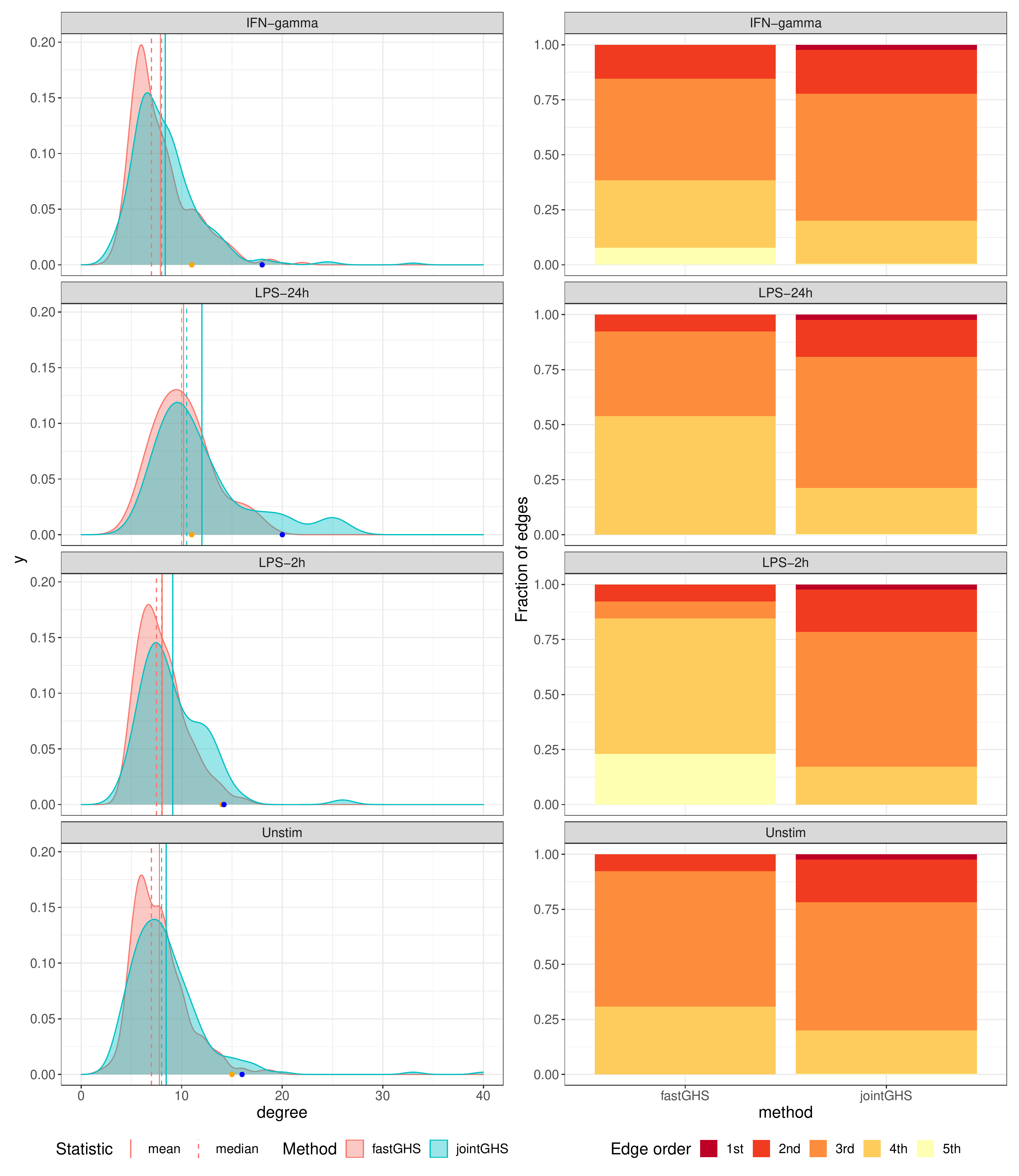}
    \caption{Comparison of the results from fastGHS and jointGHS applied to the monocyte data. The left panel compares the density of the degree distribution of the subnetwork of genes controlled by the top hotspot in each condition, showing the mean and median as solid and dashed lines respectively. The degree of LYZ and YEATS4 in the jointGHS networks are shown as orange and blue dots respectively. The right panel compares the neighbourhood status with respect to LYZ and YEATS4 of the edges common to all conditions. Only the edges found to be common in all conditions in jointGHS but not in fastGHS are shown, and vice versa. }
    \label{fig:Monocyte_density_neigh_singleVSjoint}
\end{figure}

To understand more about what differs between the single and the joint estimates, we have a closer look at the genes controlled by the top hotspot rs6581889. The left panel of Figure \ref{fig:Monocyte_density_neigh_singleVSjoint} compares the fastGHS and jointGHS density of the degree distribution of the subnetwork of genes controlled by the top hotspot in each condition. The density plots with the mean and median lines suggest that the jointGHS distribution is more shifted to the right in all conditions, implying that amongst the genes controlled by the top hotspot, the joint method identifies an overall higher activity level in terms of associations to other genes. This agrees with what we would expect, with the hotspot triggering substantial gene activity (\cite{ruffieux2020global}). 

We next investigate the role of \emph{LYZ} and \emph{YEATS4} in the networks inferred by the two methods. In particular, we want to investigate if their role is common to all four conditions. To this end, we compare the neighbourhood status (with respect to the two genes) of the edges common to all conditions, namely, how close an edge is to one of the two genes. If an edge is between \emph{LYZ} or \emph{YEATS4} and another gene, that edge has order $1$. If an edge is between two other genes where at least one of them is a neighbour of \emph{LYZ} or \emph{YEATS4}, that edge has order $2$, and so on. Only the edges found to be common in all conditions in jointGHS but not in fastGHS are shown, and vice versa. 
The results from the comparison are shown in the right panel of Figure \ref{fig:Monocyte_density_neigh_singleVSjoint}. The edges identified by jointGHS but not fastGHS are in much closer proximity to \emph{LYZ} or \emph{YEATS4}, implying that the joint method identifies a stronger effect from the two genes, which agrees with previous findings (\cite{ruffieux2020global}). Similarly, there is not a single edge between \emph{LYZ} or \emph{YEATS4} and other genes identified by fastGHS but not jointGHS, while the opposite is not true. This again implies that the joint method is able to identify more of the hotspot-mediated relationships.     

\begin{figure}
    \centering
    \hspace*{-1cm}
    \includegraphics[scale=0.6]{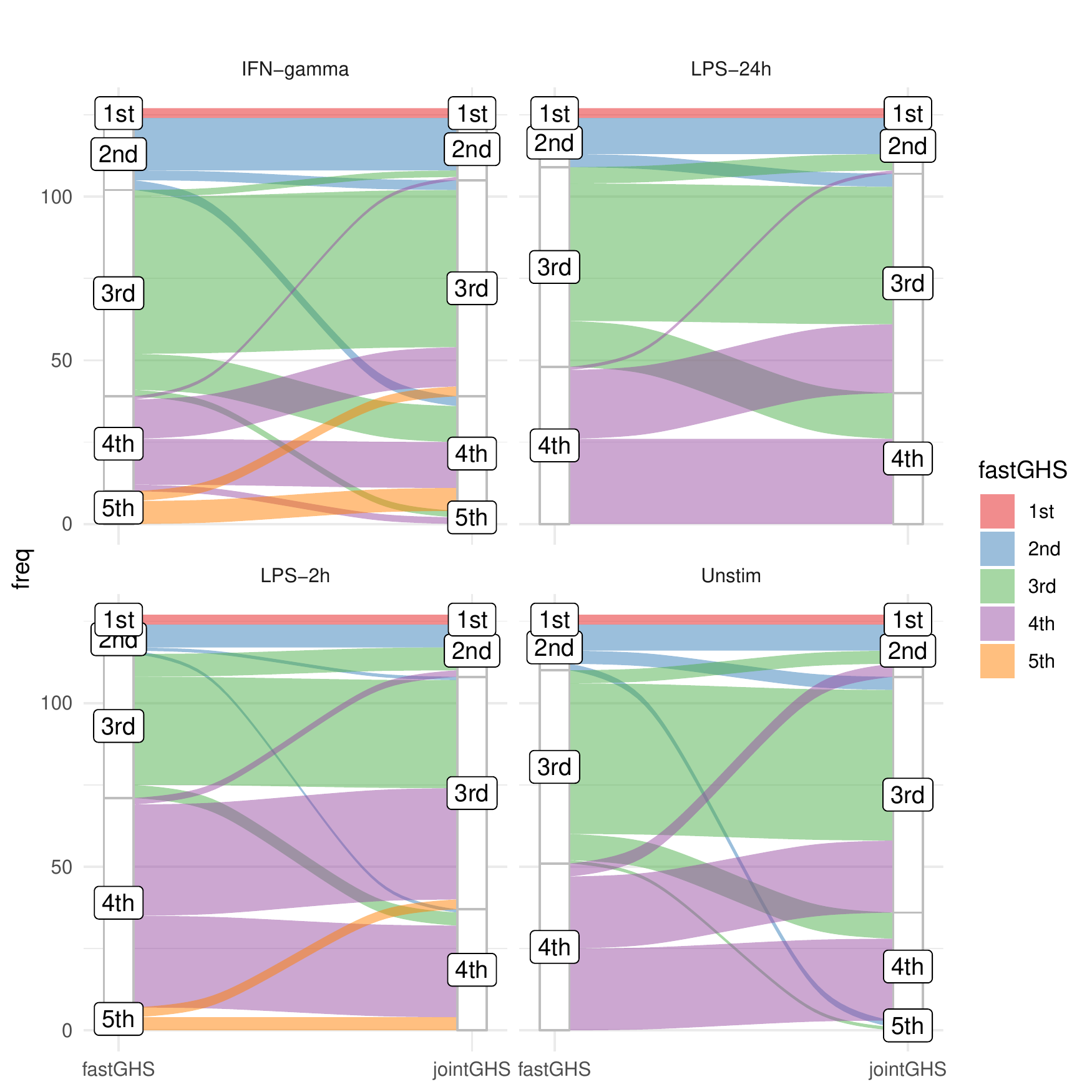}
    \caption{Comparison of the order of the edges in the fastGHS estimated graphs to the jointGHS estimated graphs, in terms of distance to LYZ and YEATS4.}
    \label{fig:alluvial}
\end{figure}

We have now investigated what differs between the networks informed by fastGHS and jointGHS, but what about the edges the two methods agree on?
The alluvial diagram in figure \ref{fig:alluvial} compares the order to the edges common to all conditions, and that fastGHS and jointGHS agree on, using the same definition of order as above. It shows that the two methods agree completely on all first order edges, meaning the direct effect of \emph{LYZ} and \emph{YEATS4} on the other genes in this case has sufficient evidence in the data to be captured even with the single-network approach. We also observe that the edges tend to have a lower order in the jointGHS estimates, implying a stronger influence from \emph{LYZ} and \emph{YEATS4} has been captured. This strong effect of the two genes observed in the jointGHS estimates agrees more with previous findings in this data set (\cite{ruffieux2020global}). In general, we see that two methods disagree more on the higher-order edges, where the effect of the \emph{LYZ} and \emph{YEATS4} can be said to be weaker. In such a setting, a joint approach that gains statistical power by sharing information between conditions can be highly useful, helping us gain more insight not only into the effects of \emph{LYZ} and \emph{YEATS4} on other genes but also role of the affected genes. In particular, many genes directly associated with \emph{LYZ} and/or \emph{YEATS4} have a very high degree, highlighting their interplay with other genes as potentially relevant for the disease-driving mechanisms.


\subsection{Application of GemBag}
\label{supp:monocyte_gembag}

\subsubsection{Data analysis details}

Gembag is applied to the monocyte data, with parameters as suggested by \cite{yang2021gembag}, described in \ref{sec:simstudy}. 

\subsubsection{Resulting networks}

\begin{figure}
    \centering
    \includegraphics[scale=0.4]{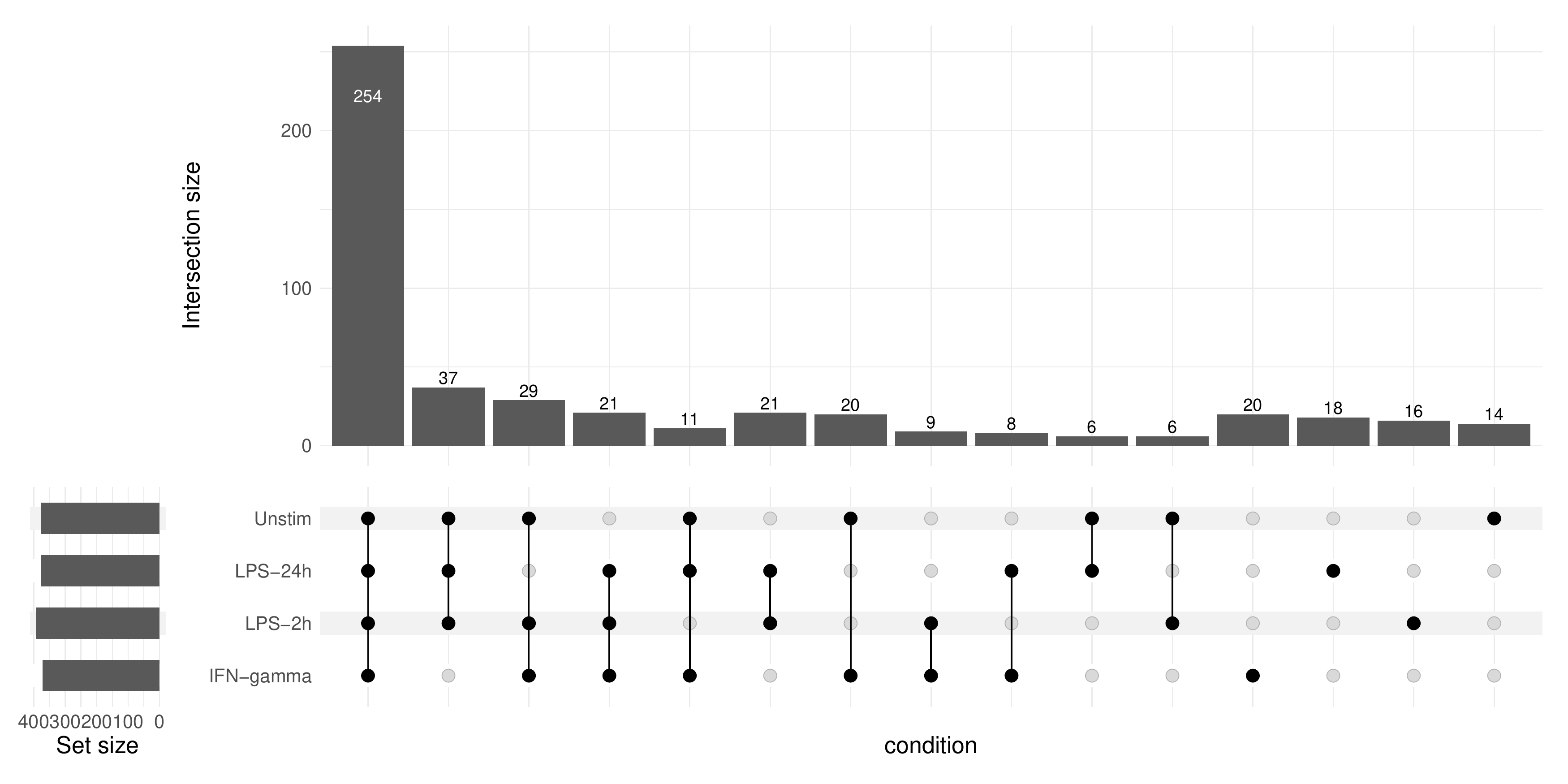}
    \caption{Upset plot of the GemBag graphs of the monocyte data, an alternative to a venn diagram showing the number of edges solely shared between conditions (\cite{conway2017upsetr}). For each intersection, the number of edges shared only by the corresponding conditions is shown. The total number of edges for each condition is represented on the left barplot. Every possible intersection is represented by the bottom plot, and their occurence is shown in the top barplot.}
    \label{fig:Monocyte_intersection_GemBag}
\end{figure}

The resulting GemBag networks all have sparsity $0.005$, which notably is much sparser than the jointGHS networks. Figure \ref{fig:Monocyte_intersection_GemBag} shows the number of edges solely shared between the inferred GemBag networks of each condition. As we see, mainly common edges for all four conditions are identified, and very few condition-specific edges. This stands in strong contrast to the jointGHS estimates, where a large number of condition-specific edges were found. 

Comparing the GemBag networks to those found by jointGHS, we find that all of the edges that were identified by GemBag were also identified by jointGHS, making the edges identified by GemBag a subset of the edges identified by jointGHS. 

\subsubsection{Hub genes}

Table \ref{Supp:table:monocyte_hubs_Gembag} shows the number of genes with node degree larger than the $90^{\text{th}}$ percentile (hubs) in the respective networks of the different conditions, for the jointGHS and the GemBag networks. The number of hubs common to all four conditions is shown as well, for both methods. As we see, while the number of hubs in each condition is comparable for the two methods, there is a far bigger overlap in top hubs in the GemBag networks (24 common hubs) than the jointGHS networks (8 common hubs). This again illustrates how GemBag identifies less condition specific interactions and traits than jointGHS. 

\begin{table}
	\centering
	\caption{The number of genes with node degree larger than the $90^{\text{th}}$ percentile (hubs) in the respective networks of the different conditions, for the jointGHS and the GemBag networks. The number of hubs shared by all four conditions is shown as well, for both methods.} 
	\renewcommand{\arraystretch}{1.4}
	\hspace*{0cm}
	\resizebox{0.8\textwidth}{!}{%
	\begin{tabular}{l r r r r r}
	    \toprule
		  & IFN-gamma & LPS-2h & LPS-24h & Unstimulated & All conditions \\
		\hline
		jointGHS & 38 & 31 & 35 & 34 & 8 \\
		GemBag & 34 & 36 & 34 & 29 &  24 \\
		[0.1cm] 
		\hline 
	\end{tabular}}
	\label{Supp:table:monocyte_hubs_Gembag}
\end{table}

\subsubsection{Hotspot control}

\begin{figure}
    \centering
    \includegraphics[scale=0.5]{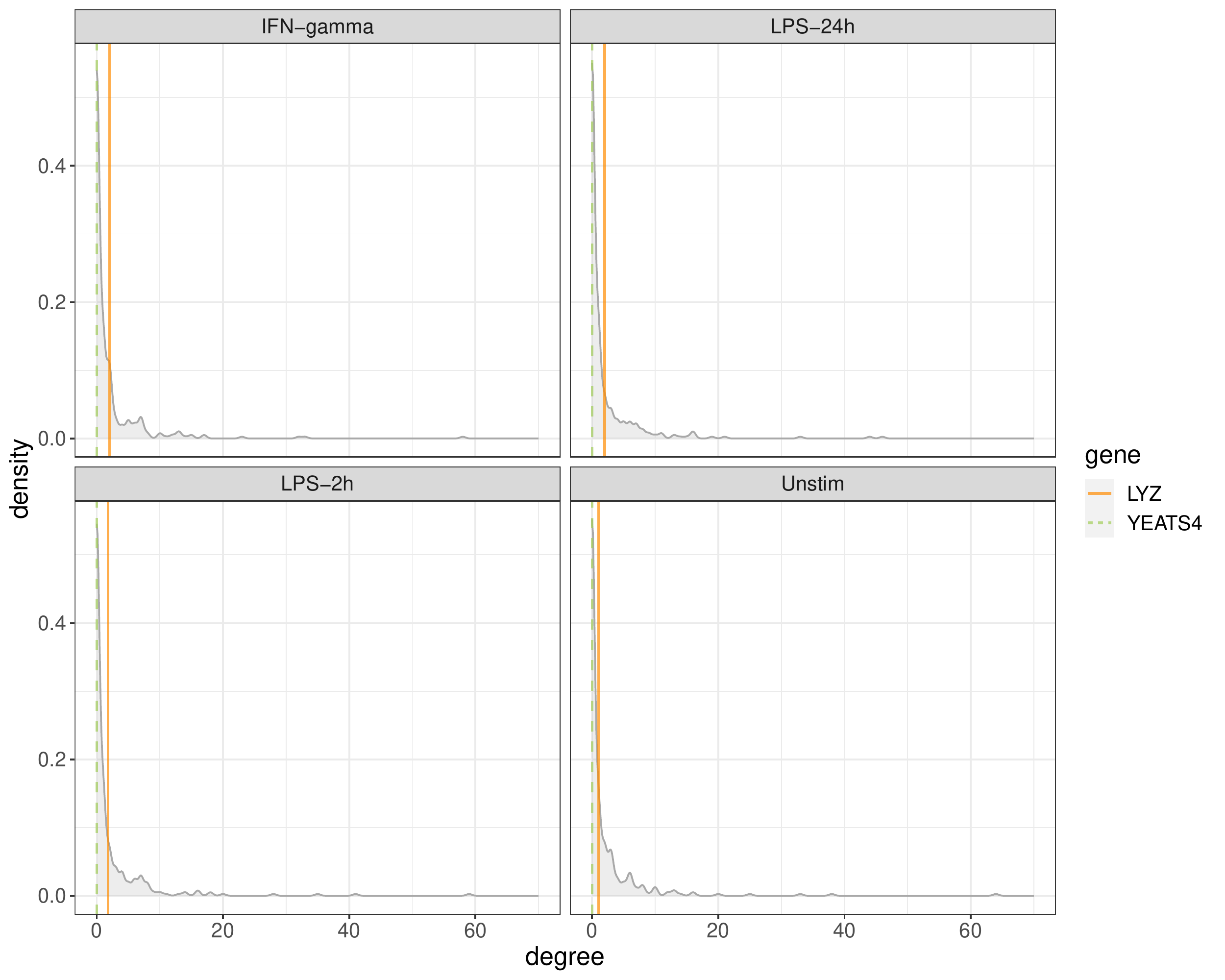}
    \caption{Density plot of the degree distribution of the GemBag graph of each condition, with the edge degree of LYZ and YEATS4 shown as vertical lines.}
    \label{Supp:fig:Monocyte_density_cis_gembag}
\end{figure}

Figure \ref{Supp:fig:Monocyte_density_cis_gembag} shows the density of the degree distribution of the GemBag graph of each condition, as well as the degree of \emph{LYZ} and \emph{YEATS4}. As we see, the two genes 
 have a very low node degree in all four conditions, unlike in the jointGHS networks where they are central (Figure \ref{fig:Monocyte_density_cis}). While the GemBag networks are sparser, and thus fewer edges can be expected, \emph{LYZ} and \emph{YEATS4} still have a very low node degree relative to the overall degree distribution. 

In fact, \emph{YEATS4} has node degree $0$ in all four conditions. Thus, the GemBag networks give no evidence for the mediation of the hotspot effect on other genes via this gene. Similarly, GemBag gives no evidence for an effect via \emph{LYZ}. While these two genes were central in the jointGHS networks, with a large number of neighbours, many central neighbours (i.e. hubs) and many top hotspot controlled neighbours, the GemBag networks do not capture this hotspot mediated effect which had been reported by \cite{fairfax2012genetics} and \cite{ruffieux2021epispot}.

\section{More on the choice of global shrinkage parameter}
\label{supp:selecting_tau}

This section supplements Sections 2 and 4 in the main manuscript.  

\subsection{The deflation issue of the global shrinkage}

\begin{figure}
    \centering
    \hspace*{-1cm}
    \includegraphics[scale=0.5]{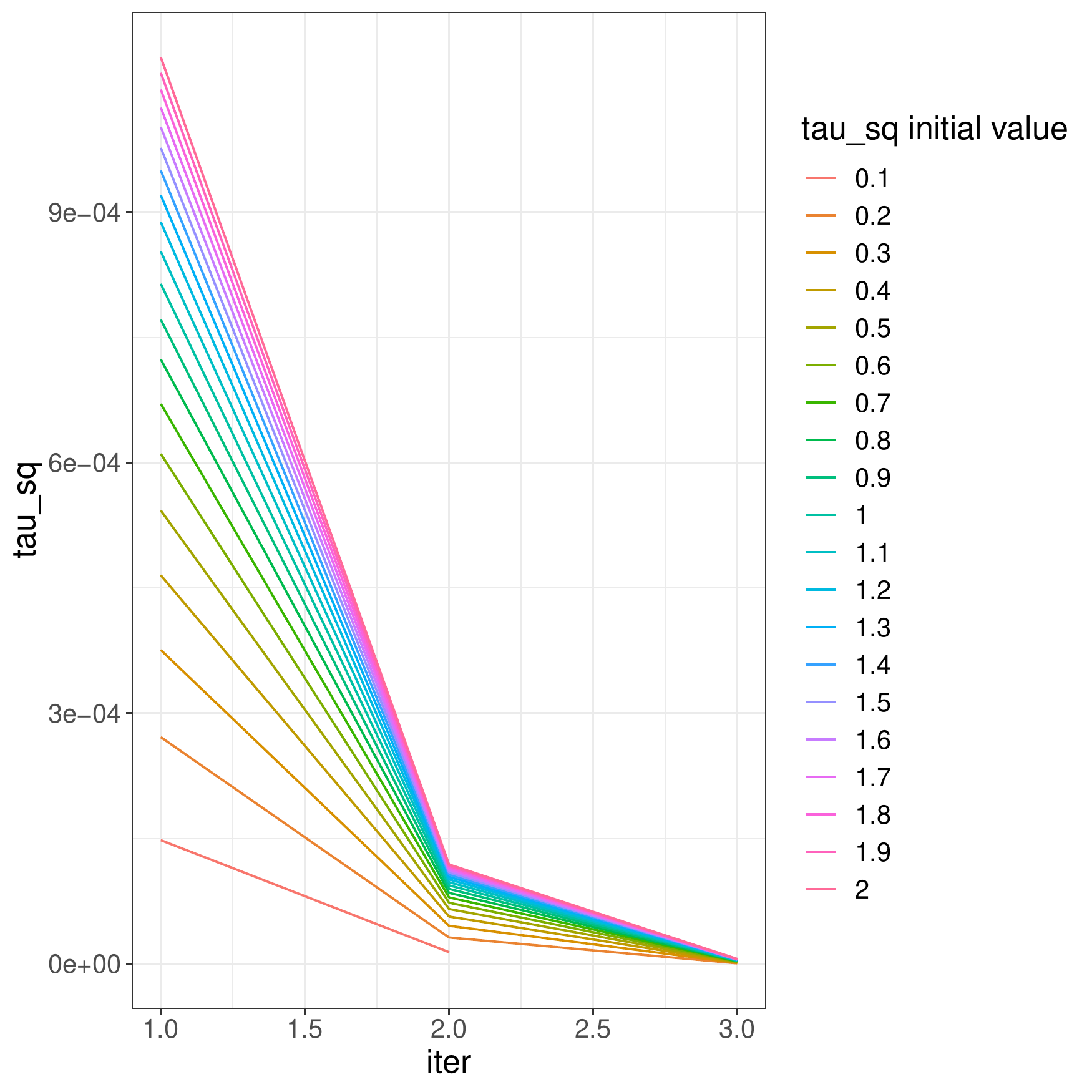}
    \caption{The value of the global shrinkage parameter $\tau^2$ at each iteration when it is not fixed, but updated as part of the ECM graphical horseshoe scheme. The results are shown for different initial values of $\tau^2$ for a setting with $p=50$ and $n=500$.}
    \label{fig:tau_shrinking}
\end{figure}

Figure \ref{fig:tau_shrinking} illustrates how the global shrinkage parameter shrinks to zero after just a few iterations, regardless of its initial value, when it is updated in the ECM algorithm. The Gaussian graphical data is generated using the procedure described in Section \ref{sec:simstudy}, with $p=50$ nodes and $n=500$ observations in each data set.

\subsection{The issue of under-selection}

\begin{figure}
    \centering
    \hspace*{-0.5cm}
    \includegraphics[scale=0.6]{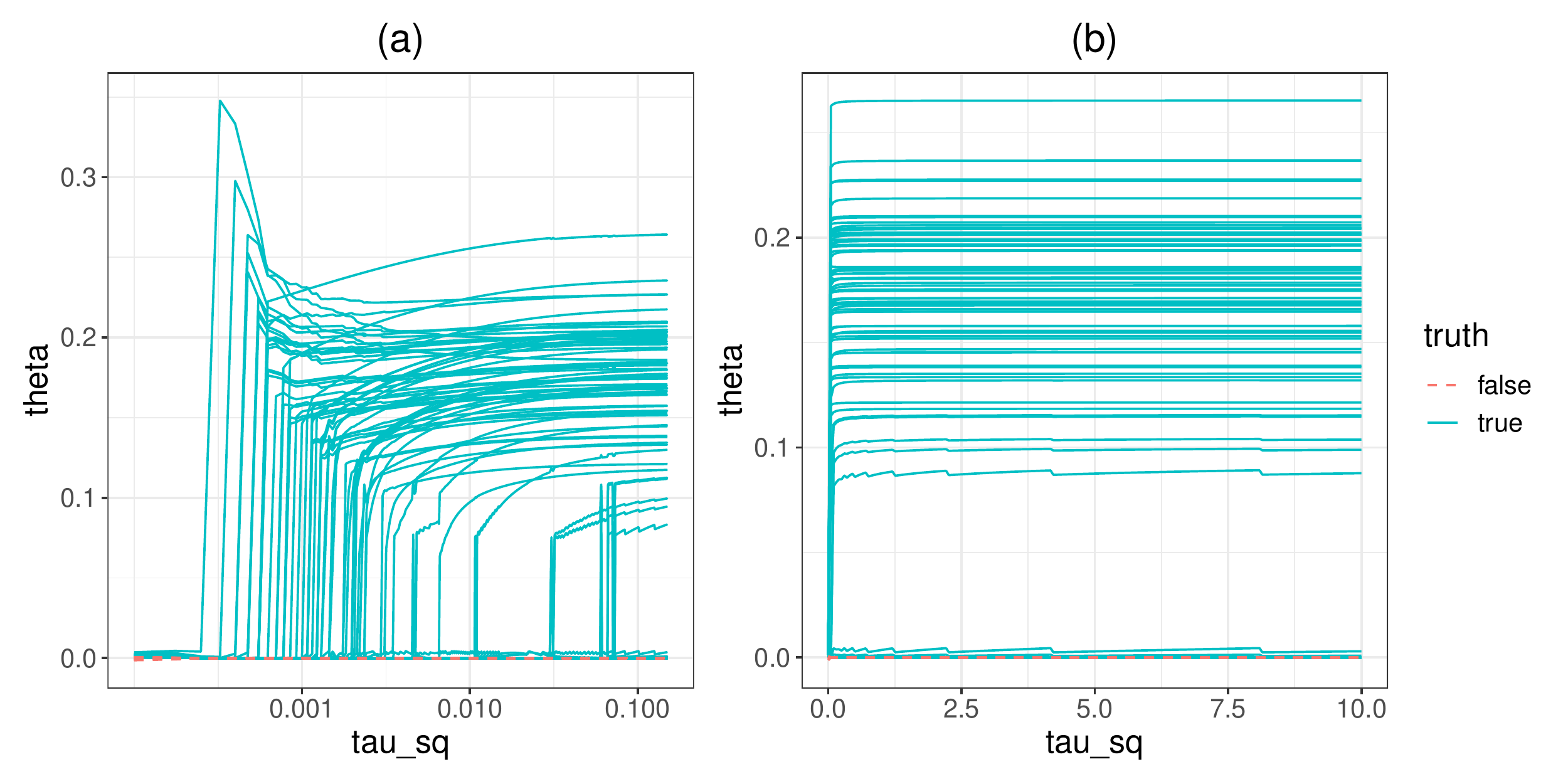}
    \caption{Trace plots showing size of the scaled precision matrix elements estimated by the ECM graphical horseshoe change for (a) smaller $\tau^2$ plotted on a logarithmic scale and (b) larger $\tau^2$ on a normal scale, for a Gaussian graphical data set with $p=150$ nodes and $n=500$ observations. Each line represents one precision matrix element, and the lines are colored according to whether the corresponding edge is present in the true network or not. The non-zero scaled precision matrix elements are all equal to $0.2$ in the simulated network.}
    \label{fig:traceplot}
\end{figure}

Our ECM approach does not tend to over-select edges. Figure \ref{fig:traceplot} shows two trace plots for a Gaussian graphical data set with $p=150$ nodes and $n=500$ observations, one considering only small values of $\tau^2$ and the other larger. Since we ultimately are interested in the partial correlations, we disregard the scale and look at the size of the scaled precision matrix elements $\theta_{ij}/\sqrt{\theta_{ii}\theta_{jj}}$. By comparing how these elements change as $\tau^2$ increases, the plot illustrates that over-selection of edges is not a concern. This can in part be explained by the fact that the local scale parameters $\lambda_{ij}$ adapt to the global shrinkage parameter, making the model flexible with respect to the choice of $\tau^2$ as long as it is large enough to avoid overshrinkage to zero. Indeed, for small values of $\tau^2$ we see that very few edges are captured, and the estimates deviate more from their true value of $0.2$. It is clear that we should take care to ensure the global scale parameter is not too small. If high computational efficiency is desired, fixing it to a large value is possible. However, for general use we propose a data-driven way to select the parameter.

\subsection{Avoiding under-selection with the AIC}

\begin{figure}
    \centering
    \hspace*{-0.5cm}
    \includegraphics[scale=0.7]{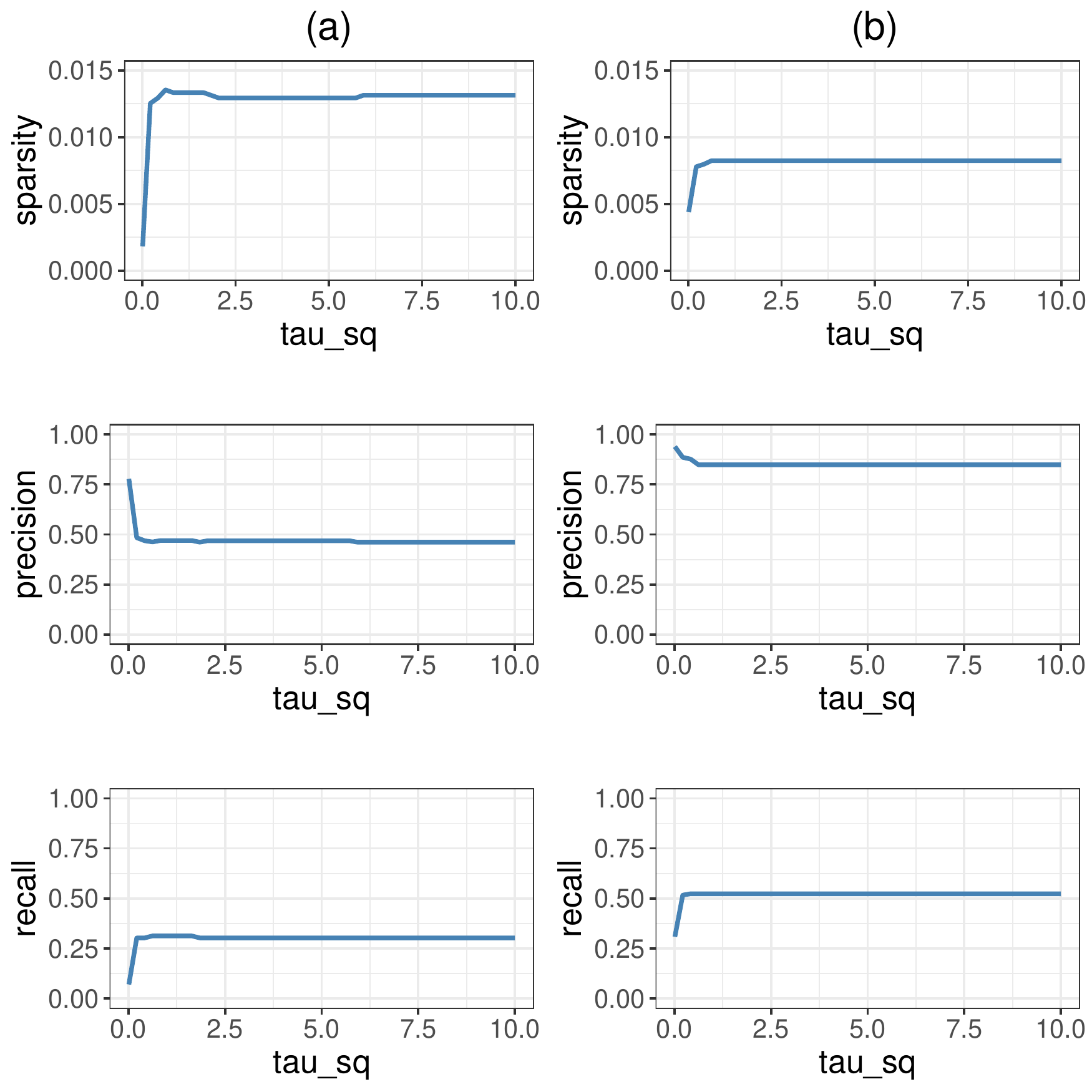}
    \caption{Sparsity, precision and recall of ECM graphical horseshoe network estimates for different values of the fixed global shrinkage parameter $\tau^2$ for data sets with (a) $p=100$ and $n=100$ with true graph sparsity $0.02$ and (b) $p=150$ and $n=200$ with true graph sparsity $0.013$.}
    \label{fig:measures_vs_tau}
\end{figure}

Figure \ref{fig:measures_vs_tau} shows plots of the sparsity, precision and recall of ECM graphical horseshoe network estimates for different values of the fixed global shrinkage parameter $\tau^2 \in (0,10]$, and for two data sets with (a) $p=100$ and $n=100$ with true graph sparsity $0.02$ and (b) $p=150$ and $n=200$ with true graph sparsity $0.013$. The estimated sparsity grows with $\tau^2$, but only to a certain point. For large enough values of the global shrinkage parameter, the results are the same in terms of precision and recall. On the other hand, too small values of $\tau^2$ can lead to severe under-selection of edges and very variable results. 

\begin{figure}

    \centering
    \hspace*{-0.5cm}
    \includegraphics[scale=0.5]{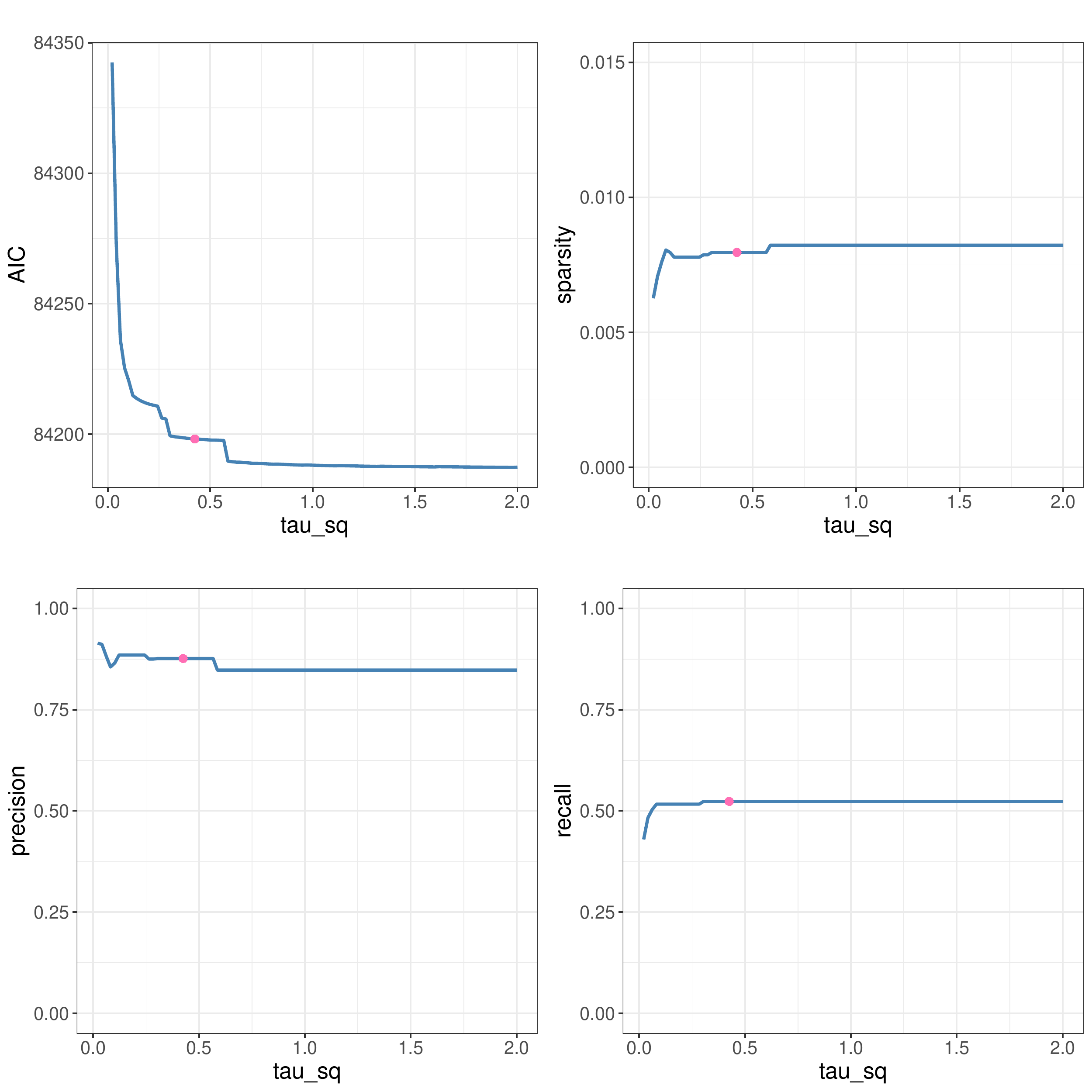}
    \caption{Sparsity, precision and recall of ECM graphical horseshoe network estimates for different values of the fixed global shrinkage parameter $\tau^2$ for a data set with $p=150$ and $n=200$ observations, and a true graph sparsity of $0.013$. The values corresponding to the $\tau_{\text{AIC}}^2$ found by our AIC selection approach are marked as points.}
    \label{fig:AIC_measures_vs_tau}
\end{figure}

Figure \ref{fig:AIC_measures_vs_tau} shows the second setting (b) for $\tau^2 \in (0,2]$, where the values corresponding to the $\tau_{\text{AIC}}^2$ found by our AIC selection approach are marked as points. This rule leads to satisfactory results in terms of both precision and recall. A smaller value of $\tau^2$ would result in reduced accuracy, and a larger value would not improve the estimate further.

\section{Posterior checks with the Bayesian bootstrap}

While the joint graphical horseshoe adapts well to the level of similarity between networks, issues may arise when the joint analysis is performed on a set with many highly similar networks and a few unrelated or less similar networks. In such a case, the highly similar networks might dominate the analysis through the common latent variables $\nu_{ij}$. We have seen that, due to the heavy tail of the horseshoe, a small $\nu_{ij}^{-1}$ does not detoriate inference since the local scales still can escape and identify edges individually on the network level. However, a large $\nu_{ij}^{-1}$ tends to lead to non-zero precision matrix elements for all networks. A setting with many highly similar and a few unrelated or less similar network can result in large $\nu_{ij}^{-1}$ for the edges $(i,j)$ common to the highly similar networks, risking wrongly included edges and thus reduced accuracy for the less similar ones. 

To account for this possibility, after performing a joint network analysis it should be possible to check whether the edges of the joint network estimates strongly contradict the single network estimates. This could indicate that a joint approach is not suitable.

For this purpose, we propose an approach for performing a posterior check for the suitability of a joint analysis. This check is implemented in our R package \texttt{jointGHS}. This extra step is more computationally demanding, but is optional and only necessary if there is doubt about the suitability of a joint approach.

\subsection{Bayesian bootstrap procedure}

We propose to use Bayesian bootstrapping (\cite{rubin1981bayesian}) to assess the suitability of a joint analysis. Specifically, for each network $k$ and each inferred edge $(i,j)$ found in the joint analysis, we assess whether the corresponding precision matrix element $\theta_{ijk}$ from the joint analysis is plausibly non-zero given its posterior Bayesian bootstrap distribution in the single network version.

In order to obtain a bootstrap sample ${\theta_{ijk}}^{(b)}$ for $b=1,\ldots,B$, we start by drawing $n_k$ weights from the Dirichlet distribution with $n_k$ categories and parameters $\alpha=(1,\ldots,1)$, i.e., the flat Dirichlet distribution. The resulting weight vector $\boldsymbol{w}^{(b)}$ is used to weigh the observations in the $n_k$ by $p$ observation matrix $\boldsymbol{X}_k$. For notational simplicity, we will from now refer to $n_k$ as $n$, $\boldsymbol{X}_k$ as $\boldsymbol{X}$ and so on. 

Using an expression for the Bayesian bootstrap sample of the covariance matrix (\cite{rodriguez2021}), we can obtain a weighted scatter matrix estimate for each bootstrap sample $b$, using the sampled weights $\boldsymbol{w}^{(b)}$

\begin{align}
\label{Supp:eq:BBscatter}
    \boldsymbol{S}_w^{(b)} &= \left(n-1\right)\left[1-\sum_{i=1}^{n} (w_i^{(b)})^2\right]^{-1} {\boldsymbol{X}_w^{(b)}}^T \boldsymbol{X}_w^{(b)}, 
\end{align}
where $\boldsymbol{X}_w^{(b)}$ is the weighted observation matrix $\boldsymbol{X}_w^{(b)} = \boldsymbol{X}\circ {\boldsymbol{w}^*}^{(b)} \boldsymbol{1}_p^T$. Here $\circ$ denotes the Hadamard product, ${\boldsymbol{w}^*}^{(b)}$ is the length $n$ vector with elements $\sqrt{w_i^{(b)}}$ and $\boldsymbol{1}_p$ is a length $p$ vector of ones. 

Replacing the unweighted scatter matrix $\boldsymbol{S}$ in the ECM graphical horseshoe algorithm by (\ref{Supp:eq:BBscatter}), we can obtain Bayesian bootstrap samples of $\theta_{ijk}$. Repeating this $B$ times, we get a set of Bayesian bootstrap samples $\{{\theta_{ijk}}^{(b)} \}_{b=1}^B$ that we can use to describe the posterior of the precision matrix elements in the single network model.

Using this sampled posterior, we can investigate whether the estimated non-zero $\theta_{ijk}$ from the joint network conflicts with its single network distribution. Since we ultimately are interested in the partial correlations, we perform comparisons regardless of scale by looking at the distribution of the scaled precision matrix elements $\theta_{ijk}/\sqrt{\theta_{iik}\theta_{jjk}}$. If a scaled precision matrix element exceeds the empirical $95^{\text{th}}$ percentile of its bootstrap distribution in absolute value, this suggests that it may have been overestimated and thus that the edge $(i,j)$ may have been wrongly included in the joint model. 

One can expect a few edges from the joint approach to be in conflict with their single network posterior as more information is available in the joint approach, but if this occurs for a large portion of edges this could suggest that the network is being forced towards networks it bears little similarity to. This could indicate that a joint approach is not suitable. 

In our R package \texttt{jointGHS} (\url{github.com/Camiling/jointGHS}), we provide a functionality for performing the Bayesian bootstrap as part of the joint graphical horseshoe, allowing assessment of all or specific edges. This also includes a plot and a print function. This optional functionality allows users to assess the suitability of a joint analysis. Notably, this tool is meant to guide users to make a decision, not to make the decision for them. For more details, we refer to the package documentation. 

\subsection{Examples}
\subsubsection*{Two similar networks}

\begin{figure}
    \centering
    \includegraphics[scale=0.32]{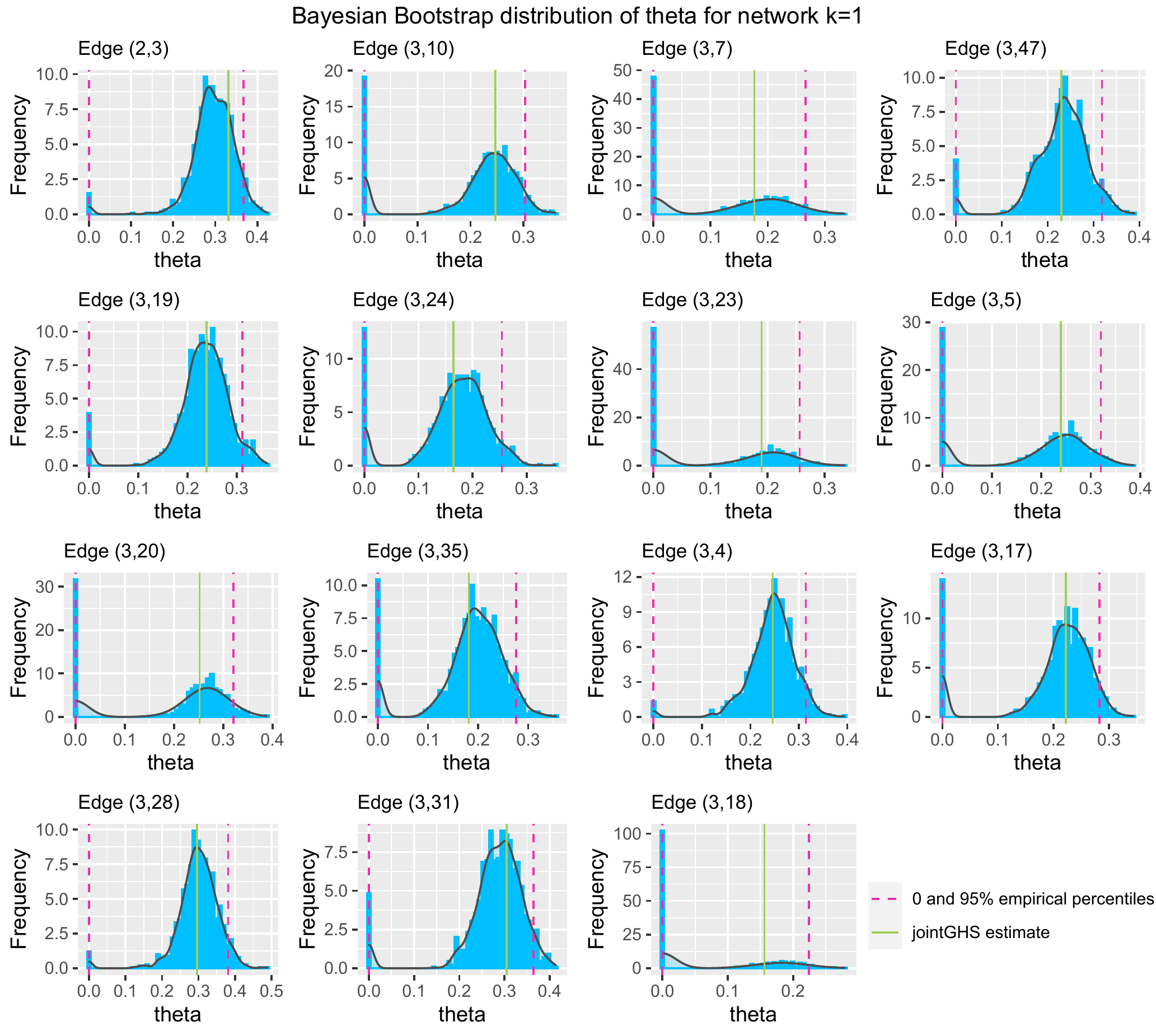}
    \includegraphics[scale=0.334]{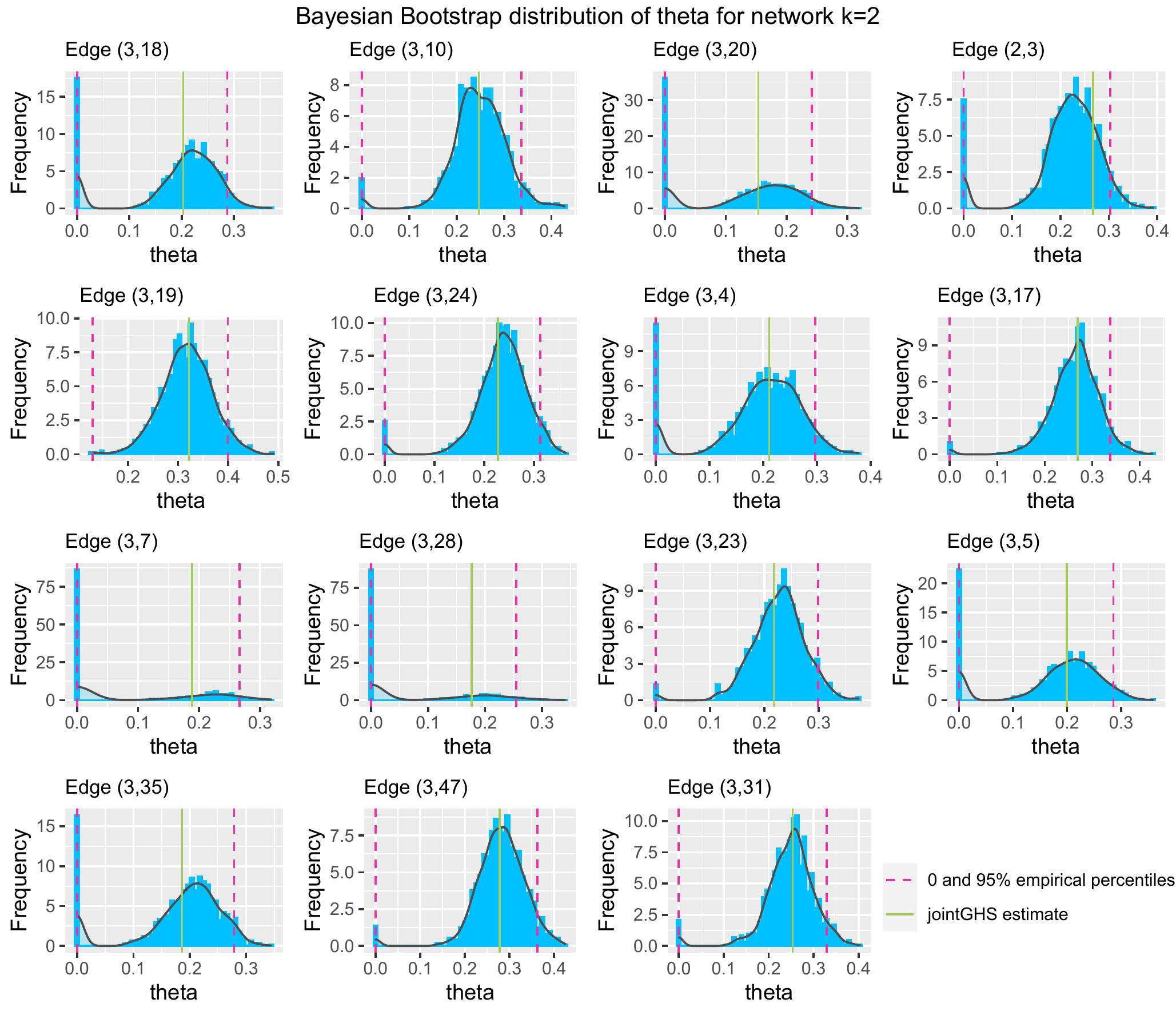}
    \caption{Output of plot function implemented in the jointGHS package, showing results of the Bayesian bootstrap procedure for assessing the suitability of a joint approach for $K=2$ networks with $p=50$ nodes. The two data sets used for inference have the same true network structure, and $n=100$ observations each. The output of the plot function is shown for both networks. For each inferred edge in the joint graphical horseshoe network of a given data set, the plot shows the Bayesian bootstrap distribution of the corresponding scaled precision matrix element of the single-network graphical horseshoe model, with the corresponding empirical $0\%$ and $95\%$ percentiles shown as dashed lines. The joint graphical horseshoe estimate is shown as a solid line.}
    \label{fig:boot_2sim}
\end{figure}

The package has a plotting function which enables visualisation of the results of the Bayesian bootstrap. Figure \ref{fig:boot_2sim} shows the output for $K=2$ data sets with the same underlying network structure. The Gaussian graphical data is generated using the procedure described in Section \ref{sec:simstudy}, with $p=50$ nodes and $n=100$ observations in each data set. In this case we see that no joint graphical horseshoe estimate exceeds its Bayesian bootstrap empirical $95^{\text{th}}$ percentile in absolute value, implying that the joint estimates are not in conflict with the single network estimates. 

\subsubsection*{Two different networks}
\begin{figure}
    \centering
    \includegraphics[scale=0.4]{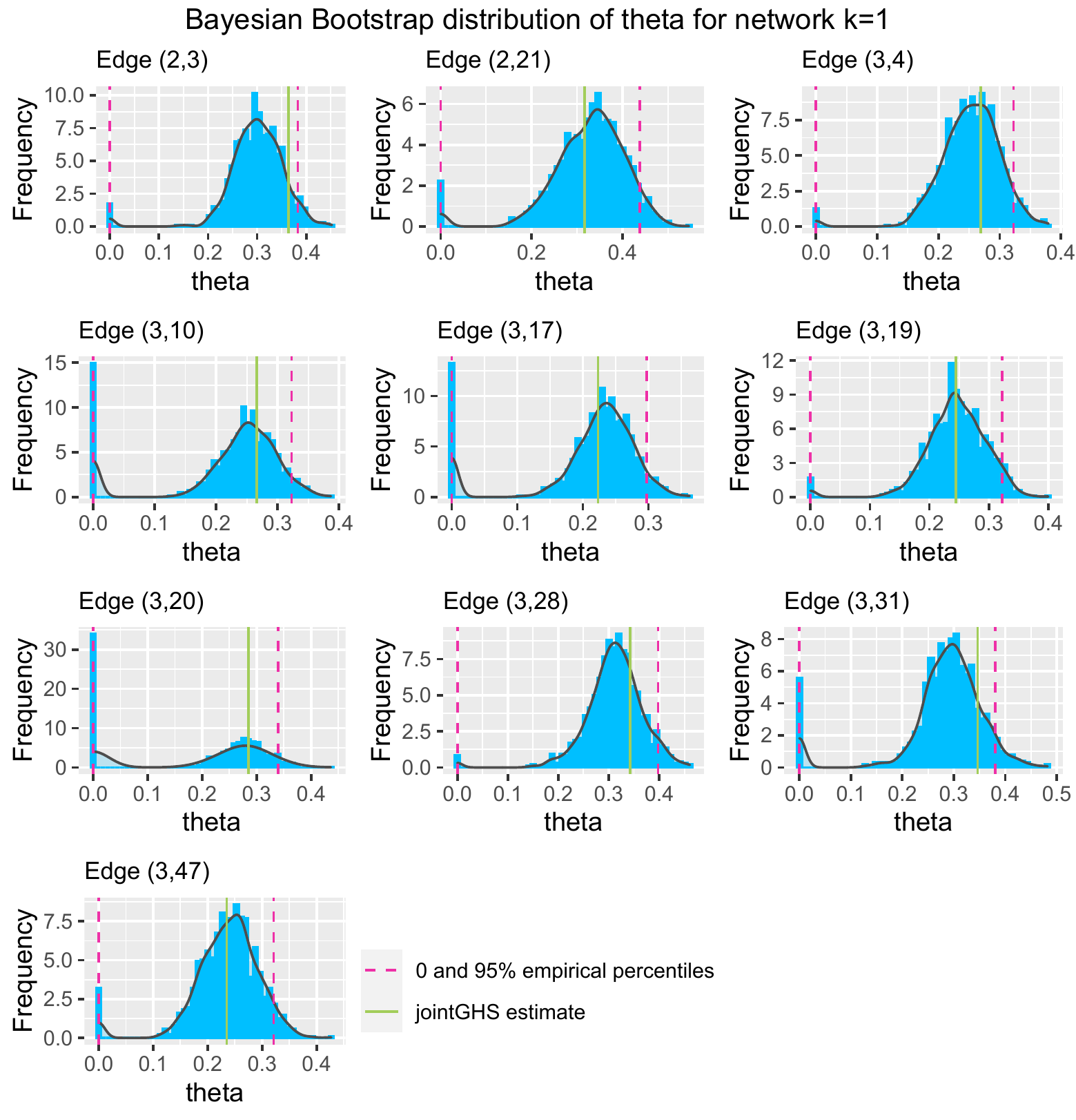}
    \includegraphics[scale=0.405]{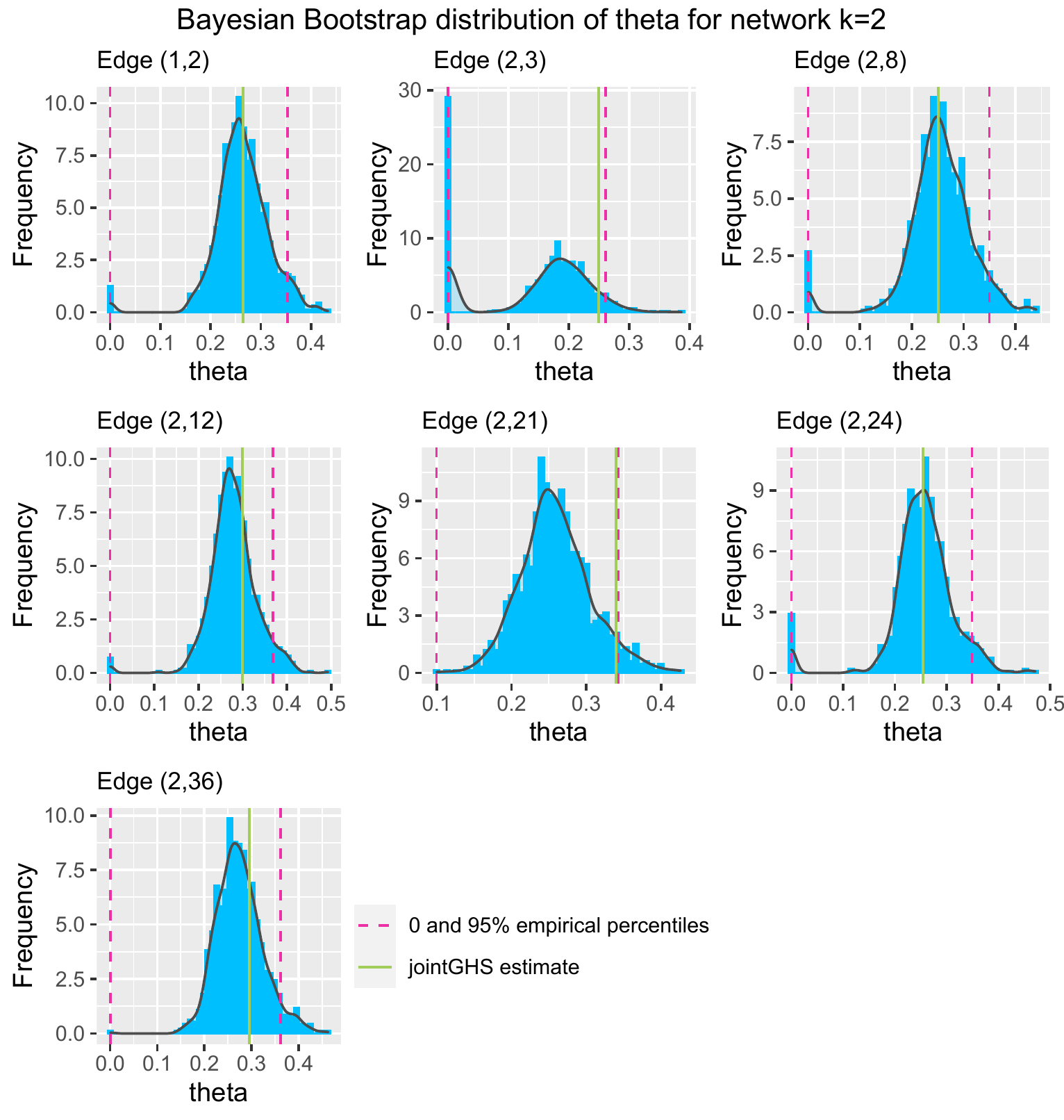}
    \caption{Output of plot function implemented in the jointGHS package, showing results of the Bayesian bootstrap procedure for assessing the suitability of a joint approach for $K=2$ networks with $p=50$ nodes. The two data sets used for the inference have unrelated true network structures, and $n=100$ and $n=200$ observations respectively. The output of the plot function is shown for both networks. For each inferred edge in the joint graphical horseshoe network of a given data set, the plot shows the Bayesian bootstrap distribution of the corresponding scaled precision matrix element of the single-network graphical horseshoe model, with the corresponding empirical $0$ and $95\%$ percentiles shown as dashed lines. The joint graphical horseshoe estimate is shown as a solid line.}
    \label{fig:boot_2diff}
\end{figure}

On the contrary, Figure \ref{fig:boot_2diff} shows the output of the plot function for $K=2$ data sets with completely unrelated network structures. The networks have $p=50$ nodes and the data sets $n=100$ and $n=200$ observations respectively. Once again, we see that there is no conflict between the joint graphical horseshoe estimates and the single-network bootstrap distributions. While this might be surprising given that the two networks are completely unrelated, it is important to note that we have a ``symmetrical'' (dis)similarity pattern and so the common latent $\nu_{ij}^{-1}$ will not be large for edges present in only one network. For this reason, a Bayesian bootstrap check is not necessary for $K=2$ networks.

\subsubsection*{Six networks with one unrelated}
\begin{figure}
    \centering
    \includegraphics[scale=0.6]{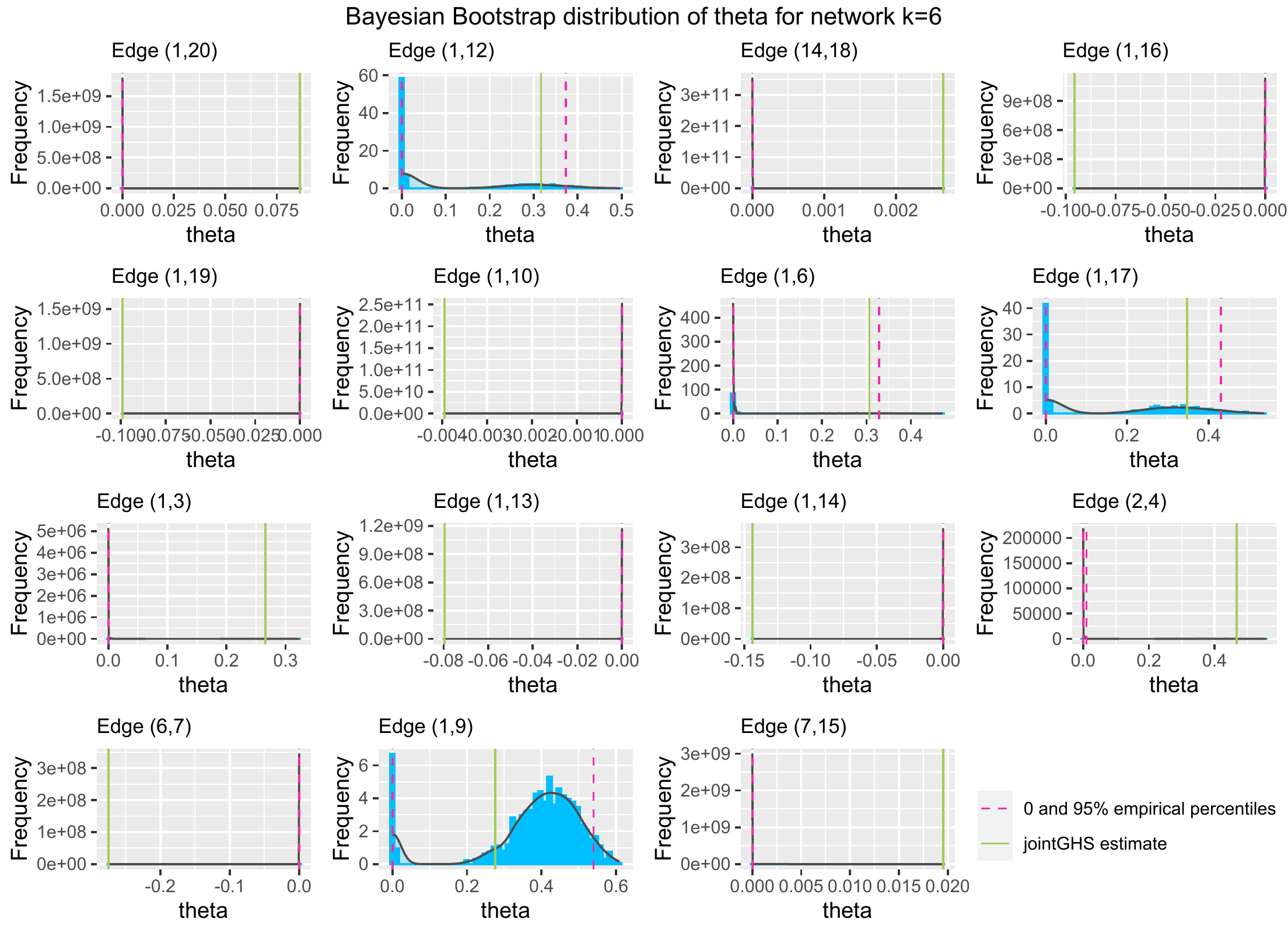}
    \caption{Output of plot function implemented in the jointGHS package, showing results of the Bayesian bootstrap procedure for assessing the suitability of a joint approach for $K=6$ networks with $p=20$ nodes. The first five data sets used for the inference have the same true network structure, and the network structure of the sixth data set in question is completely unrelated to them. All data sets have $n=150$ observations each. The output of the plot function is shown for data set $6$. For each inferred edge in the joint graphical horseshoe network of a given data set, the plot shows the Bayesian bootstrap distribution of the corresponding scaled precision matrix element of the single-network graphical horseshoe model, with the corresponding empirical $0$ and $95\%$ percentiles shown as dashed lines. The joint graphical horseshoe estimate is shown as a solid line.}
    \label{fig:boot_oneout}
\end{figure}

Figure \ref{fig:boot_oneout} shows the output of the plotting function for the last of $K=6$ data sets, where all but the last data set have the same underlying network structure. All networks have $p=20$ nodes and the data sets have $n=150$ observations each.
It is very clear from the plot that most of the scaled precision matrix elements of the inferred edges from the joint approach strongly contradict their single-network bootstrap distributions, exceeding their empirical $95^{\text{th}}$ percentiles in absolute value. Such an output indicates that the user should reconsider using a joint analysis. 

\begin{figure}
    \centering
    \includegraphics[scale=0.4]{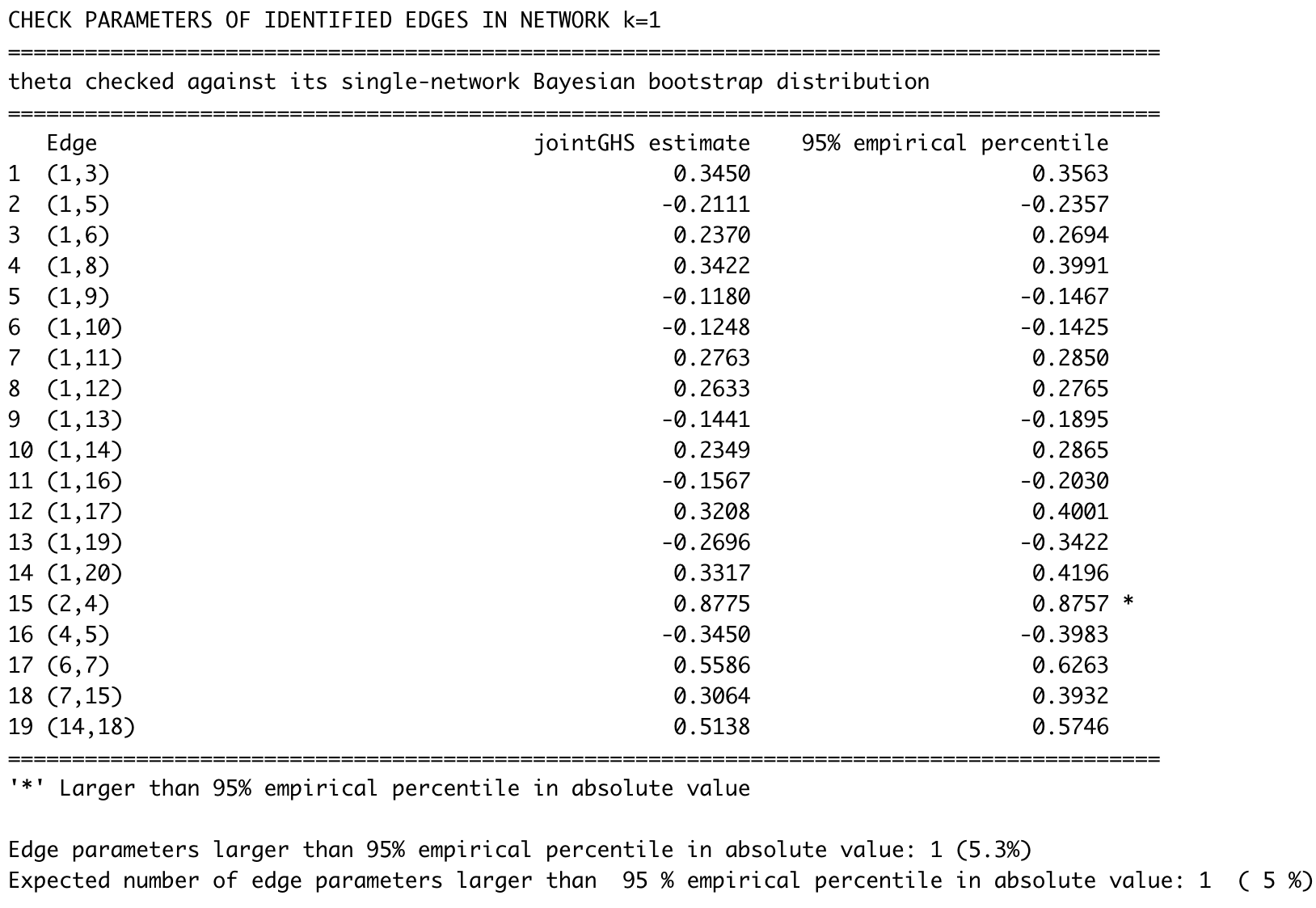}    
    \caption{Output of print function implemented in the jointGHS package, showing results of the Bayesian bootstrap procedure for assessing the suitability of a joint approach for $K=6$ networks with $p=20$ nodes. The first five data sets used for the inference have the same true network structure, and the network structure of the sixth data set in question is completely unrelated to them. All data sets have $n=150$ observations each. The bootstrap results of the first data set is shown, and for each inferred edge in the joint graphical horseshoe network its scaled precision matrix estimate as well as the empirical $95\%$ percentile of the corresponding single-network Bayesian bootstrap distribution is shown. If the estimate from the joint approach exceeds this percentile in absolute value, it is marked by a star.}
    \label{fig:boot_oneout_print_first}
\end{figure}

\begin{figure}
    \centering
 \includegraphics[scale=0.4]{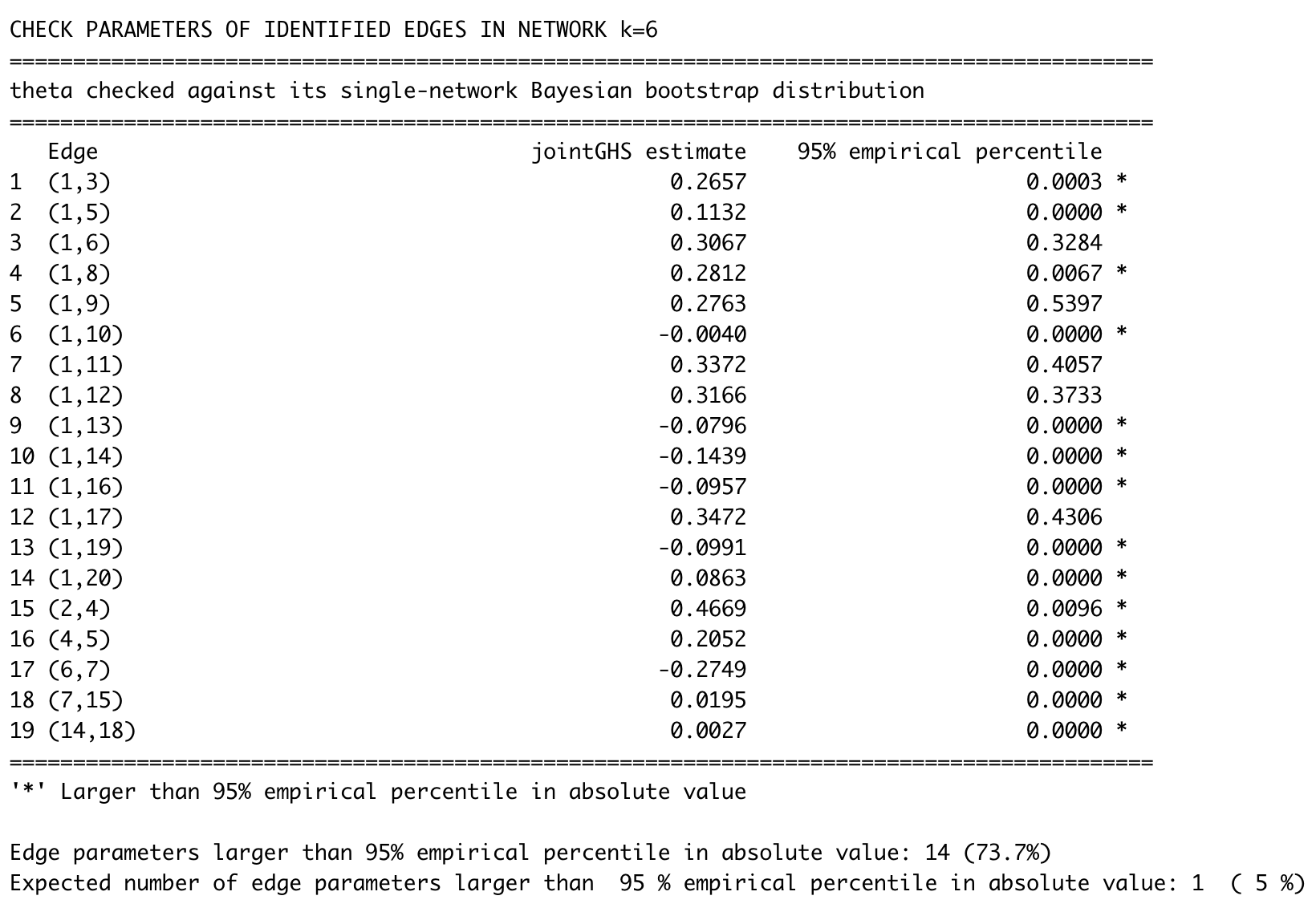}
    \caption{Output of print function implemented in the jointGHS package, showing results of the Bayesian bootstrap procedure for assessing the suitability of a joint approach for $K=6$ networks with $p=20$ nodes. The first five data sets used for the inference have the same true network structure, and the network structure of the sixth data set in question is completely unrelated to them. It is the bootstrap results of this data set that are printed. All data sets have $n=150$ observations each. For each inferred edge in the joint graphical horseshoe network, its scaled precision matrix estimate as well as the empirical $95\%$ percentile of the corresponding single-network Bayesian bootstrap distribution is shown. If the estimate from the joint approach exceeds this percentile in absolute value, it is marked by a star.}
    \label{fig:boot_oneout_print}
\end{figure}

For a summary of the Bayesian bootstrap procedure, the user can also use the print function implemented in the package. The function provides a summary of the findings, including how many edges whose corresponding scaled precision matrix element from the joint analysis exceeds its Bayesian bootstrap empirical $95^{\text{th}}$ percentile in absolute value. Figures \ref{fig:boot_oneout_print_first} and \ref{fig:boot_oneout_print} show the output of the print function for the $1^{\text{st}}$ and the $6^{\text{th}}$ data sets from the previous example respectively. This output indicates that $5.3\%$ of the edges in network $1$ exceeds this percentile, in contrast to $73.7\%$ for network $6$. When using the $95^{\text{th}}$ percentile, we can expect $5\%$ of edges to exceed this threshold. Thus, the output suggests that the joint estimate strongly contradicts the single network estimate, which agrees with the simulated truth, namely that the $6^{\text{th}}$ data set is unrelated to the others.

\section{Additional results}

\subsection{Extended simulation study}

We have repeated the simulation study from the main manuscript with different numbers of networks $K$, nodes $p$ and observations $n$. Table \ref{table:simulationjoint_2} shows the results for $K=2$ classes with $p=100$ nodes and $n_1=100$, $n_2=150$ observations. As for the setting in the main manuscript, we observe that JGL tends to over-select edges severely and thus gets a low precision but high recall. Further, we again observe that SSJGL does not capture network-specific edges even for highly unrelated networks, leading to a decline in performance as network differences increase. Once again, GemBag does not appear to adapt to the similarity level of the networks, as the disagreement level of the inferred GemBag networks remains the same for all cases. Finally, we see that jointGHS adapts well to the varying sparsity levels, with the method performing similar to SSJGL for highly related networks and outperforming it for less similar networks. 

Table \ref{table:simulationjoint_3} shows the results for $K=3$ classes with $p=100$ nodes and $n_1=100$, $n_2=150$ and $n_3=120$ observations. Notably, the average edge disagreement of the estimated networks that is displayed now only quantifies the amount of edges common to \emph{all} $K=3$ networks, not pairs. Hence this number is higher for all methods. Otherwise we observe the same tendencies as in the previous setting, with JGL severely over-selecting edges, GemBag adapting little to varying disagreement levels and SSJGL having excellent performance for identical networks, but declining accuracy for less similar networks. We see that jointGHS appears to adapt well to the varying similarity levels. 

\begin{table}
	\centering
	\renewcommand{\arraystretch}{1.4}
	\caption{Performance of the joint graphical horseshoe, (jointGHS), the spike-and-slab joint graphical lasso (SSJGL), the joint graphical lasso (JGL) and GemBag in simulations, reconstructing $K=2$ graphs with various similarity of the true graph structures. The edge disagreement between the two graphs is shown as the percentage of edges in one network not present in the other. There are $p=100$ nodes in each graph. The results are averaged over $N=100$ simulations, and shows the sparsity, precision and recall for both of the $K=2$ estimated graphs. The standard errors are shown in parentheses. The average edge disagreement of the estimated networks is shown as well. For each case the highest value of the precision is marked in bold, and so is the precision of any other method within one standard error of it. The first graph is reconstructed from $n_1=100$ observations from the corresponding Gaussian graphical distribution, and the second from $n_2=150$ observations. All graphs have true sparsity $0.02$.}
    \hspace*{-2cm}
	\begin{tabular}{r l r @{\hskip -0.1cm} r @{\hskip 0.8cm}l l l r @{\hskip 0.8cm}l l l}
        \multicolumn{9}{c}{ }\\
	    \toprule
	    &&& &\multicolumn{3}{c}{$n_1=100$}&&\multicolumn{3}{c}{$n_2=150$} \\
		\cmidrule(lr){5-7} \cmidrule(lr){9-11}
		Disagr. $\%$ & Method & $\widehat{\text{Disagr.}}$ $\%$ && Sparsity& Precision & Recall&&Sparsity& Precision & Recall \\
		\hline
        0  & JGL & 73 &&  0.201 (0.005)  &  0.08 (0.00)  &  0.79 (0.03) &&  0.161 (0.005)  &  0.11 (0.01)  &  0.85 (0.03) \\ 
        & GemBag & 43 &&  0.020 (0.003)  &  0.60 (0.06)  &  0.60 (0.05) &&  0.034 (0.006)  &  0.42 (0.04)  &  0.71 (0.04) \\ 
        & SSJGL & 0 &&  0.009 (0.001)  &  \textBF{0.90} (0.06)  &  0.42 (0.04) &&  0.009 (0.001)  &  0.90 (0.06)  &  0.42 (0.04) \\ 
        & jointGHS& 14 &&  0.008 (0.000)  &  \textBF{0.86} (0.06)  &  0.35 (0.02) &&  0.008 (0.000)  &  \textBF{0.95} (0.04)  &  0.36 (0.02) \\ 
        \hline 
        20  & JGL & 76 &&  0.196 (0.009)  &  0.08 (0.01)  &  0.80 (0.04) &&  0.151 (0.010)  &  0.11 (0.01)  &  0.81 (0.04) \\ 
        & GemBag & 54 &&  0.021 (0.007)  &  0.54 (0.09)  &  0.54 (0.05) &&  0.040 (0.016)  &  0.37 (0.07)  &  0.68 (0.05) \\ 
        & SSJGL & 0 &&  0.007 (0.001)  &  \textBF{0.91} (0.06)  &  0.32 (0.03) &&  0.007 (0.001)  &  0.92 (0.05)  &  0.32 (0.03) \\ 
        & jointGHS& 29 &&  0.008 (0.001)  &  0.79 (0.07)  &  0.32 (0.02) &&  0.006 (0.000)  &  \textBF{0.99} (0.02)  &  0.29 (0.02) \\ 
        \hline 
        40  & JGL & 77 &&  0.199 (0.007)  &  0.08 (0.01)  &  0.79 (0.04) &&  0.158 (0.007)  &  0.10 (0.01)  &  0.81 (0.03) \\ 
        & GemBag & 51 &&  0.021 (0.006)  &  0.52 (0.07)  &  0.53 (0.04) &&  0.037 (0.011)  &  0.38 (0.06)  &  0.67 (0.05) \\ 
        & SSJGL & 0 &&  0.007 (0.001)  &  \textBF{0.89} (0.06)  &  0.31 (0.03) &&  0.007 (0.001)  &  0.90 (0.06)  &  0.31 (0.03) \\ 
        & jointGHS& 30 &&  0.008 (0.000)  &  0.81 (0.06)  &  0.33 (0.02) &&  0.007 (0.000)  &  \textBF{0.96} (0.03)  &  0.34 (0.02) \\  
        \hline 
        60  & JGL & 79 &&  0.200 (0.007)  &  0.08 (0.01)  &  0.79 (0.04) &&  0.159 (0.005)  &  0.11 (0.01)  &  0.84 (0.03) \\ 
        & GemBag & 45 &&  0.022 (0.003)  &  0.46 (0.05)  &  0.50 (0.04) &&  0.036 (0.006)  &  0.38 (0.04)  &  0.66 (0.03) \\ 
        & SSJGL & 0 &&  0.006 (0.001)  &  \textBF{0.84} (0.07)  &  0.25 (0.03) &&  0.006 (0.001)  &  0.85 (0.06)  &  0.26 (0.03) \\ 
        & jointGHS& 44 &&  0.008 (0.001)  &  \textBF{0.77} (0.07)  &  0.32 (0.03) &&  0.008 (0.000)  & \textBF{0.93} (0.05)  &  0.35 (0.02) \\ 
        \hline 
        80  & JGL & 81 &&  0.200 (0.007)  &  0.08 (0.00)  &  0.79 (0.04) &&  0.156 (0.006)  &  0.11 (0.01)  &  0.84 (0.03) \\ 
        & GemBag & 48 &&  0.022 (0.005)  &  0.41 (0.05)  &  0.45 (0.04) &&  0.037 (0.01)  &  0.36 (0.05)  &  0.65 (0.05) \\
        & SSJGL & 1 &&  0.005 (0.001)  &  \textBF{0.70} (0.08)  &  0.17 (0.02) &&  0.005 (0.001)  &  0.79 (0.08)  &  0.19 (0.02) \\ 
        & jointGHS& 61 &&  0.008 (0.001)  &  \textBF{0.72} (0.06)  &  0.29 (0.02) &&  0.007 (0.000)  &  \textBF{0.94} (0.04)  &  0.32 (0.02) \\ 
        \hline 
        100  & JGL & 82 &&  0.201 (0.005)  &  0.08 (0.00)  &  0.79 (0.04) &&  0.156 (0.004)  &  0.11 (0.00)  &  0.89 (0.03) \\ 
        & GemBag & 45 &&  0.023 (0.002)  &  0.37 (0.04)  &  0.42 (0.04) &&  0.037 (0.002)  &  0.36 (0.03)  &  0.67 (0.04) \\ 
        & SSJGL & 1 &&  0.005 (0.001)  &  0.48 (0.08)  &  0.11 (0.01) &&  0.005 (0.001)  &  0.65 (0.09)  &  0.16 (0.02) \\ 
        & jointGHS& 72 &&  0.008 (0.001)  &  \textBF{0.69} (0.09)  &  0.28 (0.03) &&  0.006 (0.000)  &  \textBF{0.92} (0.05)  &  0.29 (0.02) \\ 
		\toprule
	\end{tabular}
	\label{table:simulationjoint_2}
\end{table} 

\begin{table}
	\centering
	\renewcommand{\arraystretch}{1.5}
    \vspace{-0.8cm}
	\caption{Performance of the joint graphical horseshoe, (jointGHS), the spike-and-slab joint graphical lasso (SSJGL), the joint graphical lasso (JGL) and GemBag in simulations, reconstructing $K=3$ graphs with $p=100$ nodes, with various similarity of the true graph structures. The edge disagreement between the two graphs is shown as the percentage of edges in one network not present in the other. The results are averaged over $N=100$ simulations, and shows the sparsity, precision and recall for the estimated graphs. Standard errors are shown in parentheses. The average edge disagreement of the estimated networks is also shown. For each case the highest value of the precision is in bold, and so is the precision of any other method within one standard error of it. The graphs are reconstructed from $n_1=100$, $n_2=150$ and $n_3=120$ observations. All graphs have true sparsity $0.02$.}
    \vspace{-0.2cm}
    \hspace*{-1.3cm}
    \resizebox{0.93\textwidth}{!}{
    \rotatebox{90}{
	\begin{tabular}{r l r @{\hskip 0cm} r @{\hskip 0.1cm}l l l r @{\hskip 0.1cm}l l l r @{\hskip 0.1cm}l l l}
        \multicolumn{9}{c}{ }\\
	    \toprule
	    &&& &\multicolumn{3}{c}{$n_1=100$}&&\multicolumn{3}{c}{$n_2=150$} &&\multicolumn{3}{c}{$n_2=120$} \\
		\cmidrule(lr){5-7} \cmidrule(lr){9-11} \cmidrule(lr){13-15}
		Disagr. $\%$ & Method & $\widehat{\text{Disagr.}}$ $\%$ && Sparsity& Precision & Recall&&Sparsity& Precision & Recall &&Sparsity& Precision & Recall \\
		\hline
        0  & JGL & 92 &&  0.201 (0.005)  &  0.08 (0.00)  &  0.79 (0.04) &&  0.162 (0.005)  &  0.11 (0.01)  &  0.85 (0.04) &&  0.182 (0.004)  &  0.09 (0.00)  &  0.82 (0.04) \\ 
        & GemBag & 70 &&  0.020 (0.006)  &  0.62 (0.08)  &  0.59 (0.05) &&  0.036 (0.009)  &  0.42 (0.05)  &  0.74 (0.05) &&  0.026 (0.007)  &  0.52 (0.07)  &  0.64 (0.04) \\
        & SSJGL & 33 &&  0.010 (0.001)  &  \textBF{0.98} (0.03)  &  0.49 (0.04) &&  0.010 (0.001)  &  \textBF{0.98} (0.03)  &  0.49 (0.04) &&  0.010 (0.001)  &  \textBF{0.98} (0.03)  &  0.49 (0.04) \\ 
        & jointGHS& 45 &&  0.008 (0.000)  &  0.87 (0.06)  &  0.34 (0.03) &&  0.007 (0.000)  &  \textBF{0.98} (0.03)  &  0.36 (0.02) &&  0.008 (0.000)  &  0.93 (0.05)  &  0.35 (0.02) \\ 
        \hline 
        20  & JGL & 94 &&  0.197 (0.011)  &  0.08 (0.01)  &  0.79 (0.04) &&  0.152 (0.014)  &  0.11 (0.01)  &  0.82 (0.03) &&  0.181 (0.013)  &  0.09 (0.01)  &  0.80 (0.03) \\ 
        & GemBag & 74 &&  0.018 (0.004)  &  0.59 (0.07)  &  0.53 (0.04) &&  0.035 (0.011)  &  0.39 (0.05)  &  0.66 (0.05) &&  0.023 (0.007)  &  0.52 (0.07)  &  0.57 (0.05) \\ 
        & SSJGL & 35 &&  0.008 (0.001)  &  \textBF{0.94} (0.04)  &  0.37 (0.03) &&  0.008 (0.001)  &  0.88 (0.05)  &  0.34 (0.03) &&  0.008 (0.001)  &  0.88 (0.04)  &  0.34 (0.03) \\ 
        & jointGHS& 64 &&  0.008 (0.001)  &  0.80 (0.07)  &  0.32 (0.03) &&  0.006 (0.001)  &  \textBF{0.99} (0.02)  &  0.28 (0.03) &&  0.007 (0.000)  &  \textBF{0.93} (0.04)  &  0.31 (0.02) \\ 
        \hline 
        40  & JGL & 95 &&  0.200 (0.005)  &  0.08 (0.00)  &  0.79 (0.04) &&  0.159 (0.004)  &  0.10 (0.00)  &  0.81 (0.03) &&  0.179 (0.005)  &  0.09 (0.00)  &  0.81 (0.03) \\ 
        & GemBag & 71 &&  0.018 (0.001)  &  0.56 (0.04)  &  0.52 (0.04) &&  0.033 (0.002)  &  0.40 (0.03)  &  0.66 (0.04) &&  0.023 (0.002)  &  0.49 (0.04)  &  0.56 (0.04) \\ 
        & SSJGL & 36 &&  0.008 (0.001)  &  \textBF{0.88} (0.05)  &  0.34 (0.03) &&  0.008 (0.001)  &  0.87 (0.05)  &  0.34 (0.03) &&  0.008 (0.001)  &  \textBF{0.84} (0.05)  &  0.32 (0.03) \\ 
        & jointGHS& 65 &&  0.008 (0.001)  &  0.80 (0.07)  &  0.32 (0.02) &&  0.007 (0.000)  &  \textBF{0.98} (0.02)  &  0.34 (0.02) &&  0.008 (0.000)  &  \textBF{0.87} (0.06)  &  0.34 (0.02) \\ 
        \hline 
        60  & JGL & 97 &&  0.201 (0.005)  &  0.08 (0.00)  &  0.79 (0.04) &&  0.161 (0.004)  &  0.11 (0.00)  &  0.85 (0.03) &&  0.185 (0.004)  &  0.09 (0.00)  &  0.79 (0.04) \\ 
        & GemBag & 72 &&  0.019 (0.002)  &  0.51 (0.05)  &  0.48 (0.04) &&  0.034 (0.002)  &  0.39 (0.03)  &  0.66 (0.04) &&  0.023 (0.002)  &  0.46 (0.03)  &  0.54 (0.03) \\ 
        & SSJGL & 38 &&  0.006 (0.001)  &  \textBF{0.84} (0.07)  &  0.26 (0.03) &&  0.006 (0.001)  &  0.76 (0.07)  &  0.23 (0.02) &&  0.006 (0.001)  &  0.73 (0.06)  &  0.22 (0.03) \\ 
        & jointGHS& 82 &&  0.008 (0.001)  &  0.75 (0.07)  &  0.30 (0.03) &&  0.007 (0.000)  &  \textBF{0.92} (0.04)  &  0.34 (0.02) &&  0.006 (0.000)  & \textBF{ 0.92} (0.04)  &  0.29 (0.02) \\ 
        \hline 
        80  & JGL & 97 &&  0.200 (0.007)  &  0.08 (0.01)  &  0.78 (0.04) &&  0.157 (0.007)  &  0.11 (0.01)  &  0.84 (0.04) &&  0.185 (0.008)  &  0.09 (0.01)  &  0.79 (0.04) \\ 
         & GemBag & 73 &&  0.019 (0.002)  &  0.45 (0.04)  &  0.43 (0.03) &&  0.033 (0.002)  &  0.39 (0.03)  &  0.64 (0.04) &&  0.024 (0.002)  &  0.42 (0.03)  &  0.50 (0.03) \\ 
        & SSJGL & 39 &&  0.005 (0.001)  &  \textBF{0.71} (0.08)  &  0.17 (0.02) &&  0.005 (0.001)  &  0.78 (0.07)  &  0.19 (0.02) &&  0.005 (0.001)  &  0.69 (0.08)  &  0.17 (0.02) \\ 
        & jointGHS& 87 &&  0.008 (0.001)  & \textBF{0.72} (0.06)  &  0.29 (0.02) &&  0.007 (0.000)  &  \textBF{0.95} (0.04)  &  0.32 (0.02) &&  0.005 (0.001)  &  \textBF{0.94} (0.05)  &  0.25 (0.03) \\ 
        \hline 
        100  & JGL & 98 &&  0.200 (0.005)  &  0.08 (0.00)  &  0.78 (0.04) &&  0.156 (0.004)  &  0.11 (0.00)  &  0.89 (0.03) &&  0.178 (0.005)  &  0.10 (0.00)  &  0.88 (0.04) \\ 
        & GemBag & 76 &&  0.025 (0.002)  &  0.37 (0.04)  &  0.45 (0.04) &&  0.039 (0.003)  &  0.34 (0.03)  &  0.67 (0.04) &&  0.032 (0.002)  &  0.38 (0.03)  &  0.61 (0.04) \\ 
        & SSJGL & 38 &&  0.007 (0.001)  &  0.40 (0.07)  &  0.13 (0.02) &&  0.007 (0.001)  &  0.44 (0.06)  &  0.15 (0.02) &&  0.007 (0.001)  &  0.45 (0.06)  &  0.15 (0.02) \\ 
        & jointGHS& 98 &&  0.008 (0.001)  &  \textBF{0.66} (0.07)  &  0.27 (0.03) &&  0.006 (0.000)  &  \textBF{0.93} (0.05)  &  0.29 (0.02) &&  0.006 (0.001)  &  \textBF{0.83} (0.07)  &  0.27 (0.02) \\ 
		\toprule
	\end{tabular}
    }  } 
	\label{table:simulationjoint_3}
\end{table}

\subsection{Threshold-free comparison}
\label{Supp:PRCcurves}

\subsubsection{Precision-recall curves}

To allow for a threshold-free comparison, we have created precision-recall curves. The curves show the precision plotted against the recall, and illustrates the trade-off between the two measures. A challenge when creating these curves is the tendency of the joint graphical horseshoe to do sparse edge selection, where the number of selected edges halts for sufficiently large values of the global shrinkage parameter $\tau$. Further, because the ECM estimator sets precision matrix elements to be either zero (by machine precision) or clearly non-zero, thresholding partial correlation magnitudes does not allow us to exceed the number of edges selected by jointGHS. In our simulations, this means that we can not get jointGHS to exceed recall $0.27-0.30$. To make a fair comparison, we thus create cut-off precision-recall curves. 

To vary the number of selected edges, and thus the precision and recall, we vary the penalty parameter $\lambda_1$ in JGL and the effective penalty parameter $\lambda_1/\nu_0$ in SSJGL, as suggested by \cite{li2019bayesian}. To decrease the number of selected edges in jointGHS, we threshold the estimated partial correlations. In GemBag we vary the threshold for the estimated inclusion probabilities, as suggested by \cite{yang2021gembag}. Notably, with many edges being given the same edge inclusion probabilities, in this simulation GemBag tends to give relatively large recalls ($>0.4$) outside of our comparison range. When we threshold so that only edges with inclusion probabilities equal to 1 are included, we get only one point within the interval with recall around $0.1$. Thresholding by partial correlation magnitude leads to the same problem. Thus, a fair precision-recall curve comparison with GemBag is difficult as it does not allow for sparse enough estimates. We therefore only display the few points corresponding to Gembag in the interval, but do not connect them by line or calculate the area under the curve.

\begin{figure}
    \centering
    \includegraphics[width=\textwidth]{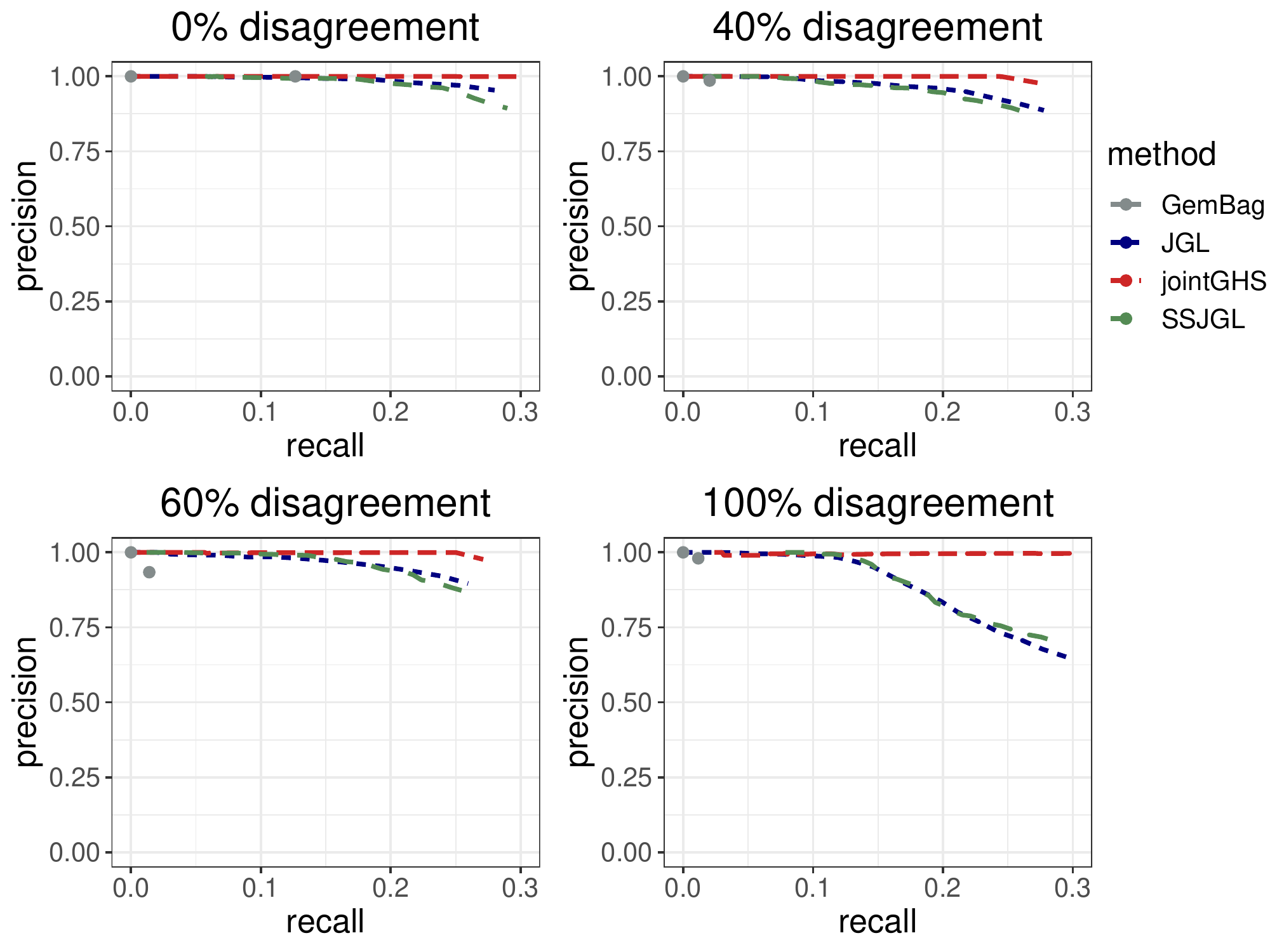}
    \caption{Cut-off precision-recall curves for the joint graphical horseshoe, (jointGHS), the spike-and-slab joint graphical lasso (SSJGL), the joint graphical lasso (JGL) and GemBag in simulations, reconstructing $K=2$ graphs with $p=50$ nodes, with various similarity of the true graph structures. The graphs are reconstructed from $n_1=80$ and $n_2=50$ observations. The edge disagreement between the two graphs is shown as the percentage of edges in one network not present in the other. The results are averaged over $N=100$ simulations and shows the results for the first estimated graph in each setting.}
    \label{fig:PRC_p50}
\end{figure}

\begin{figure}
    \centering
    \includegraphics[width=\textwidth]{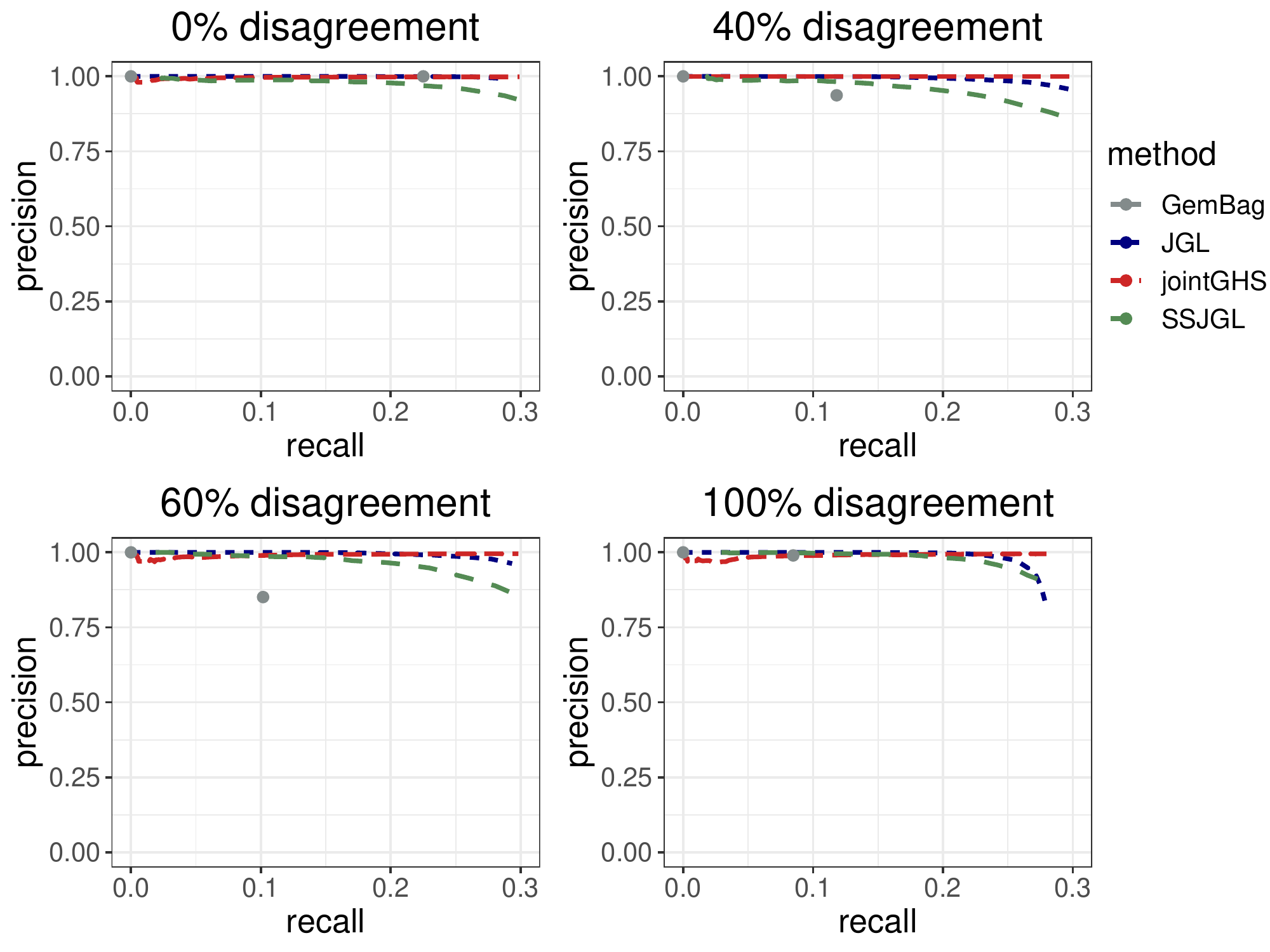}
    \caption{Cut-off precision-recall curves for the joint graphical horseshoe, (jointGHS), the spike-and-slab joint graphical lasso (SSJGL), the joint graphical lasso (JGL) and GemBag in simulations, reconstructing $K=2$ graphs with $p=100$ nodes, with various similarity of the true graph structures. The graphs are reconstructed from $n_1=150$ and $n_2=100$ observations. The edge disagreement between the two graphs is shown as the percentage of edges in one network not present in the other. The results are averaged over $N=100$ simulations and shows the results for the first estimated graph in each setting.}
    \label{fig:PRC_p100}
\end{figure}

We consider the same settings as in our comparative simulation study; all with $K=2$ graphs and either $p=50$ or $p=100$ nodes, with various similarity of the true graph structures. For $p=50$ nodes, we reconstruct networks from $n_1=80$ and $n_2=50$ observations, and for $p=100$ nodes we reconstruct networks from $n_1=15$ and $n_2=100$ observations. The results for $p=50$ nodes are shown in Figure \ref{fig:PRC_p50} and the results for $p=100$ nodes are shown in Figure \ref{fig:PRC_p100}. 

As we see, the precision of the sparse jointGHS estimates remain very high in all cases. The cut-off where the recall is around $0.3$ is exactly when jointGHS does not allow for further edge inclusion as it does not find sufficient evidence for additional edges. This is a benefit the method has in this comparison, and it should be kept in mind when noting that it outperforms the other methods in all cases. It is however evident that if sparse edge selection with very high confidence in the selected edges is desired, jointGHS is the most suitable. The other methods appear to be more suitable if one is more interested in capturing as many of the true edges as possible, and is not as concerned about false positives.

\subsubsection{AUPRC}

We have also calculated the cut-off area under the precision-recall curves (AUPRC), i.e. the area under the curves up to $\text{recall}=0.30$. We denote this the $\text{AUPRC}_{0.3}$. The closer the $\text{AUPRC}_{0.3}$ is to $0.3$, the better the estimation accuracy with fewer false positives. The results are shown in Table \ref{table:auprc}. We observe that the jointGHS had either one of the highest or the highest $\text{AUPRC}_{0.3}$ scores in all settings, with values very close to $0.3$, implying almost no false positives. The other methods also have high scores, though their scores decline more as the network disagreement increases. Again, it is evident that jointGHS is the most suitable for sparse edge selection where high confidence in the selected edges is important, while the others are more suitable for selecting more edges at the cost of lower precision.

\begin{table}
	\centering
	\renewcommand{\arraystretch}{1.4}
	\caption{Cut-off area under the precision-recall curves ($\text{AUPRC}_{0.3}$) up to recall $0.3$ for the joint graphical horseshoe, (jointGHS), the spike-and-slab joint graphical lasso (SSJGL) and the joint graphical lasso (JGL), reconstructing $K=2$ graphs with $p=200$ nodes, with various similarity of the true graph structures. The graphs are reconstructed from $n_1=15$ and $n_2=100$ observations. The edge disagreement between the two graphs is shown as the percentage of edges in one network not present in the other. The results are averaged over $N=100$ simulations and shows the results for the first estimated graph in each setting. For each setting, the highest value, and all values within $\pm0.001$ from it, is shown in bold.}
    \hspace*{-2cm}
	\begin{tabular}{r l r @{\hskip 0.8cm}r @{\hskip 0.8cm}r r }\\
	    \toprule
	    &&& \multicolumn{3}{c}{$\text{AUPRC}_{0.3}$} \\
		\cmidrule(lr){4-6}
		Disagr. $\%$ & Method && $p=50$ && $p=100$ \\
		\hline
        0  & JGL && 0.295  &&  \textBF{0.300} \\ 
        & SSJGL &&  0.292  &&  0.293  \\ 
        & jointGHS &&  \textBF{0.300}  && \textBF{0.299}\\ 
        \hline
        40  & JGL && 0.287  && 0.297  \\ 
        & SSJGL &&  0.286  &&  0.287  \\ 
        & jointGHS &&  \textBF{0.299}  && \textBF{0.300}\\ 
        \hline
        60  & JGL &&  0.285 &&  \textBF{0.298} \\ 
        & SSJGL &&  0.285  &&   0.289 \\ 
        & jointGHS && \textBF{0.299}   && \textBF{0.297}\\ 
        \hline
        100  & JGL && 0.263  &&  0.284 \\ 
        & SSJGL &&  0.267  &&  0.292 \\ 
        & jointGHS &&  \textBF{0.299}  && \textBF{0.296} \\ 
	\toprule
	\end{tabular}
	\label{table:auprc}
\end{table}

\subsection{Scalability}
\label{supp:subsec:scability}

\subsubsection{Scalability of single network inference with fastGHS}

\begin{figure}
    \centering
    \includegraphics[scale=0.4]{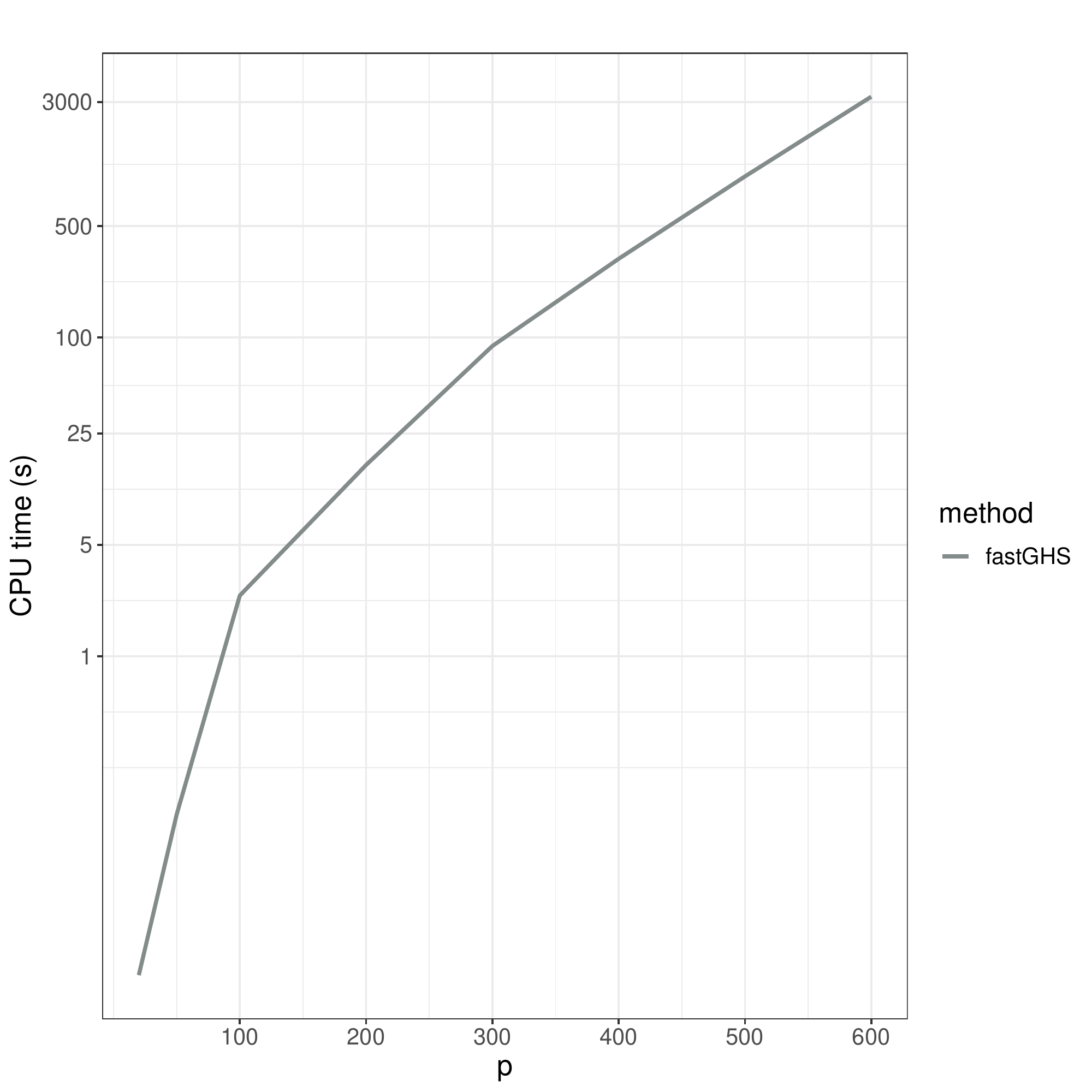}
    \caption{CPU time in seconds on a logarithmic scale used to infer a network for various numbers of nodes $p$ with $n=500$ observations, for our fast ECM implementation of the graphical horseshoe (fastGHS). The computations were performed on a 16-core Intel Xeon CPU, 2.60 GHz.}
    \label{fig:timeplot_large}
\end{figure}

Figure \ref{fig:timeplot_large} shows the CPU time used to infer a network for various numbers of nodes, for fastGHS.  


\subsubsection{Scalability of multiple network inference with jointGHS}

\begin{figure}
    \centering
    \includegraphics[width=0.5\textwidth]{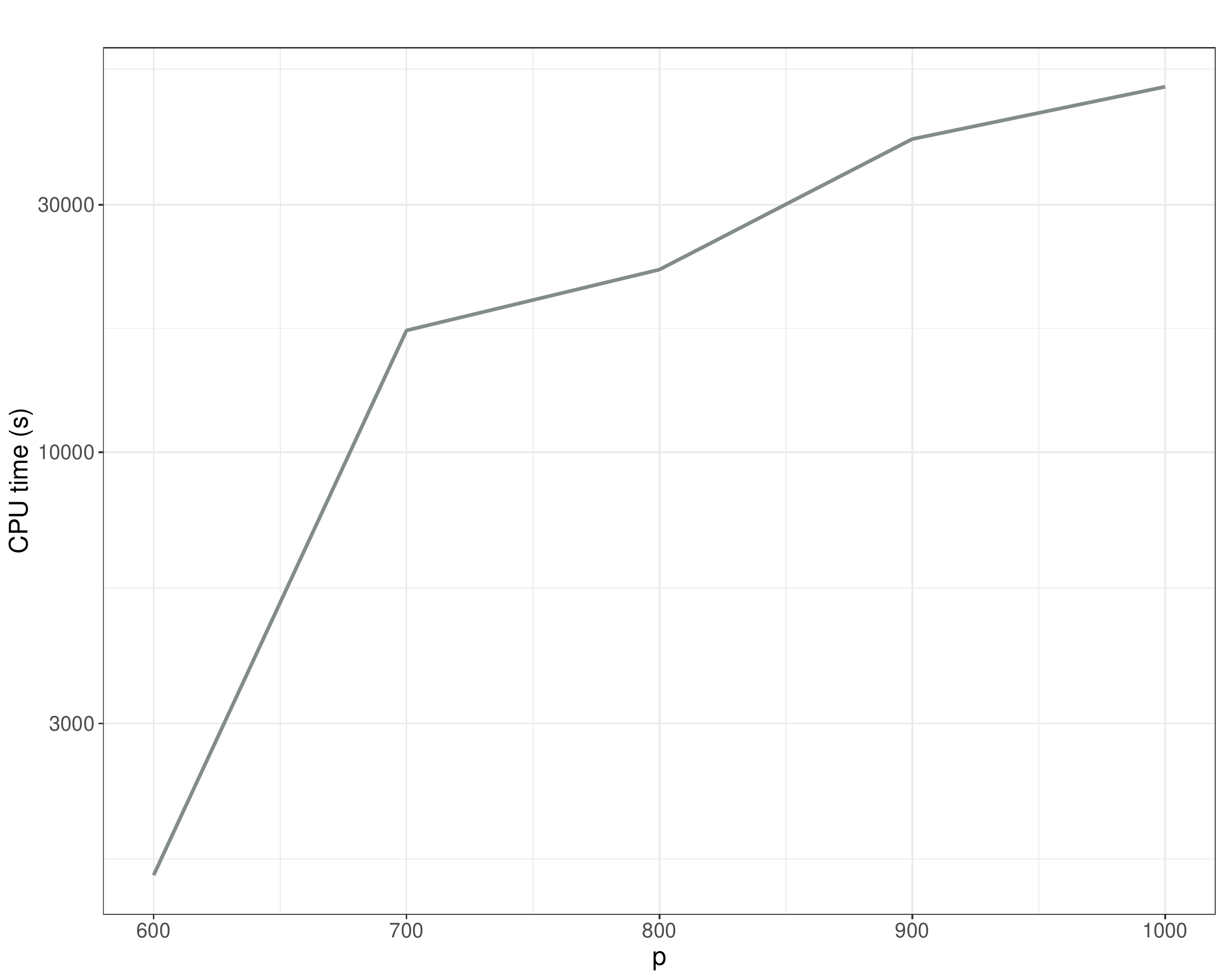}
    \caption{CPU time in seconds on a logarithmic scale used to infer $K=2$ networks for various numbers of nodes $p$ with $n=500$ observations, for our fast ECM implementation of the joint graphical horseshoe (jointGHS). The computations were performed on a 16-core Intel Xeon CPU, 2.60 GHz.}
    \label{fig:timeplot_jointGHS}
\end{figure}

Figure \ref{fig:timeplot_jointGHS} shows the CPU time used to infer $K=2$ networks with jointGHS for various numbers of nodes.


\end{document}